\documentclass[aps,floats]{revtex4-2}
\usepackage{amsmath,amssymb}
\usepackage{graphicx,epsfig}
\usepackage[greek,english]{babel}
\usepackage{bbold}

\begin{document}
\bibliographystyle {plain}

\pdfoutput=1
\def\oppropto{\mathop{\propto}} 
\def\opsimeq{\mathop{\simeq}}
\def\opoverderline{\mathop{\overline}}
\def\operarrow{\mathop{\longrightarrow}}
\def\opsim{\mathop{\sim}}

\def\opmin{\mathop{\min}} 
\def\opmax{\mathop{\max}} 
\def\oplim{\mathop{\lim}}

\title{Large deviations at level 2.5 and for trajectories observables of diffusion processes :  
 the missing parts with respect to their random-walks counterparts } 


\author{C\'ecile Monthus}
\affiliation{Universit\'e Paris-Saclay, CNRS, CEA, Institut de Physique Th\'eorique, 91191 Gif-sur-Yvette, France}


\begin{abstract}
Behind the nice unification provided by the notion of the level 2.5 in the field of large deviations for time-averages over a long Markov trajectory, there are nevertheless very important qualitative differences between the meaning of the level 2.5 for diffusion processes on one hand, and the meaning of the level 2.5 for Markov chains either in discrete-time or in continuous-time on the other hand. In order to analyze these differences in detail, it is thus useful to consider two types of random walks converging towards a given diffusion process in dimension $d$ involving arbitrary space-dependent forces and diffusion coefficients, namely (i) continuous-time random walks on the regular lattice of spacing $b$ ; (ii) discrete-time random walks in continuous space with a small time-step $\tau$. One can then analyze how the large deviations at level 2.5 for these two types of random walks behave in the limits $b \to 0$ and $\tau \to 0$ respectively, in order to describe how the fluctuations of some empirical observables of the random walks are suppressed in the limit of diffusion processes. One can then also study the limits $b \to 0$ and $\tau \to 0$ for any trajectory observable of the random walks that can be decomposed on its empirical density and its empirical flows in order to see how it is projected on the appropriate trajectory observable of the diffusion process involving its empirical density and its empirical current.

\end{abstract}

\maketitle


\section{ Introduction }

\subsection{ Large deviations at various levels for Markov processes }

The theory of large deviations has become the unifying language for statistical physics
 (see the reviews \cite{oono,ellis,review_touchette} and references therein)
 and plays nowadays a major role in the field of nonequilibrium stochastic dynamics
(see the reviews with different scopes \cite{derrida-lecture,harris_Schu,searles,harris,mft,sollich_review,lazarescu_companion,lazarescu_generic,jack_review}, 
the PhD Theses \cite{fortelle_thesis,vivien_thesis,chetrite_thesis,wynants_thesis,chabane_thesis,duBuisson_thesis} 
 and the Habilitation Thesis \cite{chetrite_HDR}).
In particular, the large deviations properties of time-averages over a long Markov trajectory
have attracted a lot of interest in many different contexts 
 \cite{peliti,derrida-lecture,sollich_review,lazarescu_companion,lazarescu_generic,jack_review,vivien_thesis,lecomte_chaotic,lecomte_thermo,lecomte_formalism,lecomte_glass,kristina1,kristina2,jack_ensemble,simon1,simon2,tailleur,simon3,Gunter1,Gunter2,Gunter3,Gunter4,chetrite_canonical,chetrite_conditioned,chetrite_optimal,chetrite_HDR,touchette_circle,touchette_langevin,touchette_occ,touchette_occupation,garrahan_lecture,Vivo,c_ring,c_detailed,chemical,derrida-conditioned,derrida-ring,bertin-conditioned,touchette-reflected,touchette-reflectedbis,c_lyapunov,previousquantum2.5doob,quantum2.5doob,quantum2.5dooblong,c_ruelle,lapolla,c_east,chabane,us_gyrator,duBuisson_gyrator,c_largedevpearson} with the construction of the corresponding Doob's conditioned processes.
Since the rate functions for these time-averaged observables are not always explicit,
an essential idea has been to consider the large deviations at higher levels 
in order to obtain explicit rate functions for Markov processes with arbitrary generators.
In particular, a very important result is that the large deviations at Level 2 concerning
the distribution of the empirical density over a large time-window $T$
are always explicit for equilibrium Markov processes satisfying detailed-balance,
i.e. when the steady currents vanish.
However, for non-equilibrium Markov processes breaking detailed-balance
where the steady currents do not vanish,
the Level 2 for the empirical density alone is usually not explicit, and one needs to introduce
the Level 2.5 concerning the joint distribution of the empirical density and of the empirical flows
in order to be able to write the corresponding explicit rate function.
The most important Markov processes where explicit rate functions at level 2.5 have been much studied
are the discrete-time Markov chains 
\cite{Hollander,fortelle_thesis,fortelle_chain,review_touchette,Polettini,c_largedevdisorder,c_reset,c_inference,carugno,c_microcanoEnsembles},
the continuous-time Markov jump processes with discrete space
\cite{fortelle_thesis,fortelle_jump,maes_canonical,maes_onandbeyond,wynants_thesis,chetrite_formal,BFG1,BFG2,chetrite_HDR,c_ring,c_interactions,c_open,barato_periodic,chetrite_periodic,c_reset,c_inference,c_LargeDevAbsorbing,c_microcanoEnsembles,c_susyboundarydriven},
the diffusion processes in continuous space
\cite{wynants_thesis,maes_diffusion,chetrite_formal,engel,chetrite_HDR,c_lyapunov,c_inference,c_susyboundarydriven}, 
as well as jump-diffusion or jump-drift processes \cite{c_reset,c_runandtumble,c_jumpdrift,c_SkewDB}.
Then the large deviations properties at any lower level can be obtained from the explicit
 level 2.5 via the operation called 'contraction' :
one needs to optimize the Level 2.5 over the empirical observables that one wishes to integrate over,
 in order to see if the rate function for the remaining empirical observables can be explicitly written
 for the specific Markov model one is interested in.
In particular, the level 2.5 is useful to analyze via contraction
the large deviations of any trajectory observable that can be rewritten in terms of empirical density and of the empirical flows.

\subsection{ Goals and organization of the present paper }

However, behind the nice unification provided by the notion of the level 2.5 recalled above,
there are nevertheless very important qualitative differences between the meaning of the level 2.5
for diffusion processes on one hand and the meaning of the level 2.5 for
 Markov chains either in discrete-time or in continuous-time on the other hand : 
 these differneces are recalled in detail in section \ref{sec_motivation}
 in order to better explain the motivations of the present work.
 
Since our goal is to understand these differences,
we will consider two types of random walks converging towards a given diffusion process in dimension $d$ involving arbitrary space-dependent forces and diffusion coefficients, namely :

 (i) continuous-time random walks on the regular lattice of spacing $b$;
 
 (ii) discrete-time random walks in continuous space with a small time-step $\tau$.
 
We will analyze what happens precisely to their level 2.5 and to their trajectory observables
in the diffusion limit $b \to 0$ and $\tau \to 0$ respectively.
For clarity, these two studies are described into two separate parts that can be read independently :

$\bullet$ The lattice regularization of lattice spacing $b$ is analyzed in the main text  
with the following organization:

- In Section \ref{sec_generator}, we introduce
 the continuous-time random walk with only nearest-neighbor jumps
on the regular lattice of spacing $b$ in dimension $d$, with appropriate rates in order to
converge towards a given Fokker-Planck dynamics in the limit of vanishing lattice spacing $b \to 0$.

- In section \ref{sec_level2.5}, we study how the large deviations at level 2.5
for the continuous-time random walk on the lattice that involve the empirical density, the empirical currents and the empirical activities behave in the limit of vanishing lattice spacing $b \to 0$,
i.e. why the empirical activities disappear in the limit diffusion processes :
the main result of this section is the rate function of rescaled empirical observables
 given in Eq. \ref{i2.5latticerescalseriesalpha}.

- In Section \ref{sec_trajobs}, we describe the consequences for the trajectory observables
of the lattice random walk that can be rewritten in terms of
the empirical density, the empirical currents and the empirical activities,
with the main result given in Eq. \ref{observableAJempilatticerescal}.

$\bullet$ The discrete-time regularization of time-step $\tau$ is analyzed in the appendices with the following organization:

- In Appendix \ref{app_PathIntegral}, we describe how the Feynman path-integral for diffusion processes
corresponds to Markov chains with the appropriate kernels for the small time-step $\tau$.

- In Appendix \ref{app_level2.5chain}, we study what happens to the Level 2.5 for the Markov chains 
with time-step $\tau$ in the limit $\tau \to 0$ towards diffusion processes : the main result of this Appendix is the rate function of Eq. \ref{rate2.5empitildewwexprescalfin}.

- In Appendix \ref{app_trajobschain}, we consider arbitrary trajectory observables of the Markov chains
with time-step $\tau$ in order to analyze the limit $\tau \to 0$ towards diffusion processes,
and we describe various examples.


\section{ Large deviations at level 2.5 for Markov chains and for diffusions}

\label{sec_motivation}

In this section, we recall the large deviations properties of Markov chains and of diffusion processes,
in order to stress their important differences that have motivated the present work.


\subsection{ Reminder on large deviations for discrete-time Markov Chain in continuous-space in dimension $d$}

\label{intro_chain}

The discrete-time Markov chain framework is the simplest to explain the essential ideas of large deviations theory \cite{Hollander,fortelle_thesis,fortelle_chain,review_touchette,Polettini,c_largedevdisorder,c_reset,c_inference,carugno,c_microcanoEnsembles}.
While one usually considers also a discrete-space of configurations ,
we will focus here instead on the case of continuous-space in dimension $d$
that will be useful in the Appendices to compare more directly with the limit of diffusion processes.

The evolution of the probability density $\rho_{t}( \vec y)  $ to be at position $\vec y$ at time $t$
\begin{eqnarray}
\rho_{t+\tau}( \vec x) =  \int d^d \vec y  \ W_{\tau}(\vec x, \vec y)  \rho_t( \vec y) 
\label{markovchainc}
\end{eqnarray}
is governed by the Markov kernel $W_{\tau}(\vec x, \vec y)  $ satisfying the normalization
\begin{eqnarray}
   \int d^d \vec x  \ W_{\tau}(\vec x, \vec y)  =1 \ \ \ \text{ for any } \vec y
\label{markovnorma}
\end{eqnarray}
so that the total normalization of the density is preserved
\begin{eqnarray}
\int d^d \vec x \rho_{t+\tau}( \vec x) =  \int d^d \vec y   \rho_t( \vec y) 
\label{markovchaincnormaconserved}
\end{eqnarray}
Whenever the dynamics of Eq. \ref{markovchainc}
converges towards some normalizable steady-state $\rho_*( \vec y) $ satisfying
\begin{eqnarray}
\rho_*( \vec x) =  \int d^d \vec y  \ W_{\tau}(\vec x, \vec y)  \rho_*( \vec y) 
\label{markovchaincsteady}
\end{eqnarray}
it is interesting to analyze 
how time-averaged observables over a large-time-window $[0,T]$ converge towards their steady values,
as recalled in the following subsections.


\subsubsection{Probability of a trajectory $x(0 \leq t \leq T) $ in terms 
of its time-empirical 2-point density $ \rho^{(2)}(\vec x, \vec y) $ with its constraints}

The probability density of the trajectory 
$\vec x(t=0,\tau,2 \tau ..,T= N \tau)$ starting at the fixed position $\vec x(0)$ at time $t=0$
\begin{eqnarray}
{\cal P}[\vec x(t=0,\tau,2 \tau ..,T= N \tau)]  && =  
  \left[ \prod_{n=1}^{N= \frac{T}{\tau}} W_{\tau}(\vec x(n \tau), \vec x(n \tau - \tau)) \right] 
\equiv e^{ \displaystyle - T {\cal I}^{Traj} [x(0 \leq t \leq T) ] }
\label{ptrajmc}
\end{eqnarray}
involves the information ${\cal I} [x(0 \leq t \leq T) ] $ of the trajectory $x(0 \leq t \leq T) $ per unit time
\begin{eqnarray}
{\cal I}^{Traj} [x(0 \leq t \leq T) ]   \equiv  - \frac{\ln {\cal P}[x(t=0,\tau,2 \tau ..,T= N \tau)] }{T} 
&& = - \frac{1}{N \tau } \sum_{n=1}^{N= \frac{T}{\tau}} \ln \left[ W_{\tau}(\vec x(n \tau), \vec x(n \tau - \tau)) \right]
\label{itrajmc}
\end{eqnarray}

The empirical 2-point density $ \rho^{(2)}(\vec x, \vec y) $ that characterizes the 
joint empirical distribution between two consecutive positions 
within this trajectory  $\vec x(0 \leq t \leq T)$
\begin{eqnarray}
  \rho^{(2)}(\vec x, \vec y) && \equiv \frac{1}{N} \sum_{n=1}^{N= \frac{T}{\tau} } \delta^{(d)} ( \vec x(n \tau)- \vec x)\delta^{(d)} ( \vec x(n \tau - \tau)- \vec y)
\label{rho2pt}
\end{eqnarray}
is useful to rewrite the trajectory information ${\cal I}^{traj} [x(0 \leq t \leq T) ]  $ of Eq. \ref{itrajmc}
as a function $ {\cal I}[  \rho^{(2)}(.,.)]  $ of the empirical 2-point density $ \rho^{(2)}(\vec x, \vec y) $ of Eq. \ref{rho2pt}
\begin{eqnarray}
{\cal I}^{traj} [x(0 \leq t \leq T) ]   
= \int d^d \vec x \int d^d \vec y  \rho^{(2)}(\vec x, \vec y) 
\left(  - \frac{\ln \left[W_{\tau}(\vec x, \vec y) \right]}{ \tau } \right)
\ \ \equiv  {\cal I}[  \rho^{(2)}(.,.)] 
\label{itrajempimc}
\end{eqnarray}

The empirical 1-point density
\begin{eqnarray}
\rho(\vec x)  \equiv \frac{1}{N} \sum_{n=1}^N  \delta^{(d)} ( \vec x(n \tau)- \vec x)
\label{rho1pt}
\end{eqnarray}
satisfying the normalization
\begin{eqnarray}
\int d^d \vec x \rho(\vec x)  = 1
\label{rho1ptnorma}
\end{eqnarray}
can be reconstructed from the integration of
the 2-point density of Eq. \ref{rho2pt} over the first position $\vec y$
\begin{eqnarray}
\int d^d \vec y \rho^{(2)}(\vec x, \vec y) 
= \frac{1}{N} \sum_{n=1}^N  \delta^{(d)} ( \vec x(n \tau)- \vec x) = \rho(\vec x) 
\label{rho2ptovery}
\end{eqnarray}
or via the integration over the second position $\vec x$
\begin{eqnarray}
\int d^d \vec x \rho^{(2)}(\vec x, \vec y) = \frac{1}{N} \sum_{n=1}^N  \delta^{(d)} ( \vec x(n \tau-\tau)- \vec y)
= \rho(\vec y) + \frac{\delta^{(d)} ( \vec x(0)- \vec y) - \delta^{(d)} ( \vec x(N \tau)- \vec y) }{N \tau } \opsimeq  \rho(\vec y) 
\label{rho2ptoverx}
\end{eqnarray}
up to a boundary term of order $1/T$ that is negligible for large duration $T \to +\infty$.


\subsubsection{ Number $  {\cal N}^{[2.5]}_T\left[ \rho^{(2)}(.,.) \right]$ of trajectories $\vec x(0 \leq t \leq T)$ of duration $T$ with a given empirical 2-point density $  \rho^{(2)}(.,.) $ }

Since all the trajectories $ x(0 \leq t \leq T)  $
 that have the same empirical 2-point density $  \rho^{(2)}(.,.) $
  have the same probability,
  the normalization over all possible trajectories 
can be rewritten as an integral over the empirical 2-point density $  \rho^{(2)}(.,.) $
\begin{eqnarray}
1&& = \int {\cal D} x(0 \leq t \leq T)  {\cal P}[x(0 \leq t \leq T)] 
=  \int {\cal D} x(0 \leq t \leq T)e^{\displaystyle  -T  {\cal I}^{Traj} \left[ x(0 \leq t \leq T) \right] }
\nonumber \\
&& = \int {\cal D}\rho^{(2)}(.,.)  {\cal N}^{[2.5]}_T\left[ \rho^{(2)}(.,.)\right] 
e^{\displaystyle  -T  {\cal I} \left[ \rho^{(2)}(.,.) \right] }
\label{normaempi}
\end{eqnarray}
where the number $  {\cal N}^{[2.5]}_T\left[ \rho^{(2)}(.,.) \right]$ of trajectories of duration $T$ 
associated to a given empirical 2-point density $  \rho^{(2)}(.,.) $
will grow exponentially with respect to the time-window $T$ in the limit $T \to +\infty$
\begin{eqnarray}
{\cal N}^{[2.5]}_T \left[ \rho^{(2)}(.,.) \right]  \opsimeq_{T \to +\infty} 
 C^{[2.5]}\left[ \rho^{(2)}(.,.) \right] \ e^{\displaystyle T S^{[2.5]}\left[ \rho^{(2)}(.,.) \right]  }
\label{omegat}
\end{eqnarray}
The prefactor $C^{[2.5]}\left[\rho^{(2)}(.,.) \right]$ 
summarizes the constitutive constraints of Eqs \ref{rho1ptnorma} \ref{rho2ptovery} \ref{rho2ptoverx}
for the empirical 2-point density 
\begin{eqnarray}
C^{[2.5]}\left[ \rho^{(2)}(.,.) \right]
  = \delta \left( \int d^d \vec x \int d^d \vec y \rho^{(2)}(\vec x, \vec y) -1 \right) 
 \prod_x \delta \left( \int d^d \vec y [\rho^{(2)}(\vec x, \vec y) - \rho^{(2)}(\vec y, \vec x)] \right)
\label{constraints2.5masteralone}
\end{eqnarray}
or more explicitly using the empirical 1-point density as intermediate
\begin{eqnarray}
 C_{2.5}  [  \rho(.) ; \rho^{(2)}(.,.) ]
  = \delta \left( \int d^d \vec x \rho(\vec x)  - 1 \right) 
  \left[ \prod_{\vec x} \delta \left(   \int d^d \vec y \rho^{(2)}(\vec x, \vec y) - \rho(\vec x) \right) \right]
 \left[\prod_{\vec y} \delta \left(   \int d^d \vec x \rho^{(2)}(\vec x, \vec y) - \rho(\vec y) \right)  \right]
\label{constraints2.5chain}
\end{eqnarray}
The function $S^{[2.5]}\left[ \rho^{(2)}(.,.) \right]  = \frac{\ln {\cal N}^{[2.5]}_T\left[ \left[ \rho^{(2)}(.,.) \right] \right] }{ T }  $ represents the 
Boltzmann intensive entropy of the set of trajectories of long duration $T$ with given empirical 2-point density $ \rho^{(2)}(.,.)$.
This entropy can be evaluated without any combinatorial computation 
from the normalization of Eq. \ref{normaempi} for large $T$ using Eq. \ref{omegat}
\begin{eqnarray}
1  \opsimeq_{T \to +\infty} 
\int {\cal D}\rho^{(2)}(.,.) C^{[2.5]}\left[ \rho^{(2)}(.,.) \right] 
e^{\displaystyle  T \left(S^{[2.5]}\left[ \rho^{(2)}(.,.) \right] - {\cal I} \left[ \rho^{(2)}(.,.) \right] \right) }
\label{normaempit}
\end{eqnarray}
When the empirical 2-point density $ \rho^{(2)}(.,.)$ 
takes its steady value $\rho^{(2)}_*(.,.) $ associated to the steady state $\rho_*(.)  $ of Eq. \ref{markovchaincsteady}
\begin{eqnarray}
\rho^{(2)}_*(\vec x, \vec y)  =W_{\tau}(\vec x, \vec y)  \rho_*( \vec y) 
\label{rho2typ}
\end{eqnarray}
then the exponential behavior in $T$ of Eq. \ref{normaempit}
should exactly vanish,
i.e. the entropy $S^{[2.5]}\left[ \rho^{(2)}_*(.,.) \right] $ should exactly compensate the 
information ${\cal I} \left[ \rho^{(2)}_*(.,.) \right] $ of Eq. \ref{itrajempimc}
\begin{eqnarray}
S^{[2.5]}\left[ \rho^{(2)}_*(.,.) \right]  =  {\cal I}[  \rho^{(2)}_*(.,.)] 
&&= - \frac{1}{ \tau } \int d^d \vec x \int d^d \vec y  \rho^{(2)}_*(\vec x, \vec y) \ln \left[W_{\tau}(\vec x, \vec y) \right]
\nonumber \\
&& = \int d^d \vec y \rho_*( \vec y)
\left(  - \frac{1}{ \tau } \int d^d \vec x  W_{\tau}(\vec x, \vec y)    \ln \left[W_{\tau}(\vec x, \vec y) \right] \right) \equiv h_{KS}
\label{itrajempimctyp}
\end{eqnarray}
which corresponds to the Kolmogorov-Sinai entropy $h_{KS}$ of the trajectories :
the normalization of Eq. \ref{normaempit} is dominated by a number $e^{T S^{[2.5]}\left[ \rho^{(2)}_*(.,.) \right]} =e^{T h_{KS}}$ of trajectories
that all have the same probability $e^{- T  {\cal I}[  \rho^{(2)}_*(.,.)] } = e^{-T h_{KS}} $.

To obtain the entropy $S^{[2.5]}\left[ \rho^{(2)}(.,.) \right]  $ for any other value of the empirical 2-point density $\rho^{(2)}(.,.) $,
one just needs to introduce the modified Markov kernel ${\tilde W}(.,.)$ that would  
have the empirical 2-point density $\rho^{(2)}(.,.)$ as steady value via Eq. \ref{rho2typ}
\begin{eqnarray}
{\tilde W}_{\tau}(\vec x, \vec y) && \equiv \frac{ \rho^{(2)}(\vec x, \vec y)  }{\rho( \vec y)  } 
\ \ \ {\rm where } \ \ \rho( \vec y)=\int d^d \vec x  \rho^{(2)}(\vec x, \vec y)
\label{modeleeff}
\end{eqnarray}
and to use Eq. \ref{itrajempimctyp} for this modified Markov generator kernel ${\tilde W}(.,.)$ to obtain
the entropy for arbitrary $\rho^{(2)}(.,.) $
\begin{eqnarray}
S^{[2.5]}\left[ \rho^{(2)}(.,.) \right]  && = 
 - \frac{1}{ \tau } \int d^d \vec x \int d^d \vec y  \rho^{(2)}(\vec x, \vec y) \ln \left[{\tilde W}_{\tau}(\vec x, \vec y) \right]
 =  - \frac{1}{ \tau } \int d^d \vec x \int d^d \vec y  \rho^{(2)}(\vec x, \vec y) \ln \left[\frac{ \rho^{(2)}(\vec x, \vec y)  }{\rho( \vec y)  }  \right]
 \nonumber \\
 && =  \frac{1}{ \tau } \left(  \int d^d \vec y \rho( \vec y)\ln \left[\rho( \vec y) \right]
 - \int d^d \vec x \int d^d \vec y  \rho^{(2)}(\vec x, \vec y) \ln \left[ \rho^{(2)}(\vec x, \vec y)    \right] \right)
\label{entropyempi}
\end{eqnarray}
where the last rewriting involves the Shannon entropies 
associated to the empirical 1-point density $\rho( .) $ 
and the 2-point density $ \rho^{(2)}(., .)$ respectively.


\subsubsection{ Large deviations at level 2.5 for the probability distribution $P_T^{[2.5]} [   \rho^{(2)}(.,.) ] $ of the empirical 2-point density $\rho^{(2)}(.,.) $  }

Eq \ref{normaempi} can be interpreted as the normalization
 \begin{eqnarray}
 1  = \int {\cal D} \rho^{(2)}(.,.) P^{[2.5]}_T \left[ \rho^{(2)}(.,.) \right]
\label{normaprobaempi}
\end{eqnarray}
for the probability distribution $P_T^{[2.5]} [  \rho^{(2)}(.,.) ]  $ of the empirical 2-point density $\rho^{(2)}(.,.) $
involving Eq. \ref{omegat}
\begin{eqnarray}
P^{[2.5]}_T \left[ \rho^{(2)}(.,.) \right] &&
\equiv  {\cal N}^{[2.5]}_T\left[ \rho^{(2)}(.,.)\right] e^{\displaystyle  -T  {\cal I} \left[ \rho^{(2)}(.,.) \right] }
\nonumber \\
&& \opsimeq_{T \to + \infty}
 C^{[2.5]}\left[ \rho^{(2)}(.,.) \right] \ e^{\displaystyle T \left( S^{[2.5]}\left[ \rho^{(2)}(.,.) \right] - {\cal I} \left[ \rho^{(2)}(.,.) \right] \right)}
 \equiv C^{[2.5]}\left[ \rho^{(2)}(.,.) \right] \ e^{\displaystyle -T  I_{2.5}\left[ \rho^{(2)}(.,.) \right] }
\label{pnormaempi}
\end{eqnarray}
where the rate function $I_{2.5}\left[ \rho^{(2)}(.,.) \right] $ at level 2.5 is simply the difference
\begin{eqnarray}
  I_{2.5} [ \rho^{(2)}(.,.) ]
  = {\cal I} \left[ \rho^{(2)}(.,.) \right] - S^{[2.5]}\left[ \rho^{(2)}(.,.) \right]
\label{rate2.5chaindiff}
\end{eqnarray}
between the trajectory information ${\cal I} \left[ \rho^{(2)}(.,.) \right] $ 
and the entropy $S^{[2.5]}\left[ \rho^{(2)}(.,.) \right] $ of trajectories associated to $ \rho^{(2)}(.,.) $.
The difference of Eq. \ref{rate2.5chaindiff}
can be rewritten in the compact form
\begin{eqnarray}
  I_{2.5} [ \rho(.), \rho^{(2)}(.,.) ]
  =    \frac{1}{ \tau }
 \int d^d \vec x \int d^d \vec y \rho^{(2)}(\vec x, \vec y) \ln \left( \frac{\rho^{(2)}(\vec x, \vec y)}{ W_{\tau}(\vec x, \vec y)  \rho(\vec y)}  \right) 
\label{rate2.5chain}
\end{eqnarray}
that makes obvious the vanishing for the steady values  $[ \rho_*(.), \rho^{(2)}_*(.,.) ]$ of Eq. \ref{rho2typ}.


These explicit large deviations at Level 2.5 
allow to analyze via contraction all the large deviations properties of lower levels,
but these contraction are not always explicit.
For instance, the level 2 concerning the distribution
of the empirical density $\rho(.) $ alone
can be obtained from the integration over Eq. \ref{pnormaempi} 
over the empirical 2-point density $\rho^{(2)}(.,.)$
using the constraints of Eq. \ref{constraints2.5chain}
\begin{eqnarray}
 P^{[2]}_T \left[  \rho(.)  \right] 
 && \opsimeq_{T \to + \infty} 
  \delta \left( \int d^d \vec x \rho(\vec x)  - 1 \right) 
\nonumber \\
&&  \int {\cal D} \rho^{(2)}(.,.)  
  \left[ \prod_{\vec x} \delta \left(   \int d^d \vec y \rho^{(2)}(\vec x, \vec y) - \rho(\vec x) \right) \right]
 \left[\prod_{\vec y} \delta \left(   \int d^d \vec x \rho^{(2)}(\vec x, \vec y) - \rho(\vec y) \right)  \right]
 e^{\displaystyle -T  I_{2.5}\left[ \rho^{(2)}(.,.) \right] }
\nonumber \\
  &&
 \opsimeq_{T \to + \infty}
 \delta \left( \int d^d \vec x \rho(\vec x)  - 1 \right)   e^{\displaystyle -T  I_{2}\left[  \rho(.)   \right] }
\label{p2rho}
\end{eqnarray}
so that the rate function $I_{2}\left[  \rho(.)   \right] $ at level 2 for the empirical density alone
corresponds to the optimization of the rate function $I_{2.5}\left[ \rho^{(2)}(.,.) \right] $ at level 2.5 
over the empirical 2-point density $\rho^{(2)}(.,.) $ satisfying the appropriate constraints.
One can also analyze via contraction the statistics of trajectory observables as described in the next subsection.


\subsubsection{ Trajectory observables that can be rewritten in terms of the empirical 2-point density $ \rho^{(2)}(.,.) $   }

An observable ${\cal O}^{Traj} [x(0 \leq t \leq T) ]$ of the trajectory $x(t=0,\tau,2 \tau ..,T= N \tau) $
that can be parametrized by a function $\Omega^{(2)}(\vec x , \vec y)$
\begin{eqnarray}
{\cal O}^{Traj}\left[ x(0 \leq t \leq T) \right]  
=   \frac{1}{N  } \sum_{n=1}^{N= \frac{T}{\tau}} \Omega^{(2)}\left(\vec x(n \tau), \vec x(n \tau - \tau)\right)
\label{observablerho2}
\end{eqnarray}
can be rewritten as a function $ {\cal O} \left[ \rho^{(2)}(.,.) \right] $
 of the empirical 2-point density ${\cal O} \left[ \rho^{(2)}(.,.) \right]  $ of Eq. \ref{rho2pt}
 as
\begin{eqnarray}
{\cal O}^{Traj}\left[ x(0 \leq t \leq T) \right] 
 =   \int d^d \vec x  \int d^d \vec y \Omega^{(2)}(\vec x , \vec y) \rho^{(2)}(\vec x , \vec y) 
\equiv  {\cal O} \left[ \rho^{(2)}(.,.) \right]
\label{observablerho2empi}
\end{eqnarray}

The probability distribution $P_T({\cal O}) $ of the trajectory observable
 can be evaluated from 
  the probability $P^{[2.5]}_T \left[  \rho^{(2)}(.,.)  \right]  $ at the Level 2.5 of Eq. \ref{pnormaempi}
\begin{eqnarray}
P_T({\cal O})  && =\int {\cal D}  \rho^{(2)} (.,.)
P^{[2.5]}_T \left[  \rho^{(2)} (.,.) \right]
\delta \left( {\cal O} - {\cal O} \left[  \rho^{(2)}(.,.)  \right] \right)
\nonumber \\
&& \opsimeq_{T \to +\infty} 
\int {\cal D}   \rho^{(2)}(.,.)
C^{[2.5]}\left[  \rho^{(2)}(.,.) \right] 
\delta \left( {\cal O} - {\cal O} \left[  \rho^{(2)} (.,.) \right] \right)
e^{- T I^{[2.5]}\left[  \rho^{(2)} (.,.)\right] }
\nonumber \\
&& \opsimeq_{T \to +\infty} e^{- T I({\cal O})}
\label{largedevadditiveproba}
\end{eqnarray}
so that the rate function $I({\cal O})$ for the trajectory observable ${\cal O} $
corresponds to the optimization of the rate function $I^{[2.5]}\left[ {\cal O} \right]  $
over the empirical 2-point density $\rho^{(2)} (.,.) $ 
satisfying the constitutive constraints $C^{[2.5]}\left[ \rho^{(2)} (.,.) \right]  $
and the supplementary constraint $\delta \left( {\cal O} - {\cal O} \left[ \rho^{(2)} (.,.)  \right] \right)$ reproducing the correct value ${\cal O}$ of the trajectory observable.


\subsection{ Reminder on large deviations for continuous-time Markov jump processes in discrete space }

\label{intro_jump}

In this subsection, we recall the large deviations for continuous-time Markov jump processes in discrete space
\cite{fortelle_thesis,fortelle_jump,maes_canonical,maes_onandbeyond,wynants_thesis,chetrite_formal,BFG1,BFG2,chetrite_HDR,c_ring,c_interactions,c_open,barato_periodic,chetrite_periodic,c_reset,c_inference,c_LargeDevAbsorbing,c_microcanoEnsembles,c_susyboundarydriven}.
Since we have already explained the main ideas in the previous subsection
concerning discrete-time Markov chains,
we will not repeat all the arguments, but we will stress the technical changes
in the various observables.

The evolution for the probability $P_t(x) $ to be at configuration $x$ at time $t$
\begin{eqnarray}
\partial_t P_t(x) =    \sum_{y }   w(x,y)  P_t(y) 
\label{mastereq}
\end{eqnarray}
is governed by the Markov matrix $w(x,y) $ :
the off-diagonal $x \ne y$ positive matrix element $w(x,y) \geq 0 $  represents the jump rate 
per unit time from configuration $y$ towards the other configuration $x \ne y$,
while the negative diagonal elements can be computed in terms of the off-diagonal elements  
\begin{eqnarray}
w(y,y)   =  - \sum_{x \ne y} w(x,y) 
\label{wdiag}
\end{eqnarray}
in order to conserve the normalization of the total probability
\begin{eqnarray}
\partial_t \left[ \sum_x P_t(x) \right] =  0
\label{markovchaincnormaconservedj}
\end{eqnarray}
Whenever the dynamics converges towards some normalizable steady-state $P_*(y)$ 
satisfying Eq. \ref{mastereq}
\begin{eqnarray}
0 = \partial_t P_*(x)=    \sum_{y }   w(x,y)  P_*(y) 
\label{mastereqst}
\end{eqnarray}
it is interesting to analyze 
how time-averaged observables over a large-time-window $[0,T]$ converge towards their steady values,
as recalled in the following subsection.


\subsubsection{Probability of a trajectory $x(0 \leq t \leq T) $ in terms of its empirical probability $P(.) $ and its empirical flows $  Q(.,.)  $} 

The probability ${\cal P }\left[ x(0 \leq t \leq T) \right]  $ of the trajectory $x(0 \leq t \leq T) $
\begin{eqnarray}
{\cal P }\left[ x(0 \leq t \leq T) \right]  
 \equiv e^{ \displaystyle  - T {\cal I}^{Traj}\left[ x(0 \leq t \leq T) \right]}
\label{pwtrajjump}
\end{eqnarray}
involves the information ${\cal I}^{Traj}\left[ x(0 \leq t \leq T) \right] $ of the trajectory $x(0 \leq t \leq T) $ per unit time 
that can be written as follows in terms of the Markov matrix $w(.,.)$
\begin{eqnarray}
{\cal I}^{Traj}\left[ x(0 \leq t \leq T) \right] && \equiv - \frac{ \ln \left( {\cal P}\left[ x(0 \leq t \leq T) \right] \right) }{T}
\nonumber \\
&&=  - \frac{1}{T}  \int_0^T dt  w(x(t) , x(t) )  
 - \frac{1}{T} \sum_{t \in [0,T] : x(t^+) \ne x(t^-) } \ln ( w(x(t^+) , x(t^-) ) ) 
\label{informationjump}
\end{eqnarray}

So here it is useful to introduce the following relevant empirical observables :

(a) the empirical probability
\begin{eqnarray}
  P(x)  \equiv \frac{1}{T} \int_0^T dt \  \delta_{x(t),x}  
 \label{rho1pj}
\end{eqnarray}
measures the fraction of the time spent by the trajectory $x(0 \leq t \leq T) $ in each 
discrete configuration $x$ 
with the normalization to unity
\begin{eqnarray}
\sum_x  P(x)   = 1
\label{rho1ptnormaj}
\end{eqnarray}

(b) the empirical flow 
\begin{eqnarray}
Q(x,y) \equiv  \frac{1}{T} \sum_{t \in [0,T] : x(t^+) \ne x(t^-)} \delta_{x(t^+),x} \delta_{x(t^-),y} 
\label{jumpempiricaldensity}
\end{eqnarray}
represents the density of jumps
from the configuration $y$ towards the other configuration $x \ne y$ 
seen during the trajectory $x(0 \leq t \leq T) $.
For large $T$, these empirical flows satisfy the following stationarity constraints : 
for any configuration $x$, the total incoming flow into $x$ is equal to the total outgoing flow out of $x$
(up to boundary terms of order $1/T$ that involve only the initial configuration at $t=0$ and the final configuration at time $t=T$)
\begin{eqnarray}
\sum_{y \ne x} Q(x,y)= \sum_{y \ne x} Q(y,x) \ \ \ \text{ for any $x$}
\label{contrainteq}
\end{eqnarray}

The trajectory information ${\cal I}^{Traj}\left[ x(0 \leq t \leq T) \right]  $ of Eq. \ref{informationjump}
can then be written as the following function $ {\cal I} \left[P(.)  ;Q(.,.) \right] $
of the empirical density $P(.)$ of Eq. \ref{rho1pj}
and the empirical flows $Q(.,.)$ of Eq. \ref{jumpempiricaldensity}
\begin{eqnarray}
{\cal I}^{Traj}\left[ x(0 \leq t \leq T) \right] 
&& = - \sum_{y  } P(y) w(y,y)  -       \sum_{y  }\sum_{ x \ne y  }Q(x,y)  \ln ( w(x,y) )     
\nonumber \\
&& =  \sum_{y  }P(y) \sum_{ x \ne y  }  w(x,y)  -       \sum_{y  }\sum_{ x \ne y  }Q(x,y)  \ln ( w(x,y) )  
\equiv  {\cal I} \left[ P(.)  ; Q(.,.) \right]
\label{informationjumpempi}
\end{eqnarray}


\subsubsection{Replacing the empirical flows $  Q(.,.)  $ by the empirical currents $  J(.,.)  $
and the empirical activities $  A(.,.)  $ }

In the present paper, we focus on cases where 
the jumps between two configurations $x \ne y$ are either both possible $w(x,y) w(y,x)>0$
or both impossible $w(x,y) =0=w(y,x)$.
On each link $x \ne y$ where the two flows are possible $w(x,y) w(y,x)>0$,
it is convenient to choose an order $x>y$, and to
 replace the two empirical flows $Q(x,y)$ and $Q(y,x)$ 
by their antisymmetric and symmetric parts called respectively 
the empirical current $J(x,y)  $ and the empirical activity $A(x,y) $ of the link
\begin{eqnarray}
J(x,y) && \equiv Q(x,y) -  Q(y,x)= - J(y,x)
\nonumber \\
A(x,y) && \equiv Q(x,y) +  Q(y,x)=  A(y,x)
\label{jafromq}
\end{eqnarray}
The advantage of this decomposition into currents and activities 
is that the stationarity constraint of Eq. \ref{contrainteq}
only involve the empirical currents $J(.,.) $
\begin{eqnarray}
0 = \sum_{y \ne x} J(x,y) \ \ \ \text{ for any $x$}
\label{contrainteqj}
\end{eqnarray}
while the empirical activities $A(.,.) $ do not appear at all in the constitutive constraints.

The trajectory information $ {\cal I} \left[P(.)  ;Q(.,.) \right] $
of Eq. \ref{informationjumpempi} becomes
\begin{eqnarray}
&& {\cal I} \left[ P(.)  ; J(.,.) ; A(.,.)\right]
 = - \sum_{y  } P(y) w(y,y)  -       \sum_{y  }\sum_{ x > y  }
 \left[\frac{ A(x,y) + J(x,y)}{2}   \ln ( w(x,y) ) +\frac{ A(x,y) - J(x,y)}{2}   \ln ( w(y,x) ) \right]
\nonumber \\
&& =  \sum_{y  } \sum_{ x > y  } \left[ w(x,y) P(y)+  w(y,x) P(x)
- \frac{  J(x,y)}{2}   \ln \left( \frac{ w(x,y) }{ w(y,x) } \right) 
- \frac{ A(x,y) }{2}   \ln ( w(x,y) w(y,x) ) \right] 
\label{informationjumpempiaj}
\end{eqnarray}


\subsubsection{ Number $  {\cal N}^{[2.5]}_T\left[ P(.)  ; Q(.,.) \right]$ of trajectories of duration $T$ with given empirical observables $\left[ P(.)  ; Q(.,.) \right] $ }

As in Eq. \ref{omegat},
the number $  {\cal N}^{[2.5]}_T\left[ P(.)  ; Q(.,.) \right]$ of trajectories of duration $T$ 
associated to given empirical observables $\left[ P(.)  ; Q(.,.) \right] $
will grow exponentially with respect to the time-window $T$ in the limit $T \to +\infty$
\begin{eqnarray}
{\cal N}^{[2.5]}_T \left[ P(.)  ; Q(.,.) \right]  \opsimeq_{T \to +\infty} 
 C^{[2.5]}\left[ P(.)  ; Q(.,.) \right] \ e^{\displaystyle T S^{[2.5]}\left[ P(.)  ; Q(.,.) \right]  }
\label{omegatjump}
\end{eqnarray}
The prefactor $C^{[2.5]}\left[ P(.)  ; Q(.,.) \right]$ 
denotes the constitutive constraints for the empirical observables 
Eqs \ref{rho1ptnormaj} and \ref{contrainteq}
\begin{eqnarray}
C^{[2.5]}\left[ P(.)  ; Q(.,.) \right]
  = \delta \left( \sum_x  P(x) - 1 \right) 
 \prod_x \delta \left( \sum_{y \ne x} [ Q(x,y)-  Q(y,x) ] \right)
\label{constraints2.5master}
\end{eqnarray}
that can be rewritten using Eq. \ref{contrainteqj}
in terms of the empirical currents as
\begin{eqnarray}
C^{[2.5]}\left[ P(.)  ; J(.,.) \right]
  = \delta \left( \sum_x  P(x) - 1 \right)  \prod_x \delta \left( \sum_{y \ne x }  J(x,y)  \right)
\label{constraints2.5masterj}
\end{eqnarray}

The entropy $S^{[2.5]}\left[ P(.)  ; Q(.,.) \right]  = \frac{\ln {\cal N}^{[2.5]}_T\left[ P(.)  ; Q(.,.) \right] }{ T }  $ of trajectories with given intensive empirical observables $\left[ P(.)  ; Q(.,.) \right] $ can be evaluated without any combinatorial computation
as explained after Eq. \ref{normaempit} of the previous subsection,
with the following adaptations :

(i) When the empirical observables $\left[ P(.)  ; Q(.,.) \right] $ 
take their typical values corresponding to the steady state values $\left[ P_*(.)  ; Q_*(.,.) \right] $ 
with
\begin{eqnarray}
Q_*(x,y) = w(x,y) P_*(y)  
\label{qtypjump}
\end{eqnarray}
 the entropy $S^{[2.5]}\left[ P^*(.)  ; Q^*(.,.) \right] $ coincides with the information $  {\cal I}\left[ P^*(.)  ; Q^*(.,.) \right] $ of Eq. \ref{informationjumpempi}
\begin{eqnarray}
S^{[2.5]}\left[ P^*(.)  ; Q^*(.,.) \right]=   
 {\cal I}\left[ P^*(.)  ; Q^*(.,.) \right]
=- \sum_{y  } P^*(y)  w(y,y)  -       \sum_{y  }\sum_{ x \ne y  } Q^*(x,y)  \ln ( w(x,y) )    \equiv    h_{KS}
\label{compensationj}
\end{eqnarray}
which corresponds to the Kolmogorov-Sinai entropy $h_{KS}$ of the trajectories as in Eq. \ref{itrajempimctyp}.

(ii) To obtain the entropy $S^{[2.5]}\left[ P(.)  ; Q(.,.) \right]  $ for other given values of the empirical observables,
one just needs to introduce the modified Markov matrix ${\tilde w}$ that would make 
the empirical values $\left[ P(.)  ; Q(.,.) \right]$ steady via Eq. \ref{qtypjump} and \ref{wdiag}
\begin{eqnarray}
 {\tilde w}(x,y) && \equiv \frac{  Q(x,y) }{P(y) } \ \ \ {\rm for } \ \ x \ne y
 \nonumber \\
  {\tilde w}(y,y) && \equiv  - \sum_{x \ne y}  {\tilde w}(x,y) =   - \sum_{x \ne y} \frac{  Q(x,y) }{P(y) } 
\label{modeleeffjump}
\end{eqnarray}
and to use Eq. \ref{compensationj} for this modified Markov matrix ${\tilde w}$ to obtain
the entropy for arbitrary empirical observables
\begin{eqnarray}
S^{[2.5]}\left[ P(.)  ; Q(.,.) \right]  && = 
  \sum_{y  }P(y) \sum_{ x \ne y  } {\tilde w}(x,y)
- \sum_{y  } P(y)  {\tilde w}(y,y)   -       \sum_{y  }\sum_{ x \ne y  } Q(x,y)  \ln (  {\tilde w}(x,y)  )     
\nonumber \\
&& = \sum_{y  } \sum_{x \ne y} \left[ Q(x,y) 
  -      Q(x,y)  \ln \left(  \frac{  Q(x,y) }{P(y) }  \right)  \right]
\label{entropyempijump}
\end{eqnarray}

The replacement of the empirical flows $Q(.,.) $  by the currents $J(.,.) $ and the activities $A(.,.)$
leads to the entropy 
\begin{eqnarray}
S^{[2.5]}\left[ P(.)  ; J(.,.) ; A(.,.) \right]    
 = \sum_{y  } \sum_{x > y} 
 \left[ A(x,y) 
  -    \frac{ A(x,y) }{2}   \ln \left(  \frac{  A^2(x,y) - J^2(x,y) }{ 4 P(y)P(x) }  \right)  
   +      \frac{  J(x,y)}{2}  \ln \left(  \frac{ [ A(x,y) - J(x,y) ]  P(y) }{ [ A(x,y) + J(x,y) ]  P(x) } \right)     
   \right]
   \nonumber \\
\label{entropyempiajjump}
\end{eqnarray}


\subsubsection{ Joint probability distribution $P^{[2.5]}_T \left[ P(.)  ; Q(.,.) \right] $ of the empirical observables $\left[ P(.)  ; Q(.,.) \right] $ }

As in Eq. \ref{pnormaempi}, the joint probability distribution $P^{[2.5]}_T \left[ P(.)  ; Q(.,.) \right] $ of the empirical observables $\left[ P(.)  ; Q(.,.) \right] $ follows the large deviation form for large $T$
\begin{eqnarray}
P^{[2.5]}_T \left[ P(.)  ; Q(.,.) \right] \opsimeq_{T \to +\infty} 
C^{[2.5]}\left[ P(.)  ; Q(.,.) \right] e^{- T I^{[2.5]}\left[ P(.)  ; Q(.,.) \right] }
\label{level2.5master}
\end{eqnarray}
where the constitutive constraints $C^{[2.5]}\left[ P(.)  ; Q(.,.) \right]$ have been written in Eq. \ref{constraints2.5master},
while the rate function at Level 2.5 reduces to the difference between the trajectory information 
of Eq. \ref{informationjumpempi}
and the entropy of Eq. \ref{entropyempi}
\begin{eqnarray}
I^{[2.5]}\left[ P(.)  ; Q(.,.) \right]
 = {\cal I} \left[ P(.)  ; Q(.,.) \right]  - S^{[2.5]} \left[ P(.)  ; Q(.,.) \right]
\label{rate2.5masterdiff}
\end{eqnarray}
that can be rewritten in the compact form
\begin{eqnarray}
I^{[2.5]}\left[ P(.)  ; Q(.,.) \right]
&&  = \sum_{y } \sum_{x \ne y} 
\left[ Q(x,y)  \ln \left( \frac{ Q(x,y)  }{  w(x,y) P(y)  }  \right) 
 - Q(x,y)  + w(x,y)  P(y)  \right]
\label{rate2.5master}
\end{eqnarray}
where the vanishing for the steady values  $[ P_*(.), Q_*(.,.) ]$ of Eq. \ref{qtypjump} is obvious.

The replacement of the empirical flows $Q(.,.) $ via of Eq. \ref{jafromq}
leads to the rate function in terms of 
the empirical observables $\left[ P(.)  ; J(.,.) ; A(.,.)\right] $
\begin{eqnarray}
 I^{[2.5]}\left[ P(.)  ; J(.,.) ; A(.,.) \right]
 &&= {\cal I} \left[ P(.)   ; J(.,.) ; A(.,.) \right]  - S^{[2.5]} \left[ P(.)   ; J(.,.) ; A(.,.) \right]
 \label{rate2.5masteraj}
 \\
&& =
 \sum_{y  } \sum_{ x > y  } \bigg[ w(x,y) P(y)+  w(y,x) P(x)  - A(x,y) 
 \nonumber \\
&&  +    \frac{ A(x,y) }{2}   \ln \left(  \frac{  A^2(x,y) - J^2(x,y) }{ 4 w(x,y) w(y,x)P(y)P(x) }  \right)  
   +      \frac{  J(x,y)}{2}  
   \ln \left(  \frac{ [ A(x,y) + J(x,y) ] w(y,x) P(x) }{ [ A(x,y) - J(x,y) ] w(x,y) P(y) } \right)     
   \bigg]
\nonumber
\end{eqnarray}
that governs the joint probability to see the empirical variables
\begin{eqnarray}
P^{[2.5]} \left[ P(.)  ;  J(.,.) ; A(.,.) \right] \opsimeq_{T \to +\infty} 
C^{[2.5]}\left[ P(.)  ; J(.,.) \right] e^{- T I^{[2.5]}\left[ P(.)  ; J(.,.) ; A(.,.) \right] }
\label{level2.5masteraj}
\end{eqnarray}

These explicit large deviations at Level 2.5 
allow to analyze via contraction all the large deviations properties of lower levels,
as described in the following section for the example of the level 2.25 that is always explicit for Markov jump processes.


\subsubsection{ Explicit level 2.25 for the joint distribution of the empirical probability $P(.) $ and the empirical currents $ J(.,.) $ }

\label{subsec_contractionLD}

The joint probability $P^{[2.25]} \left[ P(.)  ;  J(.,.) \right] $ of the empirical probability $P(.) $ and currents $ J(.,.) $ can be obtained from the integration of Eq. \ref{level2.5masteraj}
over the empirical activities $A(.,.) $ that do not appear in the constraints
\begin{eqnarray}
P^{[2.25]}_T \left[ P(.)  ;  J(.,.) \right]
&& = \int {\cal D}  A(.,.) P^{[2.5]}_T \left[ P(.)  ;  J(.,.) ; A(.,.) \right] \opsimeq_{T \to +\infty} 
C^{[2.5]}\left[ P(.)  ; J(.,.) \right] \int {\cal D}  A(.,.) e^{- T I^{[2.5]}\left[ P(.)  ; J(.,.) ; A(.,.) \right] }
\nonumber \\
&& =C^{[2.5]}\left[ P(.)  ; J(.,.) \right]  e^{- T I^{[2.25]}\left[ P(.)  ; J(.,.) \right] }
\label{level2.25masteraj}
\end{eqnarray}
where the rate function $I^{[2.25]}\left[ P(.)  ; J(.,.) \right] $ can be obtained via the saddle-point evaluation of the integral as follows.
The optimization of rate function $I^{[2.5]}\left[ P(.)  ; J(.,.) ; A(.,.) \right] $ of Eq. \ref{rate2.5masteraj}
over the activity $A(x,y)$
\begin{eqnarray}
0 && = \frac{ \partial  I^{[2.5]}\left[ P(.)  ; J(.,.) ; A(.,.) \right] }{ \partial A(x,y)} 
  =    \frac{ 1 }{2}   \ln \left(  \frac{  A^2(x,y) - J^2(x,y) }{ 4 w(x,y) w(y,x)P(y)P(x) }  \right)  
 \label{rate2.5masterajderiA}
\end{eqnarray}
leads to the optimal empirical activity $ $
as a function of the other empirical variables
\begin{eqnarray}
  A^{opt}(x,y) = \sqrt{  J^2(x,y) + 4 w(x,y) w(y,x)P(y)P(x) }  
 \label{activityoptimal2.25}
\end{eqnarray}
that can be plugged into Eq. \ref{rate2.5masteraj}
to obtain the explicit rate function at level 2.25
\begin{eqnarray}
I^{[2.25]}\left[ P(.)  ; J(.,.) \right] && =
I^{[2.5]}\left[ P(.)  ; J(.,.) ; A^{opt}(.,.) \right]
\nonumber
 \\
&& =
 \sum_{y  } \sum_{ x > y  } \bigg[ w(x,y) P(y)+  w(y,x) P(x)  - \sqrt{  J^2(x,y) + 4 w(x,y) w(y,x)P(y)P(x) }  
 \nonumber \\
&&   
 +       J(x,y)
   \ln \left(  \frac{  \sqrt{  J^2(x,y) + 4 w(x,y) w(y,x)P(y)P(x) } + J(x,y)  }
   { 2 w(x,y) P(y) } \right)      
   \bigg]
\label{rate2.25masteraj}
\end{eqnarray}


\subsubsection{ Trajectory observables that can be written in terms of the empirical observables 
 $\left[ P(.)  ; Q(.,.)  \right] $}

An observable ${\cal O}^{Traj} [x(0 \leq t \leq T) ]$ of the trajectory $x(0 \leq t \leq T) $
that can be parametrized by two functions $\Omega(x)$ and $\Upsilon(x , y)$
\begin{eqnarray}
{\cal O}^{Traj}\left[ x(0 \leq t \leq T) \right]  
=   \frac{1}{T}  \int_0^T dt  \Omega( x(t)) +  
  \frac{1}{T} \sum_{t \in [0,T] : x(t^+) \ne x(t^-) } \Upsilon(x(t^+) , x(t^-)  ) 
\label{observableQ}
\end{eqnarray}
can be rewritten as a function $ {\cal O} \left[ P(.)  ; Q(.,.) \right] $
 of the empirical probability $P(.) $ of Eq. \ref{rho1pj}
 and of the empirical flows $ Q(.,.) $ of Eq. \ref{jumpempiricaldensity}
 as
\begin{eqnarray}
{\cal O}^{Traj}\left[ x(0 \leq t \leq T) \right] 
&& =  \sum_{y  } P(y)  \Omega( y)  +       \sum_{y  }\sum_{ x \ne y  }Q(x,y)   \Upsilon(x , y)  
\equiv  {\cal O} \left[ P(.)  ; Q(.,.) \right]
\label{observableQempijump}
\end{eqnarray}
or equivalently in terms of the empirical currents $J(x,y)  $ and the empirical activities $A(x,y) $
of the links using Eq. \ref{jafromq}
to replace the empirical flows $Q(.,.)$
\begin{eqnarray}
  {\cal O} \left[ P(.)  ; J(.,.) ; A(.,.) \right]
&& =  \sum_{y  } P(y)  \Omega( y)  +       \sum_{y  }\sum_{ x > y  }
\left[ \frac{ A(x,y) + J(x,y)}{2}   \Upsilon(x , y)  
+ \frac{ A(x,y) - J(x,y)}{2}   \Upsilon(y , x)  
\right] 
\nonumber \\
&& =  \sum_{y  } P(y)  \Omega( y)  +       \sum_{y  }\sum_{ x > y  }
\left[ J(x,y) \Lambda(x,y) 
+A(x,y) \Gamma(x,y)
\right] 
\label{observableAJempi}
\end{eqnarray}
with the coefficients
\begin{eqnarray}
\Lambda(x,y) && = \frac{ \Upsilon(x , y) - \Upsilon(y , x)}{2} 
\nonumber \\
\Gamma(x,y) && =\frac{ \Upsilon(x , y) + \Upsilon(y , x)}{2} 
\label{deflambdagamma}
\end{eqnarray}

The probability distribution $P_T({\cal O}) $ of the trajectory observable ${\cal O} $
 can be evaluated from 
  the joint probability $P^{[2.5]}_T \left[ P(.)  ;  Q(.,.)  \right]  $ at the Level 2.5 of Eq. \ref{level2.5master}
\begin{eqnarray}
P_T({\cal O})  && =\int {\cal D}  P(.,.)  \int {\cal D}  Q(.,.)  
P^{[2.5]}_T \left[ P(.)  ;  Q(.,.)  \right]
\delta \left( {\cal O} - {\cal O} \left[ P(.)  ; Q(.,.)  \right] \right)
\nonumber \\
&& \opsimeq_{T \to +\infty} 
\int {\cal D}  P(.,.)  \int {\cal D}  Q(.,.) 
C^{[2.5]}\left[ P(.)  ; Q(.,.) \right] 
\delta \left( {\cal O} - {\cal O} \left[ P(.)  ; Q(.,.)  \right] \right)
e^{- T I^{[2.5]}\left[ P(.)  ; Q(.,.)  \right] }
\nonumber \\
&& \opsimeq_{T \to +\infty} e^{- T I({\cal O})}
\label{largedevadditiveprobajump}
\end{eqnarray}
where the rate function $I({\cal O})$ for the trajectory observable ${\cal O} $
corresponds to the minimization of the rate function $I^{[2.5]}\left[ P(.)  ; Q(.,.) \right]  $
over the empirical observables $\left[ P(.)  ; Q(.,.)  \right] $ 
satisfying the constitutive constraints $C^{[2.5]}\left[ P(.)  ; Q(.,.) \right]  $
and the supplementary constraint $\delta \left( {\cal O} - {\cal O} \left[ P(.)  ; Q(.,.)  \right] \right)$ reproducing the correct value ${\cal O}$ of the trajectory observable.


\subsection{ Large deviations for diffusion processes : similarities and differences with the previous Markov processes }

In this subsection, we recall the large deviations properties for diffusion processes,
in order to stress the important differences with respect to the two previous subsections,
 that have motivated the present work.
 
 The Fokker-Planck dynamics in dimension $d$,
 can be written as a continuity equation for the probability density $\rho^{[t]}(\vec x)$
to be at position $\vec x $ at time $t$
\begin{eqnarray}
 \partial_t \rho^{[t]}(\vec x)  =  -   \vec \nabla . \vec j^{[t]} (\vec x)
 \equiv  -   \sum_{\mu=1}^d \frac{ \partial  j^{[t]}_{\mu} (\vec x) }{\partial x_{\mu} }
\label{fokkerplanck}
\end{eqnarray}
where the current component $j^{[t]}_{\mu} (\vec x) $ in the direction $\mu \in \{1,..,d\}$
involves the component $F_{\mu}(\vec x)  $ of the force and
the diffusion coefficient $D_{\mu}(\vec x) $
\begin{eqnarray}
 j^{[t]}_{\mu} (\vec x) \equiv  F_{\mu} (\vec x ) \rho^{[t]}(\vec x) -D_{\mu} (\vec x)  \frac{\partial \rho^{[t]}(\vec x)}{ \partial x_{\mu}}   
\label{fokkerplanckj}
\end{eqnarray}
The more general case where the current involves a diffusion matrix ${\bold D}_{\mu \nu} (\vec x) $ 
will be considered only in the Appendices
(see Eq. \ref{fokkerplanckjnondiag}).

Whenever this dynamics converges towards a normalizable steady-state $\rho^*(\vec x)$ 
with its associated steady current of Eq. \ref{fokkerplanckj}
\begin{eqnarray}
 j^*_{\mu} (\vec x) \equiv  F_{\mu} (\vec x ) \rho^*(\vec x) -D_{\mu} (\vec x)  \frac{\partial \rho^*(\vec x)}{ \partial x_{\mu}}  
  \label{fokkerplanckjsteady}
\end{eqnarray}
satisfying Eq. \ref{fokkerplanck}
\begin{eqnarray}
0= \partial_t \rho^*(\vec x)  = -     \vec \nabla . \vec j^* (\vec x) \equiv  -   \sum_{\mu=1}^d \frac{ \partial  j^*_{\mu} (\vec x) }{\partial x_{\mu} }
\label{fokkerplancksteady}
\end{eqnarray}
it is interesting to analyze 
how time-averaged observables over a large-time-window $[0,T]$ converge towards their steady values,
as we now recall.


\subsubsection{ Explicit large deviations for the joint distribution 
of the empirical density $\rho (.) $ and the empirical current
$\vec j (.) $ }

For a diffusive trajectory $\vec x (0 \leq t \leq T) $ over a large-time-window $[0,T]$, 
 one introduces the following empirical time-averaged observables :

(i) the empirical density $\rho (.) $ that measures the fraction of the time spent at the various positions $\vec x$
\begin{eqnarray}
\rho (\vec x)  \equiv \frac{1}{T} \int_0^T dt \  \delta^{(d)} ( \vec x(t)- \vec x)  
\label{rhodiff}
\end{eqnarray}
with the normalization to unity
\begin{eqnarray}
\int d^d \vec x \ \rho (\vec x)  = 1
\label{rho1ptnormadiff}
\end{eqnarray}
is the direct analog of Eqs \ref{rho1pt} and \ref{rho1pj} discussed previously.

(ii) the empirical current $\vec j (.)  $ that measures the time-averages 
of the velocity $\frac{d  \vec x (t)}{dt} $
seen at the various positions $\vec x$ within in the Stratonovich mid-point interpretation
\begin{eqnarray} 
 \vec j (\vec x) \equiv   \frac{1}{T} \int_0^T dt \ \frac{d  \vec x (t)}{dt}   \delta^{(d)}( \vec x(t)- \vec x)  
\label{diffj}
\end{eqnarray}
should satisfy the vanishing-divergence constraint at any position $\vec x$
\begin{eqnarray}
 \vec \nabla . \vec j (\vec x)   =0 \ \ \ \text{ at any }  \vec x
 \label{divergencenulle}
\end{eqnarray}

For large $T$, the joint probability distribution 
of the normalized empirical density $\rho(.)$ of Eq. \ref{rhodiff}
and of the divergence-less empirical current $\vec j(.)$ of Eq. \ref{diffj}
satisfies the large deviation form \cite{wynants_thesis,maes_diffusion,chetrite_formal,engel,chetrite_HDR,c_lyapunov,c_inference}
\begin{eqnarray}
 p^{[2.25]}_T[ \rho(.),  \vec j(.)]   \opsimeq_{T \to +\infty}   c_{2.25} [ \rho(.),  \vec j(.)]
e^{- \displaystyle T i_{2.25} [ \rho(.),  \vec j(.)]    }
\label{ld2.5diff}
\end{eqnarray}
where $ c_{2.25} [ \rho(.),  \vec j(.)] $ summarizes the constitutive constraints of Eqs \ref{rho1ptnormadiff} and \ref{divergencenulle}
\begin{eqnarray}
  c_{2.25} [ \rho(.),  \vec j(.)]  =\delta \left(\int d^d \vec x 
\rho(\vec x) -1  \right)
\left[ \prod_{\vec x }  \delta \left(   \vec \nabla . \vec j (\vec x)  \right) \right] 
  \label{c2.25diff}
\end{eqnarray}
while the explicit rate function reads
\begin{eqnarray}
  i_{2.25} [ \rho(.),  \vec j(.)]  =
  \int d^d \vec x \sum_{\mu=1}^d \frac{\left[ j_{\mu}(\vec x) -   F_{\mu}(\vec x) \rho(\vec x)
  +D_{\mu} (\vec x) \frac{\partial \rho(\vec x)}{\partial x_{\mu} }  \right]^2}{ 4 D_{\mu} (\vec x) \rho(\vec x) } 
  \label{rate2.25diff}
\end{eqnarray}

While this result is usually called the level 2.5 of large deviations for diffusion processes,
it will be clearer to call it
the level 2.25 in the present paper where we wish to analyze in detail the correspondence
with the Markov jump processes described in the previous subsection, 
where the joint distribution of the empirical density and currents was called the level 2.25
in Eq. \ref{level2.25masteraj}),
while the level 2.5 of Eq. \ref{level2.5masteraj}
contains in addition the empirical activities $A(.,.)$.

The comparison with the two previous subsections shows that the empirical observables $ [ \rho(.),  \vec j(.)]$
of the diffusion process with the single-point current $\vec j(\vec x)$ of Eq. \ref{diffj} contain much less information than:

(i) the whole empirical 2-point density $\rho^{(2)}(\vec x,\vec y)$ of Eq. \ref{rho2pt}
for discrete-time Markov chains;

(ii) the empirical observables $ [ P(.), Q(.,.)]$ 
with the whole empirical 2-point flows $Q(x,y)$ of Eq. \ref{jumpempiricaldensity}
for continuous-time Markov chains.

So the main goal of the present paper will be to understand precisely what 
happens to the relevant empirical observables
and to their large deviations at level 2.5 
for the continuous-space random walks with finite time-step $\tau$
and for the continuous-time random walks on the lattice of spacing $b$ 
when they converge towards diffusion processes in the limits $\tau \to 0$ and $b \to 0$ respectively.


\subsubsection{ Trajectory observables that can be written in terms of the 
empirical density $\rho (.) $ and the empirical current
$\vec j (.) $ }

The large deviations at level 2.25 of Eq. \ref{rate2.25diff} allows to analyze the large deviations 
of the observables $ O^{Traj}[\vec x(0\leq t \leq T)] $ 
of the diffusive trajectories $\vec x(0\leq t \leq T) $ that 
that can be parametrized by $(1+d)$
functions $\omega( \vec x) $ and $\lambda_{\mu=1,2,..,d}(\vec x)$ as follows
\begin{eqnarray}
 O^{Traj}[\vec x (0 \leq t \leq T) ]   && \equiv \frac{1}{T} \int_0^T dt \left[ \omega(\vec x(t)) 
+ {\vec  \lambda}(\vec x) . \frac{d  \vec x (t)}{dt}\right]  
\label{Obsdiff}
\end{eqnarray}
since they can be rewritten as functions $O[ \rho(.),  \vec j(.)] $
of the empirical density $\rho(.)$ of Eq. \ref{rhodiff}
and of the empirical current $\vec j(.)$ of Eq. \ref{diffj}
\begin{eqnarray}
 O^{Traj}[\vec x (0 \leq t \leq T) ]  
  = \int d^d \vec x  \left[ \omega( \vec x)\rho (\vec x) 
+ {\vec   \lambda}(\vec x)  . \vec j (\vec x) \right]  \equiv O[ \rho(.),  \vec j(.)]
\label{Obsdiffempi}
\end{eqnarray}
The probability distribution $p_T(O) $ of the observable $O$ 
can be then obtained from the integration of the joint probability
of Eq. \ref{ld2.5diff}
\begin{eqnarray}
&& p_T(O)  = \int {\cal D} \rho(.) \int {\cal D}  \vec j(.) p^{[2.25]}_T[ \rho(.), \vec j(.)]  
\delta \left(O - \int d^d \vec x  \left[ \omega( \vec x)\rho (\vec x) 
+ {\vec   \lambda}(\vec x)  . \vec j (\vec x)  \right]   \right)
\nonumber \\
&&  \opsimeq_{T \to +\infty}   \int {\cal D} \rho(.) \int {\cal D}  \vec j(.)
\delta \left(\int d^d \vec x 
\rho(\vec x) -1  \right)
\left[ \prod_{\vec x }  \delta \left(  \vec \nabla . \vec j (\vec x)  \right) \right] 
\delta \left(O - \int d^d \vec x  \left[ \omega( \vec x)\rho (\vec x) 
+ {\vec   \lambda}(\vec x)  . \vec j (\vec x)  \right]   \right)
e^{- \displaystyle T i_{2.25} [ \rho(.),  \vec j(.)]    } 
\nonumber \\
&& \opsimeq_{T \to +\infty}  
e^{- \displaystyle T i(O)    }
\label{ld2.25diffcontractionO}
\end{eqnarray}
So the rate function $ i(O) $ for the trajectory observable $O$
corresponds to the optimization of the rate function $i_{2.25} [ \rho(.),  \vec j(.)]  $ 
over the empirical density $\rho(.)$ 
and over the empirical current $\vec j(.)$ satisfying all the constraints.

The comparison with the two previous subsections 
 shows that the observables $O$ of Eqs \ref{Obsdiff} \ref{Obsdiffempi} for the diffusion process
are much more limited than:

(i) the trajectory observables ${\cal O}$ of Eqs \ref{observablerho2} \ref{observablerho2empi} for discrete-time Markov chains that involve any function of two consecutive positions.

(ii) the trajectory observables ${\cal O}$ of Eqs \ref{observableQ} \ref{observableAJempi} 
for continuous-time Markov chains,
that involve any function of two positions separated by elementary jumps. 

So the second goal of the present paper will be
to analyze what 
happens to the general trajectory observables 
of the continuous-space random walks with finite time-step $\tau$
and of the continuous-time random walks on the lattice of spacing $b$
when they converge towards diffusion processes in the limits $\tau \to 0$ and $b \to 0$ respectively.


\subsubsection{ What about the trajectory information ${\cal I}^{Traj}[\vec x(0 \leq t \leq T)] $
that determines the trajectory probability ${\cal P}[\vec x(0 \leq t \leq T)] $ ?}

In the two previous subsections concerning Markov chains in discrete-time and in continuous-time, 
we have explained why the most important trajectory observable
is the trajectory information ${\cal I}^{Traj}[\vec x(0 \leq t \leq T)] $
that determines the probability ${\cal P}[\vec x(0 \leq t \leq T) $ of the trajectory 
\begin{eqnarray}
{\cal I}^{Traj} [\vec x(0 \leq t \leq T) ]   \equiv  - \frac{\ln {\cal P}[\vec x(0 \leq t \leq T)]}{T} 
\label{itrajdiffusion}
\end{eqnarray}
Note that Ruelle thermodynamic formalism for stochastic trajectories 
can be rephrased as the theory of large deviations theory for this trajectory information
as discussed in detail in \cite{c_ruelle}.
We have then stressed the essential role played by the explicit expressions
of the trajectory information ${\cal I}^{Traj}[\vec x(0 \leq t \leq T)] $
in terms of the Markov generators (see Eqs \ref{itrajmc} and \ref{informationjump})
and their translations in terms of the relevant empirical observables (see Eqs \ref{itrajempimc} and \ref{informationjumpempi}).

Here again, the differences with diffusion processes are obvious :

(i) for a diffusion process, the trajectory probability ${\cal P}[\vec x(0 \leq t \leq T)] $
cannot been written in a fully intrinsic manner in terms of the Markov generator,
since the corresponding Feynman path-integral
requires some regularization and some choice of the time-discretization rules
(see the discussion after Eq. \ref{pathintegral} in Appendix \ref{app_PathIntegral}).

(ii) when one tries to rewrite the information ${\cal I}^{Traj} [\vec x(0 \leq t \leq T) ]$ of 
the Feynman path-integral of Eq. \ref{pathintegral} with the Lagrangian of Eq. \ref{lagrangian}
in terms of the empirical density $\rho(.)$ of Eq. \ref{rhodiff}
and of the empirical current $\vec j(.)$ of Eq. \ref{diffj}, 
it is possible for all terms except for the kinetic-energy term
$\sum_{\mu=1}^d \sum_{\nu=1}^d \frac{ \dot x_{\mu} (s) {\bold D}^{-1}_{\mu \nu}(\vec x) \dot x_{\nu} (s)}{ 4 }  $
which is quadratic with respect to the velocity components,
as already discussed in \cite{c_inference} in the context of inference.

So the third goal of the present paper will be
to analyze what 
happens to the trajectory information ${\cal I}^{Traj}[\vec x(0 \leq t \leq T)]  $ 
of the continuous-space random walks with finite time-step $\tau$
or of the continuous-time random walks on the lattice of spacing $b$
when they converge towards diffusion processes in the limits $\tau \to 0$ and $b \to 0$ respectively.
In particular, besides the singular contributions that will appear in the limit 
$\tau \to 0$ and $b \to 0$ since the trajectory information of a diffusive trajectory diverges, 
we wish to compare the regular contributions that emerge
for the lattice regularization
and for the finite-time discretization.

As already stressed in the Introduction, the remainder of the
main text is devoted to the analysis of lattice-regularization of lattice spacing $b$,
while the discrete-rime-regularization of parameter $\tau$ is discussed in the Appendices.


\section{ Markov jump generator on the lattice of spacing $b $ and in the limit $b \to 0$ }

\label{sec_generator}

In this section, we describe the continuous-time random walk
on the regular lattice of spacing $b $ in dimension $d$
that converges in the limit $b \to 0$ towards the Fokker-Planck dynamics of Eqs \ref{fokkerplanck}
and \ref{fokkerplanckj} that involves the forces $F_{\mu}(\vec x)  $ and
the diffusion coefficient $D_{\mu}(\vec x) $.


\subsection{ Markov jump generator on the regular lattice of spacing $b $ in dimension $d$}

We consider the regular lattice in dimension $d$ of lattice spacing $b$
with the $d$ unit vectors $e_{\mu=1,..,d}$.
A lattice site $\vec x^{site} = (b \vec n)$ is labelled by $d$ integers $\vec n=(n_1,..,n_d)$ 
\begin{eqnarray}
\vec x^{site} = b \vec n = \sum_{\mu=1}^d (b n_{\mu}) \vec e_{\mu}
\label{site}
\end{eqnarray}
To denote the link between the two sites $b \vec n $ and $(b \vec n +b \vec e_{\mu})$,
it will be convenient to use their middle-point 
\begin{eqnarray}
\vec x^{[link]} = b \vec n+\frac{b}{2} \vec e_{\mu}
\label{link}
\end{eqnarray}
The two transitions rates on this link will be parametrized 
in terms of the force $F_{\mu} \left(b \vec n+\frac{b}{2} \vec e_{\mu}\right) $
and of the diffusion coefficient $D_{\mu} \left(b \vec n+\frac{b}{2} \vec e_{\mu}\right)$ 
associated to the link as follows
\begin{eqnarray}
w(b \vec n+b \vec e_{\mu},b \vec n) &&= 
 \frac{ D_{\mu} \left(b \vec n+\frac{b}{2} \vec e_{\mu}\right) }{b^2} e^{ \frac{ b F_{\mu} \left(b \vec n+\frac{b}{2} \vec e_{\mu}\right)}{2 D_{\mu} \left(b \vec n+\frac{b}{2} \vec e_{\mu}\right)} }
 \nonumber \\
w(b \vec n,b \vec n+b \vec e_{\mu}) &&=  \frac{ D_{\mu} \left(b \vec n+\frac{b}{2} \vec e_{\mu}\right) }{b^2} e^{- \frac{ b F_{\mu} \left(b \vec n+\frac{b}{2} \vec e_{\mu}\right)}{2 D_{\mu} \left(b \vec n+\frac{b}{2} \vec e_{\mu}\right)} }
\label{rates}
\end{eqnarray}

The master Eq. \ref{mastereq} for the probability $P^{[t]}(b \vec n) $ to be on site $(b \vec n)$ at time $t$ then reads
\begin{eqnarray}
&& \partial_t P^{[t]}( b \vec n)   =w(b \vec n,b \vec n ) P^{[t]}( b \vec n)
+ \sum_{\mu=1}^d  \left[ w(b \vec n,b \vec n - b \vec e_{\mu}) P^{[t]}( b \vec n-b \vec e_{\mu}) 
+ w(b \vec n,b \vec n + b \vec e_{\mu}) P^{[t]}( b \vec n+b \vec e_{\mu}) \right]
\nonumber \\
&& =w(b \vec n,b \vec n ) P^{[t]}( b \vec n)
+ \sum_{\mu=1}^d  \left[ \frac{ D_{\mu} \left(b \vec n-\frac{b}{2} \vec e_{\mu}\right) }{b^2} 
  e^{ \frac{ b F_{\mu} \left(b \vec n-\frac{b}{2} \vec e_{\mu}\right)}{2 D_{\mu} \left(b \vec n-\frac{b}{2} \vec e_{\mu}\right)} } P^{[t]}( b \vec n-b \vec e_{\mu}) 
+ \frac{ D_{\mu} \left(b \vec n+\frac{b}{2} \vec e_{\mu}\right) }{b^2}    e^{- \frac{ b F_{\mu} \left(b \vec n+\frac{b}{2} \vec e_{\mu}\right)}{2 D_{\mu} \left(b \vec n+\frac{b}{2} \vec e_{\mu}\right)} } P^{[t]}( b \vec n+b \vec e_{\mu}) \right]
 \ \ \ \ 
\label{masterlattice}
\end{eqnarray}
where the diagonal matrix elements $w(b \vec n,b \vec n ) $ are fixed by Eq. \ref{wdiag}
\begin{eqnarray}
w(b \vec n,b \vec n ) &&  =  - \sum_{\mu=1}^d  \left[ 
w(b \vec n - b \vec e_{\mu},b \vec n) 
+ w(b \vec n + b \vec e_{\mu},b \vec n) \right]
\nonumber \\
&& =- \sum_{\mu=1}^d  \left[ \frac{ D_{\mu} \left(b \vec n-\frac{b}{2} \vec e_{\mu}\right) }{b^2}
e^{- \frac{ b F_{\mu} \left(b \vec n-\frac{b}{2} \vec e_{\mu}\right)}{2 D_{\mu} \left(b \vec n-\frac{b}{2} \vec e_{\mu}\right)} }
 + \frac{ D_{\mu} \left(b \vec n+\frac{b}{2} \vec e_{\mu}\right) }{b^2}
 e^{ \frac{ b F_{\mu} \left(b \vec n+\frac{b}{2} \vec e_{\mu}\right)}{2 D_{\mu} \left(b \vec n+\frac{b}{2} \vec e_{\mu}\right)} }
  \right]
\label{wdiaglattice}
\end{eqnarray}

To see more clearly the similarity with the Fokker-Planck Eq. \ref{fokkerplanck} in continuous-space,
it is useful to rewrite the master Eq. \ref{masterlattice}
as the discrete continuity equation
\begin{eqnarray}
 \partial_t P^{[t]}( b \vec n)    = - \sum_{\mu=1}^d \left[ 
   J^{[t]}_{\mu}\left(b \vec n+\frac{b}{2} \vec e_{\mu}\right)
 - J^{[t]}_{\mu}\left(b \vec n-\frac{b}{2} \vec e_{\mu} \right) 
  \right]
\label{master}
\end{eqnarray}
where the current $J^{[t]}\left(b \vec n+\frac{b}{2} \vec e_{\mu}\right) $ associated to the link $\left(b \vec n+\frac{b}{2} \vec e_{\mu}\right) $
between the two sites $b \vec n $ and $b \vec n+b \vec e_{\mu} $ involves the two transition rates of Eq. \ref{rates}
\begin{eqnarray}
J^{[t]}_{\mu} \left(b \vec n+\frac{b}{2} \vec e_{\mu}\right) 
&& = w(b \vec n+b \vec e_{\mu},b \vec n) P^{[t]}( b \vec n) 
-  w(b \vec n,b \vec n+b \vec e_{\mu}) P^{[t]}( b \vec n+b \vec e_{\mu})
\nonumber \\
&& =  \frac{ D_{\mu} \left(b \vec n+\frac{b}{2} \vec e_{\mu}\right) }{b^2} 
\left[ e^{ \frac{ b F_{\mu} \left(b \vec n+\frac{b}{2} \vec e_{\mu}\right)}{2 D_{\mu} \left(b \vec n+\frac{b}{2} \vec e_{\mu}\right)} } P^{[t]}( b \vec n) 
-    e^{- \frac{ b F_{\mu} \left(b \vec n+\frac{b}{2} \vec e_{\mu}\right)}{2 D_{\mu} \left(b \vec n+\frac{b}{2} \vec e_{\mu}\right)} } P^{[t]}( b \vec n+b \vec e_{\mu}) \right]
\label{jlink}
\end{eqnarray}

It is convenient to introduce the finite-difference operators 
$e^{ \pm \frac{b}{2} \frac{\partial }{ \partial x_{\mu}} }$ 
with their action on any test function $g(\vec x)$
\begin{eqnarray}
e^{ \pm \frac{b}{2}  \frac{\partial }{ \partial x_{\mu}} } g(\vec x) = \sum_{p=0}^{+\infty} \frac{ \left( \pm \frac{b}{2} \right)^p }{p! } \left( \frac{\partial }{ \partial x_{\mu}}\right)^p g(\vec x)
= g \left( \vec x \pm \frac{b}{2} \vec e_{\mu}\right)
\label{finitedefop}
\end{eqnarray}
in order to rewrite the discrete continuity Eq. \ref{master} as
\begin{eqnarray}
 \partial_t P^{[t]}( \vec x = b \vec n)    = - \sum_{\mu=1}^d \left[ 
   e^{  \frac{b}{2}  \frac{\partial }{ \partial x_{\mu}} }
 -  e^{ - \frac{b}{2}  \frac{\partial }{ \partial x_{\mu}} }
  \right] J^{[t]}_{\mu} ( \vec x )
\label{masterdiffop}
\end{eqnarray}
and to rewrite the current of Eq. \ref{jlink} as
\begin{eqnarray}
J^{[t]}_{\mu} \left(\vec x = b \vec n+\frac{b}{2} \vec e_{\mu}\right) 
&& =  \frac{ D_{\mu} (\vec x)  }{b^2} 
\left[ e^{ \frac{ b F_{\mu} (\vec x)}{2 D_{\mu} (\vec x)} } P^{[t]} \left(\vec x - \frac{b}{2} \vec e_{\mu}\right)
-    e^{- \frac{ b F_{\mu} (\vec x)}{2 D_{\mu} (\vec x)} } P^{[t]}\left(\vec x + \frac{b}{2} \vec e_{\mu}\right)
\right]
\nonumber \\
&& =  \frac{ D_{\mu} (\vec x)  }{b^2} 
\left[ e^{ \frac{ b F_{\mu} (\vec x)}{2 D_{\mu} (\vec x)} } e^{ - \frac{b}{2}  \frac{\partial }{ \partial x_{\mu}} }
-    e^{- \frac{ b F_{\mu} (\vec x)}{2 D_{\mu} (\vec x)} } e^{  \frac{b}{2}  \frac{\partial }{ \partial x_{\mu}} } \right]
P^{[t]}(\vec x)
\label{jlinkdiffop}
\end{eqnarray}


\subsection{ Limit of vanishing lattice spacing $b \to 0$ towards the Fokker-Planck generator for diffusion     }

The probability $P^{[t]}( b \vec n)$ to be at time $t$ on the lattice site $(b \vec n)$
is associated to the density $\rho^{[t]}(\vec x=b \vec n)$ per volume $b^d$
\begin{eqnarray}
P^{[t]}( b \vec n) = b^d \rho^{[t]}(\vec x=b \vec n)
\label{latticedensity}
\end{eqnarray}
with the normalization
\begin{eqnarray}
1= \sum_{\vec n} P^{[t]}( b \vec n) = \sum_{\vec n} b^d \rho^{[t]}(\vec x=b \vec n) = \int d^d x \rho^{[t]}(\vec x)
\label{latticedensitynorma}
\end{eqnarray}

Plugging Eq. \ref{latticedensity} into Eq. \ref{masterdiffop}
\begin{eqnarray}
 \partial_t \rho^{[t]}( \vec x )    = - \sum_{\mu=1}^d \frac{ \left[ 
   e^{  \frac{b}{2}  \frac{\partial }{ \partial x_{\mu}} } -  e^{ - \frac{b}{2}  \frac{\partial }{ \partial x_{\mu}} }  \right]}{b} \  \left( \frac{ J^{[t]}_{\mu} ( \vec x ) }{ b^{d-1} } \right)
\label{masterdiffopb}
\end{eqnarray}
and into Eq. \ref{jlinkdiffop}
\begin{eqnarray}
 J^{[t]}_{\mu}(\vec x ) 
 = b^{d-1}  D_{\mu} (\vec x)  
\frac{ \left[ e^{ \frac{ b F_{\mu} (\vec x)}{2 D_{\mu} (\vec x)} } e^{ - \frac{b}{2}  \frac{\partial }{ \partial x_{\mu}} }
-    e^{- \frac{ b F_{\mu} (\vec x)}{2 D_{\mu} (\vec x)} } e^{  \frac{b}{2}  \frac{\partial }{ \partial x_{\mu}} } \right] }{b} 
\rho^{[t]}(\vec x)
\label{jlinkdiffopb}
\end{eqnarray}
yields that the appropriate rescaled current for $b \to 0$ is
\begin{eqnarray}
 j^{[t]}_{\mu}( \vec x ) && \equiv \frac{J^{[t]}_{\mu}\left( \vec x \right) }{b^{d-1}}  
 = D_{\mu} (\vec x)  
 \left[ \frac{  F_{\mu} (\vec x)}{ D_{\mu} (\vec x)}  -   \frac{\partial }{ \partial x_{\mu}} 
+o(b) \right]  
\rho^{[t]}(\vec x)
 = F_{\mu} ( \vec x)   \rho^{[t]}(x) - D_{\mu} (\vec x)  \frac{\partial \rho^{[t]}(x) }{ \partial x_{\mu}}    +o(b)
\label{jrescal}
\end{eqnarray}
so that Eq. \ref{masterdiffopb} 
corresponds to
 the Fokker-Planck dynamics of Eqs \ref{fokkerplanck} and \ref{fokkerplanckj} in the limit $b \to 0$
\begin{eqnarray}
 \partial_t \rho^{[t]}( \vec x )  &&  = - \sum_{\mu=1}^d  \frac{\partial }{ \partial x_{\mu}}  j^{[t]}_{\mu}( \vec x )   +o(b)
 \nonumber \\
 && = - \sum_{\mu=1}^d  \frac{\partial }{ \partial x_{\mu}}  \left[ F_{\mu} ( \vec x)   \rho^{[t]}(x) - D_{\mu} (\vec x)  \frac{\partial \rho^{[t]}(x)}{ \partial x_{\mu}}  \right]   +o(b)
\label{fokkerplanckb}
\end{eqnarray}

In preparation for the next sections concerning large deviations, it is useful to mention 
the scaling with $b$ of the link activity, corresponding to sum of the two flows on a given link
instead of the difference that corresponds to the current of Eq. \ref{jlink} 
\begin{eqnarray}
A^{[t]}_{\mu} \left(b \vec n+\frac{b}{2} \vec e_{\mu}\right) 
&& \equiv  
w(b \vec n+b \vec e_{\mu},b \vec n) P^{[t]}( b \vec n) 
+  w(b \vec n,b \vec n+b \vec e_{\mu}) P^{[t]}( b \vec n+b \vec e_{\mu})
\nonumber \\
&& =  \frac{ D_{\mu} \left(b \vec n+\frac{b}{2} \vec e_{\mu}\right) }{b^2} 
\left[ e^{ \frac{ b F_{\mu} \left(b \vec n+\frac{b}{2} \vec e_{\mu}\right)}{2 D_{\mu} \left(b \vec n+\frac{b}{2} \vec e_{\mu}\right)} } P^{[t]}( b \vec n) 
+    e^{- \frac{ b F_{\mu} \left(b \vec n+\frac{b}{2} \vec e_{\mu}\right)}{2 D_{\mu} \left(b \vec n+\frac{b}{2} \vec e_{\mu}\right)} } P^{[t]}( b \vec n+b \vec e_{\mu}) \right]
\label{alink}
\end{eqnarray}
so that the rewriting with finite-difference operators $e^{ \pm \frac{b}{2}  \frac{\partial }{ \partial x_{\mu}} }$
analogous to Eq. \ref{jlinkdiffop}
\begin{eqnarray}
A^{[t]}_{\mu} \left(\vec x = b \vec n+\frac{b}{2} \vec e_{\mu}\right) 
&& =  \frac{ D_{\mu} (\vec x)  }{b^2} 
\left[ e^{ \frac{ b F_{\mu} (\vec x)}{2 D_{\mu} (\vec x)} } P^{[t]} \left(\vec x - \frac{b}{2} \vec e_{\mu}\right)
+    e^{- \frac{ b F_{\mu} (\vec x)}{2 D_{\mu} (\vec x)} } P^{[t]}\left(\vec x + \frac{b}{2} \vec e_{\mu}\right)
\right]
\nonumber \\
&& =  \frac{ D_{\mu} (\vec x)  }{b^2} 
\left[ e^{ \frac{ b F_{\mu} (\vec x)}{2 D_{\mu} (\vec x)} } e^{ - \frac{b}{2}  \frac{\partial }{ \partial x_{\mu}} }
+    e^{- \frac{ b F_{\mu} (\vec x)}{2 D_{\mu} (\vec x)} } e^{  \frac{b}{2}  \frac{\partial }{ \partial x_{\mu}} } \right]
P^{[t]}(\vec x)
\label{alinkdiffop}
\end{eqnarray}
yields that the appropriate rescaled activity when $P^{[t]}(\vec x) $ is rewritten 
in terms of the probability density $\rho^{[t]}(\vec x) $ of Eq. \ref{latticedensity} 
\begin{eqnarray}
a^{[t]}_{\mu}( \vec x) && \equiv \frac{ A^{[t]}_{\mu} \left( \vec x \right) }{b^{d-2} } 
=  D_{\mu} (\vec x)  
\left[ e^{ \frac{ b F_{\mu} (\vec x)}{2 D_{\mu} (\vec x)} } e^{ - \frac{b}{2}  \frac{\partial }{ \partial x_{\mu}} }
+    e^{- \frac{ b F_{\mu} (\vec x)}{2 D_{\mu} (\vec x)} } e^{  \frac{b}{2}  \frac{\partial }{ \partial x_{\mu}} } \right]
\rho^{[t]}(\vec x)
\nonumber \\
&& = 2 D_{\mu} (\vec x)  \rho^{[t]}(x)   +o(b)
\label{alinkrescal}
\end{eqnarray}
does not involve the same rescaling factor of $b$ as the rescaled current of Eq. \ref{jrescal}.
These different rescaling factors have important consequences as explained in the next sections
concerning large deviations.


\subsection{ Link with non-hermitian quantum mechanics in electromagnetic potentials 
on the lattice and for $b \to 0$} 

\label{sec_magnetic}

In this subsection, we describe the link with the
non-hermitian quantum mechanics in electromagnetic potentials
that will also be useful for the time-discretized regularization considered in Appendix \ref{app_PathIntegral}.


\subsubsection{ Markov jump generator as a non-hermitian electromagnetic quantum Hamiltonian on the lattice    }

The master Eq. \ref{masterlattice} can be interpreted as an euclidean Schr\"odinger Equation on the lattice
\begin{eqnarray}
 \partial_t \vert P^{[t]} \rangle   = - {\cal H} \vert P^{[t]} \rangle
\label{schrodingerlattice}
\end{eqnarray}
where the quantum Hamiltonian reads using standard bra-ket notations
\begin{eqnarray}
&& {\cal H}  = - \sum_{\vec n} w(b \vec n,b \vec n )  \vert b \vec n \rangle \langle b \vec n   \vert
-  \sum_{\vec n} \sum_{\mu=1}^d
\bigg(  w(b \vec n,b \vec n + b \vec e_{\mu})  \vert b \vec n \rangle \langle b \vec n + b \vec e_{\mu}  \vert
+ w(b \vec n+ b \vec e_{\mu} ,b \vec n )  \vert b \vec n + b \vec e_{\mu}\rangle \langle b \vec n   \vert \bigg)
\nonumber \\
&& = - \sum_{\vec n} w(b \vec n,b \vec n )  \vert b \vec n \rangle \langle b \vec n   \vert
-  \sum_{\vec n} \sum_{\mu=1}^d \frac{ D_{\mu} \left(b \vec n+\frac{b}{2} \vec e_{\mu}\right) }{b^2}
\bigg(  e^{- \frac{ b F_{\mu} \left(b \vec n+\frac{b}{2} \vec e_{\mu}\right)}{2 D_{\mu} \left(b \vec n+\frac{b}{2} \vec e_{\mu}\right)} }  \vert b \vec n \rangle \langle b \vec n + b \vec e_{\mu}  \vert
+   e^{ \frac{ b F_{\mu} \left(b \vec n+\frac{b}{2} \vec e_{\mu}\right)}{2 D_{\mu} \left(b \vec n+\frac{b}{2} \vec e_{\mu}\right)} }  \vert b \vec n + b \vec e_{\mu}\rangle \langle b \vec n   \vert \bigg)
\nonumber \\
&& \equiv - \sum_{\vec n} w(b \vec n,b \vec n )  \vert b \vec n \rangle \langle b \vec n   \vert
-  \sum_{\vec n} \sum_{\mu=1}^d \frac{ D_{\mu} \left(b \vec n+\frac{b}{2} \vec e_{\mu}\right) }{b^2}
\bigg(  e^{- b {\cal A}_{\mu} \left(b \vec n+\frac{b}{2} \vec e_{\mu}\right)}   \vert b \vec n \rangle \langle b \vec n + b \vec e_{\mu}  \vert
+   e^{ b {\cal A}_{\mu} \left(b \vec n+\frac{b}{2} \vec e_{\mu}\right)}  \vert b \vec n + b \vec e_{\mu}\rangle \langle b \vec n   \vert \bigg)
\ \ \ \ 
\label{Hamiltonianlattice}
\end{eqnarray}
On the last line, we have introduced the notation
\begin{eqnarray}
{\cal A}_{\mu}  \left(b \vec n+\frac{b}{2} \vec e_{\mu}\right) \equiv \frac{  F_{\mu} \left(b \vec n+\frac{b}{2} \vec e_{\mu}\right)}{2 D_{\mu} \left(b \vec n+\frac{b}{2} \vec e_{\mu}\right)}
\label{vectorpotlattice}
\end{eqnarray}
that can be interpreted as a vector potential associated to the link $\left(b \vec n+\frac{b}{2} \vec e_{\mu}\right) $
that makes the quantum Hamiltonian of Eq. \ref{Hamiltonianlattice} non-hermitian,
while the standard hermitian quantum electromagnetic Hamiltonian would correspond to the case where 
${\cal A}_{\mu}  \left(b \vec n+\frac{b}{2} \vec e_{\mu}\right) $ is imaginary.

To each plaquette of four points 
$[b \vec n ; 
b \vec n+b \vec e_{\mu} ; 
b \vec n+b \vec e_{\nu};
b \vec n+b \vec e_{\mu} +b \vec e_{\nu}]$ with $\mu<\nu$
that can be parametrized by their middle-point 
\begin{eqnarray}
\vec x^{[plaquette]} = b \vec n+\frac{b}{2} \vec e_{\mu}+\frac{b}{2} \vec e_{\nu}
\label{plaquette}
\end{eqnarray}
one can associate the magnetic flux $\Phi_{\mu \nu}$ corresponding to the circulation of the vector potential
of Eq. \ref{vectorpotlattice} around the four links of the plaquette
\begin{eqnarray}
\Phi_{\mu \nu} \left(\vec x  = b \vec n+\frac{b}{2} \vec e_{\mu}+\frac{b}{2} \vec e_{\nu} \right)
&& \equiv b {\cal A}_{\mu}  \left(b \vec n+\frac{b}{2} \vec e_{\mu}\right)
+ b{\cal A}_{\nu}  \left(b \vec n+b \vec e_{\mu}+\frac{b}{2} \vec e_{\nu}\right)
- b {\cal A}_{\mu}  \left(b \vec n+b \vec e_{\nu}+\frac{b}{2} \vec e_{\mu}\right)
- b {\cal A}_{\nu}  \left(b \vec n+\frac{b}{2} \vec e_{\nu}\right)
\nonumber \\
&& =
b {\cal A}_{\mu}  \left( \vec x - \frac{b}{2} \vec e_{\nu}\right)
+ b{\cal A}_{\nu}  \left( \vec x+\frac{b}{2} \vec e_{\mu}\right)
- b {\cal A}_{\mu}  \left( \vec x+\frac{b}{2} \vec e_{\nu}\right)
- b {\cal A}_{\nu}  \left( \vec x-\frac{b}{2} \vec e_{\mu}\right)
\label{fluxplaquette}
\end{eqnarray}

If one makes the following gauge transformation involving the function $U( b \vec n)$
\begin{eqnarray}
P^{[t]}( b \vec n)   =  e^{ - \frac{U( b \vec n)}{2} } {\hat P}^{[t]}( b \vec n)
\label{gaugeP}
\end{eqnarray}
then the new function ${\hat P}^{[t]}( b \vec n) $ evolves with the Euclidean Scr\"odinger equation governed by the 
 Hamiltonian similar to Eq. \ref{Hamiltonianlattice}
\begin{eqnarray}
 {\hat  {\cal H} }  = - \sum_{\vec n} w(b \vec n,b \vec n )  \vert b \vec n \rangle \langle b \vec n   \vert
 -  \sum_{\vec n} \sum_{\mu=1}^d \frac{ D_{\mu} \left(b \vec n+\frac{b}{2} \vec e_{\mu}\right) }{b^2}
\bigg(  e^{- b {\hat {\cal A}}_{\mu} \left(b \vec n+\frac{b}{2} \vec e_{\mu}\right)}   \vert b \vec n \rangle \langle b \vec n + b \vec e_{\mu}  \vert
+   e^{ b {\hat {\cal A}}_{\mu} \left(b \vec n+\frac{b}{2} \vec e_{\mu}\right)}  \vert b \vec n + b \vec e_{\mu}\rangle \langle b \vec n   \vert \bigg) \ \ \ \ \ \ 
\label{Hamiltonianlatticetilted}
\end{eqnarray}
but with the new vector potential
\begin{eqnarray}
{\hat {\cal A}}_{\mu}  \left(b \vec n+\frac{b}{2} \vec e_{\mu}\right) 
= {\cal A}_{\mu} \left(b \vec n+\frac{b}{2} \vec e_{\mu}\right)
+ \frac{U( b \vec n + b \vec e_{\mu}) - U( b \vec n)} { 2b} 
\label{vectorpotlatticetilde}
\end{eqnarray}
associated to the same magnetic flux $\Phi_{\mu \nu} (\vec x)  $ of Eq. \ref{fluxplaquette} for each plaquette.


\subsubsection{ Fokker-Planck generator as a non-hermitian electromagnetic quantum Hamiltonian
in continuous space     }

The Fokker-Planck dynamics of Eq. \ref{fokkerplanckb}
can be interpreted as an euclidean Schr\"odinger Equation
\begin{eqnarray}
 \partial_t \rho^{[t]}( \vec x )    = -H \rho^{[t]}( \vec x ) 
\label{schrodinger}
\end{eqnarray}
where the quantum non-hermitian Hamiltonian $H$  can be rewritten using the vector potential of Eq. \ref{vectorpotlattice}
on continuous space
\begin{eqnarray}
{\cal A}_{\mu} ( \vec x )  \equiv \frac{  F_{\mu} ( \vec x )}{2 D_{\mu} ( \vec x )}
\label{vectorpot}
\end{eqnarray}
in the electromagnetic form
\begin{eqnarray}
H && = \sum_{\mu=1}^d \bigg( -  \frac{\partial }{ \partial x_{\mu}}  D_{\mu} (\vec x)  \frac{\partial }{ \partial x_{\mu}} +  F_{\mu} ( \vec x) \partial_{\mu} 
+  \frac{\partial F_{\mu} ( \vec x)}{ \partial x_{\mu}}    \bigg)
\nonumber \\
&&  =  -  \sum_{\mu=1}^d  \bigg(  \frac{\partial }{ \partial x_{\mu}}  -  {\cal A}_{\mu} ( \vec x ) \bigg) D_{\mu} (\vec x) \bigg(  \frac{\partial }{ \partial x_{\mu}}  -  {\cal A}_{\mu} ( \vec x ) \bigg)
    +  V(\vec x)
\label{Hamiltonian}
\end{eqnarray}
where the scalar potential $V(\vec x) $ reads
\begin{eqnarray}
V(\vec x) \equiv  \sum_{\mu=1}^d \left[  D_{\mu} (\vec x) {\cal A}^2_{\mu} ( \vec x ) +  \partial_{\mu} \big(D_{\mu} (\vec x) {\cal A}_{\mu} ( \vec x ) \big) \right]
 = \sum_{\mu=1}^d \left[ \frac{ F_{\mu}^2 (\vec x) }{ 4 D_{\mu} (\vec x)} + \frac{ \partial_{\mu} F_{\mu} (\vec x)}{2} \right] 
\label{scalarpot}
\end{eqnarray}

The plaquette magnetic flux $\Phi_{\mu \nu}$ of Eq. \ref{fluxplaquette}
scales as the surface $b^2$ of the plaquette in the limit $b \to 0$
\begin{eqnarray}
\Phi_{\mu \nu} (\vec x)
&& \equiv  b \left[ {\cal A}_{\nu}  \left( \vec x+\frac{b}{2} \vec e_{\mu}\right)
-  {\cal A}_{\nu}  \left( \vec x-\frac{b}{2} \vec e_{\mu}\right) \right]
- b \left[ {\cal A}_{\mu}  \left( \vec x+\frac{b}{2} \vec e_{\nu}\right)
- {\cal A}_{\mu}  \left( \vec x - \frac{b}{2} \vec e_{\nu}\right)   \right] 
\nonumber \\
&& = b^2 \left[ \partial_{\mu} {\cal A}_{\nu} (\vec x) - \partial_{\nu} {\cal A}_{\mu} (\vec x)   + o(b) \right]
\equiv b^2 \left[  {\cal B}_{\mu \nu} (\vec x)   + o(b) \right]
\label{fluxplaquetteb}
\end{eqnarray}
where the magnetic matrix ${\cal B}_{\mu \nu} (\vec x) $
is given by the formula generalizing the three-dimensional curl the vector potential $\vec A ( \vec x) $ of Eq. \ref{vectorpot}
\begin{eqnarray}
 {\cal B}_{\mu \nu}(\vec x )  
 \equiv \partial_{\mu} {\cal A}_{\nu} (\vec x) - \partial_{\nu} {\cal A}_{\mu} (\vec x)
 \label{magneticN}
\end{eqnarray}
and is invariant under the gauge transformations : 
if one makes the gauge transformation involving the function $U(  \vec x)$ analogous to Eq. \ref{gaugeP}
\begin{eqnarray}
\rho^{[t]}(  \vec x)   =  e^{ - \frac{U(  \vec x)}{2} } {\hat \rho}^{[t]}(  \vec x)
\label{gaugerho}
\end{eqnarray}
then the new function ${\hat \rho}^{[t]}( b \vec n) $ evolves with the Euclidean Scr\"odinger equation governed by the 
 Hamiltonian similar to Eq. \ref{Hamiltonian}
\begin{eqnarray}
 {\hat  H }  = -  \sum_{\mu=1}^d  \bigg( \partial_{\mu}  -  {\hat {\cal A}}_{\mu} ( \vec x ) \bigg) D_{\mu} (\vec x) \bigg( \partial_{\mu}  -  {\hat {\cal A}}_{\mu} ( \vec x ) \bigg)
    +  V(\vec x)
\label{Hamiltoniantilted}
\end{eqnarray}
but with the new vector potential corresponding to the limit $b \to 0$ of Eq. \ref{vectorpotlatticetilde}
\begin{eqnarray}
{\hat {\cal A}}_{\mu}  (\vec x)
= {\cal A}_{\mu} (\vec x) + \frac{ \partial_{\mu} U(  \vec x) } { 2} 
\label{vectorpottilde}
\end{eqnarray}
associated to the same magnetic matrix $B_{\mu \nu} (\vec x)  $ of Eq. \ref{magneticN}.


\subsubsection{ Discussion     }

As discussed in detail in \cite{us_gyrator} on the special case of the uniform diffusion coefficients $D_{\mu} (\vec x)=\frac{1}{2} $, the magnetic matrix $B_{\mu \nu} (\vec x) $ is the relevant parameter of the irreversibility of the Fokker-Planck dynamics, and gauge transformations are very useful to analyze many properties.
On the lattice, the relevant parameters of irreversibility are the magnetic fluxes $\Phi_{\mu \nu} (\vec x)  $ of Eq. \ref{fluxplaquette} associated to plaquettes.
 This link with the
non-hermitian quantum mechanics in electromagnetic potentials
will be also very useful for the time-discretized regularization considered in Appendix \ref{app_PathIntegral}.


\section{ Level 2.5 for the Markov jump process on the lattice in the limit $b \to 0$ }

\label{sec_level2.5}

For the continuous-time random walk on the lattice described in the previous section \ref{sec_generator}, the general analysis concerning general Markov jump processes 
summarized in subsection \ref{intro_jump} can be applied as follows.

\subsection{Empirical observables $ \left[P(.)  ;  J_.(.) ; A_.(.) \right]$ for finite $b$ 
with their rescalings for $b \to 0$} 

The relevant empirical observables that determine the trajectory probabilities are :

(i) the empirical probability $P(\vec x= b \vec n) $ of Eq. \ref{rho1pj} 
defined on the sites $\vec x= b \vec n $ of the lattice satisfying the normalization of Eq. \ref{rho1ptnormaj}
\begin{eqnarray}
\sum_{\vec n}  P(b \vec n)   = 1
\label{rho1ptnormalattice}
\end{eqnarray}

(ii) the empirical flows $Q(.,.)$ of Eq. \ref{jumpempiricaldensity} 
that can be replaced via Eq. \ref{jafromq}
by the empirical currents $J_{\mu}\left(b \vec n+\frac{b}{2} \vec e_{\mu}\right)$
and by the empirical activities $A_{\mu}\left(b \vec n+\frac{b}{2} \vec e_{\mu}\right)$ of the links 
\begin{eqnarray}
J_{\mu}\left(b \vec n+\frac{b}{2} \vec e_{\mu}\right) && 
\equiv Q(b \vec n+b \vec e_{\mu},b \vec n) -  Q(b \vec n, b \vec n+b \vec e_{\mu})
\nonumber \\
A_{\mu}\left(b \vec n+\frac{b}{2} \vec e_{\mu}\right) && 
\equiv Q(b \vec n+b \vec e_{\mu},b \vec n) +  Q(b \vec n, b \vec n+b \vec e_{\mu})
\label{jafromqlattice}
\end{eqnarray}
with the stationarity constraints of Eq. \ref{contrainteqj}
\begin{eqnarray}
0 = \sum_{\mu=1}^d \left[ J_{\mu}\left(b \vec n+\frac{b}{2} \vec e_{\mu} \right) 
 -  J_{\mu}\left(b \vec n-\frac{b}{2} \vec e_{\mu}\right) \right] \ \ \ \text{ for any $\vec n$}
\label{contrainteqjlattice}
\end{eqnarray}

In summary, the constraints $C^{[2.5]}\left[ P(.)  ; J_.(.) \right] $
 of Eq. \ref{constraints2.5masterj}
read for the lattice model
\begin{eqnarray}
C^{[2.5]}\left[ P(.)  ; J_.(.) \right] = \delta \left( \sum_{\vec n}  P(b \vec n)   - 1\right)
\prod_{\vec n} \left(  \sum_{\mu=1}^d \left[ J_{\mu}\left(b \vec n+\frac{b}{2} \vec e_{\mu} \right) 
 -  J_{\mu}\left(b \vec n-\frac{b}{2} \vec e_{\mu}\right) \right]\right)
\label{C2.5lattice}
\end{eqnarray}

In the limit of vanishing lattice spacing $b \to 0$, 
the rescaling of the empirical observables $\left[ P(.)  ; J_.(.) ; A_.(.) \right] $
with respect to the lattice spacing $b$
is similar to Eqs \ref{latticedensity} \ref{jrescal} and \ref{alinkrescal}
\begin{eqnarray}
P( \vec x) && = b^d \rho(\vec x)
\nonumber \\
J_{\mu}( \vec x) && \equiv   b^{d-1} j_{\mu} ( \vec x)
\nonumber \\
A_{\mu}( \vec x) && \equiv   b^{d-2} a_{\mu} ( \vec x)
\label{empiricalRescal}
\end{eqnarray}
so that the constraints of Eq. \ref{C2.5lattice} become the constraints 
$c_{2.25} [ \rho(.),  \vec j(.)]  $ of Eq. \ref{c2.25diff} for diffusion processes in the limit $b \to 0$.


\subsection{Trajectory information ${\cal I}^{Traj}[ \vec x(0 \leq t \leq T)] = {\cal I} \left[P(.)  ;  J_.(.) ; A_.(.) \right]$ for finite $b$ and for $b \to 0$ } 

The trajectory information ${\cal I} \left[P(.)  ;  J_.(.) ; A_.(.) \right]$
of Eq. \ref{informationjumpempiaj} in terms of the empirical observables
\begin{eqnarray}
{\cal I} \left[P(.)  ;  J_.(.) ; A_.(.) \right]
&& =  \sum_{\vec n } \sum_{\mu=1}^d 
\bigg[ w(b \vec n+b \vec e_{\mu},b \vec n) P(b \vec n) 
+ w(b \vec n,b \vec n+b \vec e_{\mu})  P(b \vec n+b \vec e_{\mu}) 
\nonumber \\
&&
+ \frac{J_{\mu}\left( b \vec n+\frac{b}{2} \vec e_{\mu}\right)}{2} 
 \ln \left( \frac{  w(b \vec n,b \vec n+b \vec e_{\mu}) }
{   w(b \vec n+b \vec e_{\mu},b \vec n)  }  \right) 
  - \frac{A_{\mu}\left( b \vec n+\frac{b}{2} \vec e_{\mu}\right) }{2}  
\ln \bigg(  w(b \vec n+b \vec e_{\mu},b \vec n)w(b \vec n,b \vec n+b \vec e_{\mu})  \bigg)  
  \bigg] \ \ \ 
\label{informationjumpempiajlattice}
\end{eqnarray}
reads using the rates of Eq. \ref{rates}
\begin{eqnarray}
{\cal I} \left[P(.)  ;  J_.(.) ; A_.(.) \right]
&& =  \sum_{\vec n } \sum_{\mu=1}^d 
\bigg[\frac{ D_{\mu} \left(b \vec n+\frac{b}{2} \vec e_{\mu}\right) }{b^2} 
\left( e^{ \frac{ b F_{\mu} \left(b \vec n+\frac{b}{2} \vec e_{\mu}\right)}{2 D_{\mu} \left(b \vec n+\frac{b}{2} \vec e_{\mu}\right)} } P(b \vec n) 
+   e^{- \frac{ b F_{\mu} \left(b \vec n+\frac{b}{2} \vec e_{\mu}\right)}{2 D_{\mu} \left(b \vec n+\frac{b}{2} \vec e_{\mu}\right)} }  P(b \vec n+b \vec e_{\mu}) \right)
\nonumber \\
&&
- b J_{\mu}\left( b \vec n+\frac{b}{2} \vec e_{\mu}\right)
  \frac{  F_{\mu} \left(b \vec n+\frac{b}{2} \vec e_{\mu}\right)}{ 2 D_{\mu} \left(b \vec n+\frac{b}{2} \vec e_{\mu}\right)}  - \frac{A_{\mu}\left( b \vec n+\frac{b}{2} \vec e_{\mu}\right) }{2}  
\ln \left(   \frac{ D_{\mu}^2 \left(b \vec n+\frac{b}{2} \vec e_{\mu}\right) }{b^4}   \right) 
   \bigg] 
\label{informationjumpempiajlatticerates}
\end{eqnarray}

In the limit of vanishing lattice spacing $b \to 0$, 
the rescaling of Eq. \ref{empiricalRescal}
for the empirical observables $\left[ P(.)  ; J_.(.) ; A_.(.) \right]  $
leads to the following form for Eq. \ref{informationjumpempiajlatticerates}
\begin{eqnarray}
&& {\cal I} [ P(.)=b^d \rho(.) ; J_.(.)=b^{d-1}j_.(.) ; A_.(.) =b^{d-2}a_.(.)]
\nonumber \\
&& =   \sum_{\mu=1}^d \sum_{\vec n } b^d
\bigg[\frac{ D_{\mu} \left(b \vec n+\frac{b}{2} \vec e_{\mu}\right) }{b^2} 
\left( e^{ \frac{ b F_{\mu} \left(b \vec n+\frac{b}{2} \vec e_{\mu}\right)}{2 D_{\mu} \left(b \vec n+\frac{b}{2} \vec e_{\mu}\right)} }  \rho(b \vec n) 
+   e^{- \frac{ b F_{\mu} \left(b \vec n+\frac{b}{2} \vec e_{\mu}\right)}{2 D_{\mu} \left(b \vec n+\frac{b}{2} \vec e_{\mu}\right)} }   \rho(b \vec n+b \vec e_{\mu}) \right)
\nonumber \\
&&
-  j_{\mu}\left( b \vec n+\frac{b}{2} \vec e_{\mu}\right)
  \frac{  F_{\mu} \left(b \vec n+\frac{b}{2} \vec e_{\mu}\right)}{ 2 D_{\mu} \left(b \vec n+\frac{b}{2} \vec e_{\mu}\right)}  - \frac{ a_{\mu}\left( b \vec n+\frac{b}{2} \vec e_{\mu}\right) }{2 b^2}  
\ln \left(   \frac{ D_{\mu}^2 \left(b \vec n+\frac{b}{2} \vec e_{\mu}\right) }{b^4}   \right) 
   \bigg]  
\label{informationjumpempiajlatticeratesrescal}
\end{eqnarray}
For each $\mu$, the replacement of the sum $\sum_{\vec n } b^d $ 
by an integral $\int d\vec x $ over the link position $\vec x=b \vec n+\frac{b}{2} \vec e_{\mu} $ yields
\begin{eqnarray}
{\cal I} [ b^d \rho(.) ; b^{d-1}J_.(.) ; b^{d-2}A_.(.)]
&& =   \sum_{\mu=1}^d \int d\vec x
\bigg[\frac{ D_{\mu} (\vec x) }{b^2} 
\left( e^{ \frac{ b F_{\mu} (\vec x)}{2 D_{\mu} (\vec x)} }  \rho\left( \vec x- \frac{b}{2} \vec e_{\mu}\right)
+   e^{- \frac{ b F_{\mu} (\vec x)}{2 D_{\mu}(\vec x)} }   \rho\left( \vec x+ \frac{b}{2} \vec e_{\mu}\right) \right)
 \nonumber \\ &&
-  j_{\mu}(\vec x)
  \frac{  F_{\mu} (\vec x)}{ 2 D_{\mu} (\vec x)}  - \frac{ a_{\mu}(\vec x) }{2 b^2}  
\ln \left(   \frac{ D_{\mu}^2 (\vec x) }{b^4}   \right) 
   \bigg] 
\label{informationjumpempiajlatticeratesrescalint}
\end{eqnarray}
The series expansion for small $b$ of $e^{ \pm \frac{ b F_{\mu} (\vec x)}{2 D_{\mu} (\vec x)} } $ and $\rho\left( \vec x \pm \frac{b}{2} \vec e_{\mu}\right) $
leads to the expansion of Eq. \ref{informationjumpempiajlatticeratesrescalint} up to order $o(b)$
\begin{eqnarray}
&& {\cal I} [ b^d \rho(.) ; b^{d-1}j_.(.) ; b^{d-2}a_.(.)]
=  \sum_{\mu=1}^d \int d\vec x
\bigg[-  j_{\mu}(\vec x)
  \frac{  F_{\mu} (\vec x)}{ 2 D_{\mu} (\vec x)}  - \frac{ a_{\mu}(\vec x) }{2 b^2}  
\ln \left(   \frac{ D_{\mu}^2 (\vec x) }{b^4}   \right) 
 \nonumber \\ 
&& +\frac{ 2 D_{\mu} (\vec x) }{b^2} 
\bigg( \rho( \vec x ) \left[ 1 +\frac{ b^2 F^2_{\mu} (\vec x)}{8 D^2_{\mu} (\vec x)}  \right] 
- b^2\frac{F_{\mu} (\vec x)}{4 D_{\mu} (\vec x)} \frac{\partial \rho( \vec x ) }{\partial x_{\mu} }  
+ \frac{b^2}{8} \frac{\partial^2 \rho( \vec x ) }{\partial x_{\mu}^2 } +o(b^2)
\bigg)
   \bigg] 
   \nonumber \\
&& =  \sum_{\mu=1}^d \int d\vec x
\bigg[\frac{ 2 D_{\mu} (\vec x)  \rho( \vec x ) - \frac{ a_{\mu}(\vec x) }{2}  \ln \left(   \frac{ D_{\mu}^2 (\vec x) }{b^4}   \right)}{b^2}  
-  j_{\mu}(\vec x)  \frac{  F_{\mu} (\vec x)}{ 2 D_{\mu} (\vec x)}  
+   \frac{  F^2_{\mu} (\vec x)}{4 D_{\mu} (\vec x)}   \rho( \vec x )
- \frac{F_{\mu} (\vec x)}{2} \frac{\partial \rho( \vec x ) }{\partial x_{\mu} }  
+  \frac{D_{\mu} (\vec x)}{4} \frac{\partial^2 \rho( \vec x ) }{\partial x_{\mu}^2 } 
   \bigg]    +o(b) \nonumber \\
\label{informationjumpempiajlatticeratesrescalseries}
\end{eqnarray}

An alternative expression
can be obtained via the integration by parts of the two last terms involving derivatives of $ \rho( \vec x ) $
in order to rewrite them in terms of $ \rho( \vec x ) $
\begin{eqnarray}
&& {\cal I} [ b^d \rho(.) ; b^{d-1}J_.(.) ; b^{d-2}A_.(.)]
\nonumber \\
&& =  \sum_{\mu=1}^d \int d\vec x
\bigg[\frac{ 2 D_{\mu} (\vec x)  \rho( \vec x ) - \frac{ a_{\mu}(\vec x) }{2}  \ln \left(   \frac{ D_{\mu}^2 (\vec x) }{b^4}   \right)}{b^2}  
-  j_{\mu}(\vec x)  \frac{  F_{\mu} (\vec x)}{ 2 D_{\mu} (\vec x)}  
+   \rho( \vec x ) \left( \frac{  F^2_{\mu} (\vec x)}{4 D_{\mu} (\vec x)}  
+ \frac{1}{2} \frac{\partial F_{\mu} (\vec x) }{\partial x_{\mu} }  
+  \frac{1}{4} \frac{\partial^2 D_{\mu} (\vec x) }{\partial x_{\mu}^2 } 
\right)
   \bigg]    +o(b) \nonumber \\
\label{informationjumpempiajlatticeratesrescalseriesfinal}
\end{eqnarray}


\subsection{Entropy $S^{[2.5]}\left[ P(.)  ; J_.(.) ; A_.(.)  \right] $ of trajectories with given empirical observables $\left[ P(.)  ; J_.(.) ; A_.(.)  \right]$ }

The entropy $S^{[2.5]}\left[ P(.)  ; J_.(.) ; A_.(.)  \right] $ of trajectories with given empirical observables $\left[ P(.)  ; J_.(.) ; A_.(.)  \right] $ of Eq. \ref{entropyempiajjump}
reads
\begin{eqnarray}
 S^{[2.5]}\left[ P(.)  ; J(.,.) ; A(.,.) \right]    
&& = \sum_{\vec n } \sum_{\mu=1}^d 
\bigg[ A_{\mu}\left( b \vec n+\frac{b}{2} \vec e_{\mu}\right)  - \frac{A_{\mu}\left( b \vec n+\frac{b}{2} \vec e_{\mu}\right) }{2}  
\ln \left( \frac{ A_{\mu}^2\left( b \vec n+\frac{b}{2} \vec e_{\mu}\right) 
-J_{\mu}^2 \left( b \vec n+\frac{b}{2} \vec e_{\mu}\right) }
{4    P(b \vec n) P(b \vec n+b \vec e_{\mu}) }  \right) 
\nonumber \\
&&
- \frac{J_{\mu}\left( b \vec n+\frac{b}{2} \vec e_{\mu}\right)}{2} 
 \ln \left( \frac{ \left[ A_{\mu}\left( b \vec n+\frac{b}{2} \vec e_{\mu}\right) +J_{\mu} \left( b \vec n+\frac{b}{2} \vec e_{\mu}\right)\right]   P(b \vec n+b \vec e_{\mu})}
{ \left[A_{\mu}\left( b \vec n+\frac{b}{2} \vec e_{\mu}\right) - J_{\mu}\left( b \vec n+\frac{b}{2} \vec e_{\mu}\right) \right]
   P(b \vec n) }  \right) 
  \bigg] 
\label{entropyempiajjumplattice}
\end{eqnarray}

In the limit of vanishing lattice spacing $b \to 0$, 
the rescaling of Eq. \ref{empiricalRescal}
for the empirical observables $\left[ P(.)  ; J_.(.) ; A_.(.)  \right]  $
leads to the following form for Eq. \ref{entropyempiajjumplattice}
\begin{eqnarray}
S^{[2.5]} [ b^d \rho(.) ; b^{d-1}j_.(.) ; b^{d-2}a_.(.)]  
&& =  \sum_{\mu=1}^d \sum_{\vec n } b^d
\bigg[ \frac{a_{\mu}\left( b \vec n+\frac{b}{2} \vec e_{\mu}\right)}{b^2} 
 - \frac{a_{\mu}\left( b \vec n+\frac{b}{2} \vec e_{\mu}\right) }{2 b^2}  
\ln \left( \frac{ a_{\mu}^2\left( b \vec n+\frac{b}{2} \vec e_{\mu}\right)
 - b^2 j_{\mu}^2 \left( b \vec n+\frac{b}{2} \vec e_{\mu}\right) }
{4  b^4  \rho(b \vec n) \rho(b \vec n+b \vec e_{\mu}) }  \right) 
\nonumber \\
&&
- \frac{ j_{\mu}\left( b \vec n+\frac{b}{2} \vec e_{\mu}\right)}{2 b} 
 \ln \left( \frac{ \left[ a_{\mu}\left( b \vec n+\frac{b}{2} \vec e_{\mu}\right) +b j_{\mu} \left( b \vec n+\frac{b}{2} \vec e_{\mu}\right)\right]   \rho(b \vec n+b \vec e_{\mu})}
{ \left[a_{\mu}\left( b \vec n+\frac{b}{2} \vec e_{\mu}\right) - bj_{\mu}\left( b \vec n+\frac{b}{2} \vec e_{\mu}\right) \right]
   \rho(b \vec n) }  \right) 
  \bigg] \ \ \ 
\label{entropyempiajjumplatticerescal}
\end{eqnarray}
For each $\mu$, the replacement of the sum $\sum_{\vec n } b^d $ 
by an integral $\int d\vec x $ over the link position $\vec x=b \vec n+\frac{b}{2} \vec e_{\mu} $ yields
\begin{eqnarray}
S^{[2.5]} [ b^d \rho(.) ; b^{d-1}j_.(.) ; b^{d-2}a_.(.)]  
&& =  \sum_{\mu=1}^d \int d\vec x
\bigg[ \frac{a_{\mu}(\vec x) }{b^2} 
 - \frac{a_{\mu}(\vec x) }{2 b^2}  
\ln \left( \frac{ a_{\mu}^2(\vec x)
 - b^2 j_{\mu}^2 (\vec x) }
{4  b^4  \rho\left(  \vec x-\frac{b}{2} \vec e_{\mu}\right) \rho\left(  \vec x+\frac{b}{2} \vec e_{\mu}\right) }  \right) 
\nonumber \\
&&
- \frac{ j_{\mu}(\vec x)}{2 b} 
 \ln \left( \frac{ \left[ a_{\mu}(\vec x) +b j_{\mu} (\vec x)\right] 
   \rho\left(  \vec x+\frac{b}{2} \vec e_{\mu}\right)}
{ \left[a_{\mu}(\vec x) - b j_{\mu}(\vec x) \right]
  \rho\left(  \vec x-\frac{b}{2} \vec e_{\mu}\right) }  \right) 
  \bigg] \ \ \ 
\label{entropyempiajjumplatticerescalint}
\end{eqnarray}
The series expansion for small $b$ of $\left[a_{\mu}(\vec x) \pm b j_{\mu}(\vec x) \right] $ and $\rho\left( \vec x \pm \frac{b}{2} \vec e_{\mu}\right) $
leads to the final result
\begin{eqnarray}
&&S^{[2.5]} [ b^d \rho(.) ; b^{d-1}j_.(.) ; b^{d-2}a_.(.)]  
 =  \sum_{\mu=1}^d \int d\vec x
\bigg[ \frac{a_{\mu}(\vec x) }{b^2} 
 - \frac{a_{\mu}(\vec x) }{2 b^2} 
  \ln \left( \frac{ a_{\mu}^2(\vec x) \left[ 1 - b^2 \frac{ j_{\mu}^2 (\vec x)}{a_{\mu}^2(\vec x)}  +o(b^2) \right] }
{4  b^4  \rho^2 (  \vec x)
\left[ 1 + \frac{b^2}{4} 
\left( \frac{ \frac{\partial^2 \rho( \vec x ) }{\partial x_{\mu}^2 } }{ \rho (  \vec x)} 
-  \frac{ \left(\frac{\partial \rho( \vec x ) }{\partial x_{\mu} } \right)^2}{ \rho^2 (  \vec x)}
\right) +o(b^2)\right] }  \right) 
\nonumber \\
&&
- \frac{ j_{\mu}(\vec x)}{2 b} 
 \ln \left( \frac{ \left[ 1 +b \frac{ j_{\mu} (\vec x)}{a_{\mu}(\vec x)}  +o(b)\right] 
  \left[ \rho (  \vec x)+ \frac{b}{2} \frac{\partial \rho( \vec x ) }{\partial x_{\mu} }+ o(b) \right] }
{ \left[1 -b \frac{ j_{\mu} (\vec x)}{a_{\mu}(\vec x)}  +o(b) \right]
  \left[ \rho (  \vec x)- \frac{b}{2} \frac{\partial \rho( \vec x ) }{\partial x_{\mu} }+ o(b) \right] }  \right) 
  \bigg] 
  \nonumber \\
&& =  \sum_{\mu=1}^d \int d\vec x
\bigg[ \frac{a_{\mu}(\vec x)  - \frac{a_{\mu}(\vec x) }{2 }   \ln \left( \frac{ a_{\mu}^2(\vec x)  }{4  b^4  \rho^2 (  \vec x) } \right)}{b^2} 
  + \frac{a_{\mu}(\vec x) }{8} 
\left( \frac{ \frac{\partial^2 \rho( \vec x ) }{\partial x_{\mu}^2 } }{ \rho (  \vec x)} 
-  \frac{ \left(\frac{\partial \rho( \vec x ) }{\partial x_{\mu} } \right)^2}{ \rho^2 (  \vec x)}
\right)  
 - \frac{ j_{\mu}^2 (\vec x)}{2 a_{\mu}(\vec x)}
-  \frac{ j_{\mu}(\vec x) }{2 \rho( \vec x )}  \frac{\partial \rho( \vec x ) }{\partial x_{\mu} }  \bigg]    
  +o(b)
\label{entropyempiajjumplatticerescalintseries}
\end{eqnarray}


\subsection{Rate function $I_{2.5}\left[ P(.)  ; J_.(.) ; A_.(.) \right] $ for the joint distribution of the empirical observables $\left[ P(.)  ; J_.(.) ; A_.(.) \right]$ } 

The rate function at level 2.5 of Eq. \ref{rate2.5masterdiff}
 is simply the difference between Eqs \ref{informationjumpempiajlatticerates}
 and \ref{entropyempiajjumplattice}
\begin{eqnarray}
  I_{2.5} \left[ P(.)  ; J_.(.) ; A_.(.) \right]
  = {\cal I} \left[ P(.)  ; J_.(.) ; A_.(.) \right] -  S^{[2.5]}\left[ P(.)  ; J_.(.) ; A_.(.) \right]    
\label{rate2.5masterdifflattice}
\end{eqnarray}

In the limit of vanishing lattice spacing $b \to 0$ with
the rescaling of Eq. \ref{empiricalRescal}
for the empirical observables $\left[ P(.)  ; J_.(.) ; A_.(.) \right]  $,
the rate function corresponds to the difference between the series expansion of 
Eqs \ref{informationjumpempiajlatticeratesrescalseries}
and \ref{entropyempiajjumplatticerescalintseries}
\begin{eqnarray}
&& I_{2.5}[ b^d \rho(.) ; b^{d-1}j_.(.) ; b^{d-2}a_.(.)]  
= {\cal I} [ b^d \rho(.) ; b^{d-1}j_.(.) ; b^{d-2}a_.(.)]
- S^{[2.5]} [ b^d \rho(.) ; b^{d-1}j_.(.) ; b^{d-2}a_.(.)]  
\nonumber \\
&& =  \sum_{\mu=1}^d \int d\vec x
\left[\frac{ [2 D_{\mu} (\vec x)  \rho( \vec x ) - a_{\mu}(\vec x)]  
+  a_{\mu}(\vec x)   \ln \left(   \frac{a_{\mu}(\vec x)}{ 2  D_{\mu} (\vec x)  \rho (  \vec x) }   \right)}{b^2}  
+ \frac{a_{\mu}(\vec x) }{8} 
\left(  \frac{ \left(\frac{\partial \rho( \vec x ) }{\partial x_{\mu} } \right)^2}{ \rho^2 (  \vec x)}
 - \frac{ \frac{\partial^2 \rho( \vec x ) }{\partial x_{\mu}^2 } }{ \rho (  \vec x)} 
\right) 
   \right]   
   \nonumber \\
   && 
   +  \sum_{\mu=1}^d \int d\vec x
\left[   
  \frac{ j_{\mu}^2 (\vec x)}{2 a_{\mu}(\vec x)} 
 -  j_{\mu}(\vec x)  \frac{  F_{\mu} (\vec x)}{ 2 D_{\mu} (\vec x)}  
 \frac{ j_{\mu}(\vec x) }{2 \rho( \vec x )}  \frac{\partial \rho( \vec x ) }{\partial x_{\mu} }
+   \frac{  F^2_{\mu} (\vec x)}{4 D_{\mu} (\vec x)}   \rho( \vec x )
- \frac{F_{\mu} (\vec x)}{2} \frac{\partial \rho( \vec x ) }{\partial x_{\mu} }  
+  \frac{D_{\mu} (\vec x)}{4} \frac{\partial^2 \rho( \vec x ) }{\partial x_{\mu}^2 }  
  \right]    
  +o(b)
\label{i2.5latticerescalseries}
\end{eqnarray}

The first contribution diverges as $\frac{1}{b^2}$ for $b \to 0$
if the difference between the rescaled activity $a_{\mu}\left(  \vec x \right) $ 
and $2 D_{\mu}\left(  \vec x \right)  \rho( \vec x) $ remains finite.
So to keep the rescaled rate function of Eq. \ref{i2.5latticerescalseries}
finite for $b \to 0$,
the rescaled activity $a_{\mu}\left(  \vec x \right) $ 
has to be of the form
\begin{eqnarray}
 a_{\mu} ( \vec x)=2 D_{\mu}( \vec x)  \rho( \vec x) + b  \alpha_{\mu} ( \vec x)
\label{ascalfixed}
\end{eqnarray}
where the new function $ \alpha_{\mu} ( \vec x)$ takes into account the 
remaining possible subleading fluctuations of order $b$ around the leading value $2 D_{\mu}( \vec x)  \rho( \vec x) $.

Plugging Eq. \ref{ascalfixed}
into Eq. \ref{i2.5latticerescalseries}
\begin{eqnarray}
&& I_{2.5}[ b^d \rho(.) ; b^{d-1}j_.(.) ; b^{d-2}2 D_{\mu}( .)  \rho(.) + b^{d-1}  \alpha_{\mu} ( .)]  
\nonumber \\
&& =  \sum_{\mu=1}^d \int d\vec x
\left[   \frac{    \alpha^2_{\mu}(  \vec x ) }{4  D_{\mu}( \vec x)   \rho( \vec x)  }  
 +
  \frac{ j_{\mu}^2 (\vec x)}{4 D_{\mu}( \vec x)  \rho( \vec x)} 
 -  j_{\mu}(\vec x)  \frac{  F_{\mu} (\vec x)}{ 2 D_{\mu} (\vec x)}  
  \frac{ j_{\mu}(\vec x) }{2 \rho( \vec x )}  \frac{\partial \rho( \vec x ) }{\partial x_{\mu} }
+   \frac{  F^2_{\mu} (\vec x)}{4 D_{\mu} (\vec x)}   \rho( \vec x )
- \frac{F_{\mu} (\vec x)}{2} \frac{\partial \rho( \vec x ) }{\partial x_{\mu} }  
+ \frac{ D_{\mu}( \vec x)  }{4  \rho (  \vec x)}    \left(\frac{\partial \rho( \vec x ) }{\partial x_{\mu} } \right)^2
  \right]    
  +o(b)
  \nonumber \\
&& =  \sum_{\mu=1}^d \int d\vec x \frac{ \alpha^2_{\mu}(  \vec x ) }{4  D_{\mu}( \vec x)   \rho( \vec x)  }        
  \nonumber \\
&& +  \sum_{\mu=1}^d \int d\vec x
\frac{ \left[  j_{\mu}^2 (\vec x)
 -  2 j_{\mu}(\vec x)   F_{\mu} (\vec x)  \rho( \vec x)
   + 2  j_{\mu}(\vec x)   D_{\mu}( \vec x)  \frac{\partial \rho( \vec x ) }{\partial x_{\mu} }   
+   F^2_{\mu} (\vec x)   \rho^2( \vec x ) 
- 2 F_{\mu} (\vec x) \rho( \vec x) D_{\mu}( \vec x)    \frac{\partial \rho( \vec x ) }{\partial x_{\mu} }    
+  D_{\mu}^2( \vec x)     \left(\frac{\partial \rho( \vec x ) }{\partial x_{\mu} } \right)^2
\right]  }{4  D_{\mu}( \vec x)   \rho( \vec x)  }     
  +o(b)
   \nonumber \\
&& =  \sum_{\mu=1}^d \int d\vec x \frac{ \alpha^2_{\mu}(  \vec x ) }{4  D_{\mu}( \vec x)   \rho( \vec x)  }+  \sum_{\mu=1}^d \int d\vec x
\frac{ \left[  j_{\mu} (\vec x)
 -     F_{\mu} (\vec x)  \rho( \vec x)
   +   D_{\mu}( \vec x)  \frac{\partial \rho( \vec x ) }{\partial x_{\mu} }   
\right]^2  }{4  D_{\mu}( \vec x)   \rho( \vec x)  }     
  +o(b)
     \nonumber \\
&& \equiv  \sum_{\mu=1}^d \int d\vec x \frac{ \alpha^2_{\mu}(  \vec x ) }{4  D_{\mu}( \vec x)   \rho( \vec x)  }
+  i_{2.25} [ \rho(.),  \vec j(.)]  
  +o(b)
\label{i2.5latticerescalseriesalpha}
\end{eqnarray}
where the last contribution corresponds to the rate function $  i_{2.25} [ \rho(.),  \vec j(.)] $ of Eq. \ref{rate2.25diff} for 
for the joint distribution of empirical density $ \rho(.) $ and the empirical current $\vec j(.) $ of diffusion processes.
The first contribution involving the subleading correction $\alpha_{\mu}(  \vec x ) $
to the rescaled activity of Eq. \ref{ascalfixed}
means that $\alpha_{\mu}(  \vec x ) $ are independent Gaussian variables of zero mean and
 of variance $2  D_{\mu}( \vec x)   \rho( \vec x) $ involving the empirical density $ \rho(.) $.

 The rescalings of Eqs \ref{empiricalRescal} with Eq. \ref{ascalfixed}
\begin{eqnarray}
J_{\mu}( \vec x) && \equiv   b^{d-1} j_{\mu} ( \vec x)
\nonumber \\
A_{\mu}( \vec x) && \equiv  b^{d-2} a_{\mu} ( \vec x)=  b^{d-2} \left[ 2 D_{\mu}( \vec x)  \rho( \vec x) + b  \alpha_{\mu} ( \vec x) \right]
= b^{d-2} 2 D_{\mu}( \vec x)  \rho( \vec x) + b^{d-1}  \alpha_{\mu} ( \vec x) 
\label{empiricalRescalalpha}
\end{eqnarray}
mean that the original empirical flows on the lattice of Eq. \ref{jafromqlattice} read
\begin{eqnarray}
 Q( \vec x+\frac{b}{2} \vec e_{\mu},b \vec n-\frac{b}{2} \vec e_{\mu}) = \frac{A_{\mu}( \vec x)+J_{\mu}( \vec x)}{2}
 = b^{d-2}  D_{\mu}( \vec x)  \rho( \vec x) +  b^{d-1}\frac{\alpha_{\mu}( \vec x)+j_{\mu}( \vec x)}{2}
\nonumber \\
 Q( \vec x-\frac{b}{2} \vec e_{\mu},b \vec n+\frac{b}{2} \vec e_{\mu}) = \frac{A_{\mu}( \vec x)-J_{\mu}( \vec x)}{2}
 =   b^{d-2}  D_{\mu}( \vec x)  \rho( \vec x) +  b^{d-1}\frac{\alpha_{\mu}( \vec x)-j_{\mu}( \vec x)}{2}
\label{jafromqlatticerescal}
\end{eqnarray}
where $\alpha_{\mu} ( \vec x) $ appears together with empirical currents $j_{\mu} ( \vec x) $
at the same same order.



\subsection{Trajectory information ${\cal I}[ b^d \rho(.) ; b^{d-1}j_.(.) ; b^{d-2}2 D_{\mu}( .)  \rho(.) + b^{d-1}  \alpha_{\mu} ( .)] $ 
with the rescaled activities } 

Now that we have understood the properties of the rescaled activities,
it is useful to plug Eq. \ref{ascalfixed}
 into the trajectory information of Eq. \ref{informationjumpempiajlatticeratesrescalseriesfinal}
 \begin{eqnarray}
&& {\cal I} [ b^d \rho(.) ; b^{d-1}j_.(.) ; b^{d-2}2 D_{\mu}( .)  \rho(.) + b^{d-1}  \alpha_{\mu} ( .)]
\nonumber \\
&&  = \frac{2}{b^2}   \int d\vec x    \rho( \vec x )
\sum_{\mu=1}^d D_{\mu} (\vec x) \left[ 1 - \frac{ 1  }{2}  \ln \left(   \frac{ D_{\mu}^2 (\vec x) }{b^4}   \right)  \right]
- \frac{1}{b}  \sum_{\mu=1}^d \int d\vec x
\left[   \frac{    \alpha_{\mu} ( \vec x) }{2}  \ln \left(   \frac{ D_{\mu}^2 (\vec x) }{b^4}   \right)  \right]
\nonumber \\
&&+   \int d\vec x
\sum_{\mu=1}^d \left[-  j_{\mu}(\vec x)  \frac{  F_{\mu} (\vec x)}{ 2 D_{\mu} (\vec x)}  
+   \rho( \vec x ) \left( \frac{  F^2_{\mu} (\vec x)}{4 D_{\mu} (\vec x)}  
+ \frac{1}{2} \frac{\partial F_{\mu} (\vec x) }{\partial x_{\mu} }  
+  \frac{1}{4} \frac{\partial^2 D_{\mu} (\vec x) }{\partial x_{\mu}^2 } 
\right)
   \right]    +o(b) 
\label{informationjumpempiajlatticeratesrescalseriesfinalalpha}
\end{eqnarray}
So besides the singular contributions that diverge for $b \to 0$,
the regular finite contribution 
 \begin{eqnarray}
 {\cal I}^{Regular} [  \rho(.) ; j_.(.) ]
&& =   \int d\vec x
\sum_{\mu=1}^d \left[-  j_{\mu}(\vec x)  \frac{  F_{\mu} (\vec x)}{ 2 D_{\mu} (\vec x)}  
+   \rho( \vec x ) \left( \frac{  F^2_{\mu} (\vec x)}{4 D_{\mu} (\vec x)}  
+ \frac{1}{2} \frac{\partial F_{\mu} (\vec x) }{\partial x_{\mu} }  
+  \frac{1}{4} \frac{\partial^2 D_{\mu} (\vec x) }{\partial x_{\mu}^2 } 
\right)
   \right]  
\nonumber \\
&&  =   \int d\vec x
 \left[- \sum_{\mu=1}^d j_{\mu}(\vec x)  {\cal A}_{\mu} ( \vec x )  
+   \rho( \vec x ) \left( V(x)  
+ \frac{1}{4}  \sum_{\mu=1}^d \frac{\partial^2 D_{\mu} (\vec x) }{\partial x_{\mu}^2 } 
\right)
   \right]  
\label{informationjumpempiajlatticeratesrescalseriesfinalalphafinite}
\end{eqnarray}
involves the vector potential $ {\cal A}_{\mu} ( \vec x )$ of Eq. \ref{vectorpot}
and the scalar potential $V(x)$ of Eq. \ref{scalarpot},
while the additional scalar potential
\begin{eqnarray}
V^{add}(\vec x) = \frac{1}{4}  \sum_{\mu=1}^d \frac{\partial^2 D_{\mu} (\vec x) }{\partial x_{\mu}^2 } 
\label{vectorpotadddiag}
\end{eqnarray}
also appears in the Feynman quantum path-integral for the mid-point discretization
as discussed around Eq. \ref{vectorpotadd}.


\section{ Trajectory observables of the Markov jump process in the limit $b \to 0$ }

\label{sec_trajobs}

In this section, we consider the general trajectories observables for the Markov jump process on the lattice 
in order to analyze the limit of vanishing lattice spacing $b \to 0$ towards diffusion processes.

\subsection{Large deviations of trajectories observables on the lattice of spacing $b $ in dimension $d$} 

As explained in detail in subsection \ref{intro_jump},
the trajectories observables ${\cal O}^{Traj}\left[ x(0 \leq t \leq T) \right] $ of the form of Eq. \ref{observableQ}
can be rewritten as functions ${\cal O} \left[ P(.)  ; J_.(.) ; A_.(.) \right]
$ of the empirical observables $\left[ P(.)  ; J_.(.) ; A_.(.) \right]$
via Eq. \ref{observableAJempi}.
For the continuous-time random walk on the lattice, we are thus interested into observables 
parametrized by the functions $\Omega( .)$ and $\Lambda_.(.) $ and $\Gamma_.(.) $
\begin{eqnarray}
&& {\cal O}^{Traj}\left[ x(0 \leq t \leq T) \right] 
={\cal O} \left[ P(.)  ; J_.(.) ; A_.(.) \right]
\nonumber \\
&&  =  \sum_{ \vec n  } P( b \vec n )  \Omega( b \vec n)  
+\sum_{\vec n } \sum_{\mu=1}^d 
\bigg[  J_{\mu}\left( b \vec n+\frac{b}{2} \vec e_{\mu}\right) \Lambda_{\mu} \left( b \vec n+\frac{b}{2} \vec e_{\mu}\right)
+ A_{\mu}\left( b \vec n+\frac{b}{2} \vec e_{\mu}\right) \Gamma_{\mu} \left( b \vec n+\frac{b}{2} \vec e_{\mu}\right)
\bigg]
\label{observableAJempilattice}
\end{eqnarray}
We have recalled in Eq. \ref{largedevadditiveprobajump}
how the probability distribution $P_T({\cal O}) $ of the trajectory observable ${\cal O} $
 could be evaluated from the joint probability $P^{[2.5]}_T\left[ P(.)  ; J_.(.) ; A_.(.) \right] $
 of the empirical observables $\left[ P(.)  ; J_.(.) ; A_.(.) \right] $.
 In the next subsection, we wish to analyze what happens to Eq. \ref{observableAJempilattice}
 in the limit $b \to 0$.


\subsection{ Limit of vanishing lattice spacing $b \to 0$
 for the trajectory observable $ {\cal O} $  }

Plugging the rescaling of the empirical observables of Eqs \ref{empiricalRescal}
into the trajectory observable of Eq. \ref{observableAJempilattice}
and replacing sums $\sum_{ \vec n  } b^d  $ by integrals $\int d^d \vec x $ over sites $\vec x=b \vec n $ 
or links $\vec x = b \vec n+\frac{b}{2} \vec e_{\mu}$
\begin{eqnarray}
&& {\cal O}\left[ b^d \rho(.) ; b^{d-1}j_.(.) ; b^{d-2}a_.(.) \right] 
\nonumber \\
&&  =  \sum_{ \vec n  } b^d \rho( b\vec n)  \Omega( b \vec n)  
+\sum_{\vec n } b^d \sum_{\mu=1}^d 
\bigg[  j_{\mu}\left( b \vec n+\frac{b}{2} \vec e_{\mu}\right) 
\frac{\Lambda_{\mu} \left( b \vec n+\frac{b}{2} \vec e_{\mu} \right) }{b}
+  a_{\mu}\left( b \vec n+\frac{b}{2} \vec e_{\mu}\right) \frac{ \Gamma_{\mu} \left( b \vec n+\frac{b}{2} \vec e_{\mu}\right)}
{b^2}
\bigg]
\nonumber \\
&& =
\int d^d \vec x \rho( \vec x)  \Omega(  \vec x)
+ \sum_{\mu=1}^d \int d^d \vec x
\bigg[  j_{\mu} ( \vec x)\frac{\Lambda_{\mu} (\vec x) }{b}
+ a_{\mu}( \vec x)\frac{ \Gamma_{\mu} (\vec x)}{b^2}\bigg]
\label{observableAJempilatticerescaldeb}
\end{eqnarray}
Then the further rescaling of Eq. \ref{ascalfixed} for the activities
yields
\begin{eqnarray}
&& {\cal O}\left[ b^d \rho(.) ; b^{d-1}j_.(.) ; b^{d-2}2 D_{\mu}( .)  \rho(.) + b^{d-1}  \alpha_{\mu} ( .) \right] 
\nonumber \\
&& =
\int d^d \vec x \rho( \vec x)  \Omega(  \vec x)
+ \sum_{\mu=1}^d \int d^d \vec x
\bigg[  j_{\mu}( \vec x) \frac{\Lambda_{\mu} (\vec x) }{b}
+ \left[ 2 D_{\mu}( \vec x)  \rho( \vec x) + b  \alpha_{\mu} ( \vec x) \right]\frac{ \Gamma_{\mu} (\vec x)}{b^2}\bigg]
\nonumber \\
&& = 
\int d^d \vec x \rho( \vec x) \left[  \Omega(  \vec x) +\sum_{\mu=1}^d 2 D_{\mu}( \vec x) \frac{ \Gamma_{\mu} (\vec x)}{b^2} \right]
+ \sum_{\mu=1}^d\int d^d \vec x  
\bigg[  j_{\mu}( \vec x) \frac{\Lambda_{\mu} (\vec x) }{b}
+    \alpha_{\mu} ( \vec x)\frac{ \Gamma_{\mu} (\vec x)}{b}\bigg]
\nonumber \\
&& \equiv
\int d^d \vec x \rho( \vec x)  \omega(\vec x)
+\int d^d \vec x \sum_{\mu=1}^d 
\bigg[  j_{\mu}( \vec x)  \lambda_{\mu}(\vec x) 
+    \alpha_{\mu} ( \vec x)  b \gamma_{\mu}(\vec x)\bigg]
\label{observableAJempilatticerescal}
\end{eqnarray}
with the new appropriate coefficients for the rescaled empirical variables $[\rho( \vec x) ; j_{\mu}( \vec x) ;\alpha_{\mu} ( \vec x)]$ 
\begin{eqnarray}
 \lambda_{\mu}(\vec x) && =\frac{\Lambda_{\mu} (\vec x) }{b} 
 \nonumber \\
\gamma_{\mu}(\vec x) && = \frac{ \Gamma_{\mu} (\vec x)}{b^2}  
 \nonumber \\
 \omega(\vec x) && =\Omega(  \vec x) +\sum_{\mu=1}^d  2 D_{\mu}( \vec x) \frac{ \Gamma_{\mu} (\vec x)}{b^2}
 = \Omega(  \vec x) +\sum_{\mu=1}^d  2 D_{\mu}( \vec x)\gamma_{\mu}(\vec x)
\label{lambdagammarescal}
\end{eqnarray}

In specific examples of trajectory observables, one thus needs to analyze
how the rescaled coefficients of Eq. \ref{lambdagammarescal} 
depend on the lattice spacing in the limit $b \to 0$ in order to compute their singular and their regular contributions.
We have already discussed the example of the trajectory information in Eq. \ref{informationjumpempiajlatticeratesrescalseriesfinalalpha}.
Let us mention some other examples.


\subsubsection{ Case of trajectory observables whose functions $\Omega(\vec x) $,  $ \lambda_{\mu}(\vec x) $ and $\gamma_{\mu}(\vec x) $ remain finite when $b \to 0$   }
 
When the functions $\Omega(\vec x) $,  $ \lambda_{\mu}(\vec x) $ and $\gamma_{\mu}(\vec x) $ 
of Eq. \ref{lambdagammarescal} remain finite when $b \to 0$,
then the trajectory observable of Eq. \ref{observableAJempilatticerescal}
\begin{eqnarray}
 {\cal O}\left[ x(0 \leq t \leq T) \right] 
 \opsimeq_{b \to 0} 
\int d^d \vec x {\hat \rho}( \vec x) \omega(\vec x)
+\int d^d \vec x \sum_{\mu=1}^d   {\hat j}_{\mu}( \vec x) \lambda_{\mu}(\vec x)
\equiv O[ \rho(.),  \vec j(.)] 
\label{observableAJempilatticerescalelima}
\end{eqnarray}
can be rewritten as a function $O[ \rho(.),  \vec j(.)]  $ of the empirical density $\rho(.)$ 
and of the empirical current $\vec j(.)$ of the diffusion process,
i.e. one recovers trajectory observables of the form of Eqs \ref{Obsdiff} and \ref{Obsdiffempi}.
Note however that the coefficient $\omega(\vec x)$ involves both 
the coefficient $\Omega (\vec x)$ associated to the lattice empirical density
and the coefficient $ \Gamma_{\mu} (\vec x)$ that was associated to the empirical activities on the lattice.
This shows how trajectory observables that involve the activities on the lattice 
become observables that involve the empirical density of the diffusion process in the limit $b \to 0$.
Let us describe an example of such transmutation in the following subsection.


\subsubsection{ Example of the sum of the squares of elementary displacements along the trajectory }

On the lattice, the square of any elementary jump reduces to $[x(t_+)-x(t_-)]^2=b^2$,
so the sum of the squares of elementary jumps along the trajectory
is directly related to the total number of jumps,
i.e. to the total activity when one sums the link activities over all the links
\begin{eqnarray}
{\cal O}^{Traj}\left[ x(0 \leq t \leq T) \right]  
&& =      \frac{1}{T} \sum_{t \in [0,T] : x(t^+) \ne x(t^-) } [x(t^+) - x(t^-)]^2  
= \frac{b^2}{T} \sum_{t \in [0,T] : x(t^+) \ne x(t^-) } 1
\nonumber \\
&&= b^2 \sum_{\vec n} \sum_{\mu=1}^d A_{\mu} \left(b \vec n+\frac{b}{2} \vec e_{\mu}\right)
\label{observableQdis}
\end{eqnarray}
This observable corresponds to the form of Eq. \ref{observableAJempilattice}
with the coefficients
\begin{eqnarray}
\Gamma_{\mu} \left( b \vec n+\frac{b}{2} \vec e_{\mu}\right) = b^2
\label{gammadis}
\end{eqnarray}
so that the rescaled coefficients of Eq. \ref{lambdagammarescal} 
\begin{eqnarray}
 \lambda_{\mu}(\vec x) && =0
 \nonumber \\
\gamma_{\mu}(\vec x) && = 1  
 \nonumber \\
 \omega(\vec x) && =\sum_{\mu=1}^d  2 D_{\mu}( \vec x)
\label{lambdagammarescalex}
\end{eqnarray}
lead to the behavior of Eq. \ref{observableAJempilatticerescal}
for small $b$
\begin{eqnarray}
 {\cal O}\left[ b^d \rho(.) ; b^{d-1}j_.(.) ; b^{d-2}2 D_{\mu}( .)  \rho(.) + b^{d-1}  \alpha_{\mu} ( .) \right] 
 \opsimeq_{b \to 0} 
\int d^d \vec x \rho( \vec x) \sum_{\mu=1}^d 2 D_{\mu}( \vec x) 
\label{observableAJempilatticerescaldis}
\end{eqnarray}
that only involves the empirical density $\rho(\vec x)$ in the limit $b \to 0$.


\section{ Conclusions  }

\label{sec_conclusion}

In this paper, we have first recalled in detail why the meaning of the level 2.5 for diffusion processes is different from the meaning of the level 2.5 for Markov chains either in discrete-time or in continuous-time. In order to better understand these important differences, we have considered two types of random walks converging towards a given diffusion process in dimension $d$ involving arbitrary space-dependent forces and diffusion coefficients, namely (i) continuous-time random walks on the regular lattice of spacing $b$ in the main text ; (ii) discrete-time random walks in continuous space with a small time-step $\tau$ in the Appendices. 

We have analyzed how the large deviations at level 2.5 for these two types of random walks behave in the limits $b \to 0$ and $\tau \to 0$ respectively, in order to describe how the fluctuations of some empirical observables of the random walks are suppressed in the limit of diffusion processes :
the main results of the present paper are the rate functions given in Eqs \ref{i2.5latticerescalseriesalpha}
and \ref{rate2.5empitildewwexprescalfin} respectively.

We have then studied the limits $b \to 0$ and $\tau \to 0$ for any trajectory observable of the random walks that can be decomposed on its empirical density and its empirical flows in order to see how it is projected on the appropriate trajectory observable of the diffusion process involving its empirical density and its empirical current. As simplest example of this type of projection, we 
have considered the sum of the squares of elementary displacements along the trajectory,
that involves the empirical flows of the Markov chains as long as $b$ and $\tau$ are finite,
but that only involves the empirical density $\rho(\vec x)$ of the limiting diffusion process 
in the limit $b \to 0$ (Eq. \ref{observableAJempilatticerescaldis}) and in the limit $\tau \to 0$ (Eq. \ref{itrajempitildeww2ex}).

As a final remark, let us mention that in both cases, 
we have emphasized the link with non-hermitian quantum mechanics
which is very useful to understand many properties of non-equilibrium Markov processes, 
as already stressed in the other works \cite{us_gyrator,c_Inverse,c_SmallNoise}.


\appendix

\section { Feynman path-integral for diffusion processes with space-dependent diffusion matrices ${\bold D}_{\mu \nu}(\vec x) $}

\label{app_PathIntegral}

In this Appendix, we describe how the Feynman path-integral for diffusion processes
corresponds to Markov chains with the appropriate kernels $W_{\tau}(.,.)$ associated to the small time-step $\tau$.

\subsection{ Fokker-Planck generator involving a diffusion matrix ${\bold D}_{\mu \nu}(\vec x) $ as an electromagnetic quantum Hamiltonian }

In the Appendices, we will consider the more general case of a $d \times d$ 
symmetric diffusion matrix 
\begin{eqnarray}
{\bold D}_{\mu \nu}(\vec x) = {\bold D}_{\nu \mu}(\vec x)
\text{ instead of the diagonal case ${\bold D}_{\mu \nu}(\vec x)=D_{\mu}(\vec x) \delta_{\mu \nu} $ considered in the main text.}
\label{Dmatrixsymmetric}
\end{eqnarray}
So for the Fokker-Planck equation written as a continuity Equation \ref{fokkerplanck},
the only difference is that the current of Eq. \ref{fokkerplanckj} is replaced by
\begin{eqnarray}
 j^{[t]}_{\mu} (\vec x) \equiv  F_{\mu} (\vec x ) \rho^{[t]}(\vec x) - \sum_{\nu=1}^d {\bold D}_{\mu \nu} (\vec x)  \frac{\partial \rho^{[t]}(\vec x)}{ \partial x_{\nu}}   
\label{fokkerplanckjnondiag}
\end{eqnarray}
As a consequence, the rewriting of the Fokker-Planck generator
as an euclidean Hamiltonian quantum
yields that the electromagnetic quantum Hamiltonian of Eq. \ref{Hamiltonian}
is replaced by 
\begin{eqnarray}
H && = - \sum_{\mu=1}^d  \sum_{\nu=1}^d  \frac{\partial }{ \partial x_{\mu}}     {\bold D}_{\mu \nu} (\vec x) \frac{\partial }{ \partial x_{\nu}} +  
\sum_{\mu=1}^d \bigg( F_{\mu} ( \vec x) \frac{\partial }{ \partial x_{\mu}}
+   \frac{\partial F_{\mu} ( \vec x)}{ \partial x_{\mu}}    \bigg)
\nonumber \\
&&  =  -  \sum_{\mu=1}^d  \sum_{\nu=1}^d
\bigg( \frac{\partial }{ \partial x_{\mu}}  -  {\cal A}_{\mu} ( \vec x ) \bigg) {\bold D}_{\mu \nu} (\vec x) 
\bigg( \frac{\partial }{ \partial x_{\nu}} -  {\cal A}_{\nu} ( \vec x ) \bigg)
    +  V(\vec x)
\label{Hamiltonianmatrix}
\end{eqnarray}
where the vector potential of Eq. \ref{vectorpot} now involves the matrix elements ${\bold D}^{-1}_{\mu \nu}(\vec x)={\bold D}^{-1}_{\nu \mu} (\vec x)$ of the inverse ${\bold D}^{-1} (\vec x)$ of the 
symmetric diffusion matrix ${\bold D} (\vec x)$ of Eq. \ref{Dmatrixsymmetric}
\begin{eqnarray}
{\cal A}_{\mu} ( \vec x )  \equiv \sum_{\nu=1}^d {\bold D}^{-1}_{\mu \nu} \frac{  F_{\nu} ( \vec x )}{2 }
\label{vectorpotmatrix}
\end{eqnarray}
while the scalar potential $V(\vec x) $ of Eq. \ref{scalarpot} is replaced by
\begin{eqnarray}
V(\vec x) && \equiv  \sum_{\mu=1}^d \sum_{\nu=1}^d  {\cal A}_{\mu} ( \vec x ) {\bold D}_{\mu \nu}(\vec x)  {\cal A}_{\nu} ( \vec x ) +  \sum_{\mu=1}^d \partial_{\mu} \left( \sum_{\nu=1}^d {\bold D}_{\mu \nu} (\vec x) {\cal A}_{\nu} ( \vec x )  \right)
\nonumber \\
&& = \sum_{\mu=1}^d \sum_{\nu=1}^d \frac{ F_{\mu} (\vec x) {\bold D}^{-1}_{\mu \nu}(\vec x) F_{\nu} (\vec x)}{ 4 } + \frac{1 }{2} \sum_{\mu=1}^d  \frac{\partial F_{\mu} ( \vec x)}{ \partial x_{\mu}}
\label{scalarpotmatrix}
\end{eqnarray}


\subsection{ Feynman path-integral associated to the quantum electromagnetic Hamiltonian $H$ 
with $[{\bold D}(.) ; {\cal A}_{.}(.); V(.)]$}

\subsubsection{ Specific issues for Feynman path-integrals when the diffusion matrix ${\bold D}(\vec x) $ is space-dependent }

The general form of the Feynman path-integral for the propagator $P(\vec x,t \vert \vec x_0,0) $ associated to the quantum Hamiltonian $H$
of Eq. \ref{Hamiltonianmatrix} can be written as follows
\begin{eqnarray}
P(\vec x,t \vert \vec x_0,0) = \langle \vec x \vert e^{- t H} \vert \vec x_0 \rangle
= \int_{\vec x(s=0)=\vec x_0}^{\vec x(s=t)=\vec x} \frac{ {\cal D}   \vec x(.) }{ {\cal K}[ \vec x(.)] }
 e^{ - \displaystyle 
  \int_0^{t} ds {\cal L} (\vec x(s), \dot {\vec x}(s) )
 } 
\label{pathintegral}
\end{eqnarray}
where $ {\cal D}   \vec x(.) $ denotes the measure over the trajectories $x(0 \leq s \leq t) $, 
while $\frac{ 1 }{ {\cal K} [ \vec x(.)] } $ is some appropriate 
normalization prefactor in front of
 the exponential containing the action, i.e. the integral over time $s \in [0,t]$
of the quadratic Lagrangian with respect to the velocity components $\dot x_{\mu} (s) $
\begin{eqnarray}
 {\cal L} (\vec x(s), \dot {\vec x}(s))
  \equiv \sum_{\mu=1}^d \sum_{\nu=1}^d \frac{ \dot x_{\mu} (s) {\bold D}^{-1}_{\mu \nu}(\vec x) \dot x_{\nu} (s)}{ 4 } 
 - \sum_{\mu=1}^d \dot x_{\mu} (s)  {\cal A}^{Lag}_{\mu}( \vec x(s))
 + V^{Lag}(\vec x(s))
\label{lagrangian}
\end{eqnarray}
that involves the appropriate vector potential ${\cal A}^{Lag}_{\mu}(\vec x) $ 
and the appropriate scalar potential $V^{Lag}( \vec x)$.

The standard case considered
 in quantum mechanics \cite{feynman} corresponds to the space-independent diagonal uniform 
diffusion matrix ${\bold D}_{\mu \nu}(\vec x) =\frac{\delta_{\mu \nu}}{2m}$  
involving only the mass $m$ of the particle : 
then the prefactor $\frac{ 1 }{ {\cal K} [ \vec x(.)] } $ does not depend on the trajectory $\vec x(.) $
and can be included as a global normalization in the path-measure ${\cal D}   \vec x(.) $,
while the vector potential ${\cal A}^{Lag}(\vec x) $ and the scalar potential $V^{Lag}( \vec x)$ appearing in the Lagrangian $ $ of Eq. \ref{lagrangian}
coincide with the vector potential ${\cal A}(\vec x) $ and the scalar potential $V( \vec x)$
appearing in the Hamiltonian of Eq. \ref{Hamiltonianmatrix}
\begin{eqnarray}
\text{Usual quantum mechanics ${\bold D}_{\mu \nu}(\vec x) =\frac{\delta_{\mu \nu}}{2m}$} :
\ \  {\cal A}^{Lag}(\vec x)={\cal A}(\vec x) \ \ \ {\rm and } \ \ V^{Lag}( \vec x)=V( \vec x)
\ \ \ \ \ \ 
\label{usualQM}
\end{eqnarray}
This property is essential for the Feynman perspective on quantum mechanics \cite{feynman},
where the quantum propagator satisfying the Schr\"odinger equation
can be interpreted as a sum over many trajectories characterized by their classical actions.
For the specific case of non-hermitian quantum mechanics in electromagnetic fields,
these path-integrals with uniform diffusion coefficient ${\bold D}_{\mu \nu}(\vec x) =\frac{\delta_{\mu \nu}}{2} $
 have been used recently to analyze many 
properties of the corresponding non-equilibrium Fokker-Planck dynamics \cite{us_gyrator}.

The generalization of path-integrals to space-dependent diffusion matrix ${\bold D}(\vec x) $,
where the prefactor $\frac{ 1 }{ {\cal K} [ \vec x(.)] } $ does depend on the trajectory $\vec x(.) $,
has been considered by many authors over the years (see for instance \cite{Morette,DeWitt,GrahamPath,GrahamNoneq,weiss,langouche,dekker,jarzynski,Vivien2017,Vivien2019,curved} and the recent review \cite{Vivien2023} with its extensive list of references),
but the field has remained somewhat confusing for the following reasons :

(i) The literature contains a lot of different expressions for the vector potential ${\cal A}^{Lag}(\vec x) $ and for the scalar potential $V^{Lag}( \vec x)$ appearing in the Lagrangian
 of Eq. \ref{lagrangian} (see the recent review \cite{Vivien2023} and references therein)
 and they are usually rather different from the electromagnetic potentials $[ {\cal A}(\vec x);V( \vec x)]$
 appearing in the Hamiltonian of Eq. \ref{Hamiltonianmatrix},
 in contrast to the nice quantum-classical correspondence recalled in Eq. \ref{usualQM} 
\begin{eqnarray}
\text{Case ${\bold D}_{\mu \nu}(\vec x) $ :
 $[ {\cal A}^{Lag}(\vec x);V^{Lag}( \vec x)]$ different from $[ {\cal A}(\vec x);V( \vec x)]$
and depend on the authors  
}
\label{DxLagrangianpot}
\end{eqnarray}
  The problem is that the various authors make different choices for the time discretization
and for the separation between what should be kept in the prefactor $\frac{ 1 }{ {\cal K} [ \vec x(.)] } $
and what should  be included in the Lagrangian $ {\cal L} (\vec x(s), \dot {\vec x}(s)) $.

(ii) Most authors have chosen to turn to the vocabulary and to the notations of Riemannian geometry
based on the introduction of the metric tensor $g_{\mu \nu}={\bold D}^{-1}_{\mu \nu}(\vec x)/2  $
with its Christoffel symbols with three indices and curvature tensor with four indices,
which is of course standard for physicists working on general relativity, but is usually less familiar for statistical physicists.
In the following, we will thus instead keep the vocabulary and the notations of quantum mechanics,
since every physicist has a clear intuition about the meaning of vector potentials and scalar potentials.

(iii) Another complication is that most authors have chosen to 
derive the path-integral for diffusion processes from the Langevin Stochastic differential equations
with its various different interpretations (Ito or Stratonovith or others).
Here again, we believe that it is actually simpler to stick to the methodology of 
path-integral in quantum mechanics,
where the Feynman path-integral is derived in correspondence with the Schr\"odinger equation involving the Hamiltonian (since one does not write Langevin Stochastic differential equations in quantum mechanics).


\subsubsection{ Markov kernel $W_{\tau}(\vec x \vert \vec y) $ on the elementary small time-step $\tau$ based on the intermediate position $\left( \beta  \vec x + (1-\beta)   \vec y\right)$}

In the present work, we will consider the following Markov kernel $W^{[\beta]}_{\tau}(\vec x \vert \vec y) $
on the elementary small time-step $\tau$, 
where the quadratic Lagrangian ${\cal L} (\vec x(s), \dot {\vec x}(s))$ of Eq. \ref{lagrangian}
is replaced by $ {\cal L} \left(  \beta  \vec x + (1-\beta)   \vec y , \frac{\vec x -\vec y}{\tau} \right) $ 
involving the discretized velocity $\frac{\vec x -\vec y}{\tau} $
and the intermediate position $\left(  \beta  \vec x + (1-\beta)   \vec y  \right) $ with the parameter $\beta \in [0,1]$
that will be also used to evaluate the position-dependent prefactor $\frac{ 1 }{ K[ .] } $ associated to the small time-interval $\tau$
\begin{eqnarray}
&&  W^{[\beta]}_{\tau}(\vec x \vert \vec y) 
 \equiv \frac{ e^{ \displaystyle - \tau {\cal L}^{[\beta]} \left(  \beta  \vec x + (1-\beta)   \vec y , \frac{\vec x -\vec y}{\tau} \right) }}{ K^{[\beta]} \left( \beta  \vec x + (1-\beta)   \vec y\right) }
 \label{pathintegraltau}
 \\
&& \equiv  \frac{  e^{ - \displaystyle  \sum_{\mu=1}^d \sum_{\nu=1}^d 
\frac{ (x_{\mu}-y_{\mu}) \left[ {\bold D}^{-1}_{\mu \nu} \left( \beta  \vec x + (1-\beta)   \vec y\right) \right] (x_{\nu}-y_{\nu}) }
{ 4 \tau } 
 + \sum_{\mu=1}^d (x_{\mu} -y_{\mu} )  {\cal A}^{[\beta]}_{\mu}\left( \beta  \vec x + (1-\beta)   \vec y\right)
 - \tau V^{[\beta]}\left( \beta  \vec x + (1-\beta)   \vec y\right) }  }
 { \displaystyle \sqrt { (4 \pi \tau)^d {\rm det } \left[ {\bold D} \left( \beta  \vec x + (1-\beta)   \vec y\right) \right] }  }
\nonumber
\end{eqnarray}
The two main choices for the intermediate position $\left(  \beta  \vec x + (1-\beta)   \vec y  \right) $
parametrized by $\beta$ are 
\begin{eqnarray}
  \beta  \vec x + (1-\beta)   \vec y
  = 
 \begin{cases} \frac{ \vec x + \vec y}{2}
\text{ for the standard mid-point choice $\beta=\frac{1}{2}$ in quantum path-integrals } 
\\  \vec y
 \text{ for the initial-point choice $\beta=0$}
\end{cases}
 \label{xbetainterpretation}
\end{eqnarray}
Let us stress that most authors consider discretization rules that are more complicated
than the simple choice of Eq. \ref{pathintegraltau} 
based on the single intermediate position $\left[\beta  \vec x + (1-\beta)   \vec y \right]$
(see the recent review \cite{Vivien2023} and references therein):
note that the position at which the scalar potential $V^{[\beta]}(.) $ is evaluated is actually irrelevant as a consequence of its prefactor $\tau$, while the positions chosen to evaluate the vector potential ${\vec {\cal A}}^{[\beta]}(.) $
and the inverse diffusion matrix ${\bold D}^{-1} (.)$ in the Lagrangian,
as well as the determinant ${\rm det } [{\bold D} (.)]$ of the diffusion matrix in the denominator 
are on the contrary very relevant
and explain the many different results that can be found in the literature.
In particular, most authors choose to evaluate the determinant ${\rm det } [{\bold D} (.)]$ of the diffusion matrix in the denominator at a different position than the intermediate position used in the Lagrangian, 
although this normalization is directly related to the gaussian part of the Lagrangian involving the inverse matrix
${\bold D}^{-1} (.)$. In the following, we will focus only on the simple choice of Eq. \ref{pathintegraltau},
already considered in \cite{jarzynski} for the mid-point case $\beta=\frac{1}{2}$
 in order to discuss fluctuation theorems involving backward trajectories.

Once the choice of Eq. \ref{pathintegraltau} has been made for the form of the elementary propagator,
the appropriate vector potential ${\vec {\cal A}}^{[\beta]}(.) $ and scalar potential $V^{[\beta]}(.) $
associated to each value of the parameter $\beta$ 
can be determined via the Feynman pedestrian method explained on page 77 of the textbook \cite{feynman}
for the special case $\beta=1/2$ and for constant diffusion matrix :
the finite-$\tau$ dynamics generated by the elementary kernel $W^{[\beta]}_{\tau}(\vec x \vert \vec y) $ of Eq. \ref{pathintegraltau}
\begin{eqnarray}
\rho_{t+\tau} (\vec x)  = \int d^d y W^{[\beta]}_{\tau}(\vec x \vert \vec y)  \rho_t (\vec y) 
\label{dynquantumtau}
\end{eqnarray}
should correspond to the Schr\"odinger equation 
with the quantum Hamiltonian $H$ of Eq. \ref{Hamiltonianmatrix}
when both sides are expanded up to order $\tau$
\begin{eqnarray}
\rho_{t} (\vec x) + \tau \partial_t \rho_t (\vec x)  +O(\tau) = 
\rho_{t} (\vec x) - \tau H \rho_t (\vec x)  +O(\tau) 
\label{dynquantumtauexp}
\end{eqnarray}
i.e. the dynamics that emerges from Eq. \ref{dynquantumtau}
at order $\tau$ should coincide with the Fokker-Planck dynamics
with the Hamiltonian of Eq. \ref{Hamiltonianmatrix}
\begin{eqnarray}
\partial_t \rho_t (\vec x)  = -  H \rho_t (\vec x) 
&& =  \sum_{\mu=1}^d  \sum_{\nu=1}^d  \frac{\partial }{ \partial x_{\mu}}  
\left[   {\bold D}_{\mu \nu} (\vec x) \frac{\partial \rho_t (\vec x)}{ \partial x_{\nu}} \right]
-  \sum_{\mu=1}^d  F_{\mu} ( \vec x) \frac{\partial \rho_t (\vec x)}{ \partial x_{\mu}}
- \rho_t (\vec x) \sum_{\mu=1}^d \frac{\partial F_{\mu} ( \vec x)}{ \partial x_{\mu}}   
\label{FPexpandrho}
\end{eqnarray}
Note that this conserved Fokker-Planck dynamics at order $\tau$
 means that the Markov kernel $W^{[\beta]}_{\tau}(\vec x \vert \vec y) $ will be normalized at order $\tau$ for any $\vec y$
 \begin{eqnarray}
 \text{ For any $\vec y$} : 1+o(\tau) &&= \int d^d x W^{[\beta]}_{\tau}(\vec x \vert \vec y) 
\label{pathintegraltaunormadef}
\end{eqnarray}

So let us write the finite-$\tau$ dynamics of Eq. \ref{dynquantumtau}
for the Markov kernel of Eq. \ref{pathintegraltau}
using the change of variable from $\vec y$ to $\vec Z= \frac{\vec x - \vec y}{\sqrt \tau }$
\begin{eqnarray}
&& \rho_{t+\tau} (\vec x)  = \int d^d y W^{[\beta]}_{\tau}(\vec x \vert \vec y)  \rho_t (\vec y) 
= \int d^d \vec Z   \rho_t\left( \vec x - {\sqrt \tau } \vec Z \right) 
\nonumber \\
&& \times    
 \frac{  e^{ - \displaystyle  \sum_{\mu=1}^d \sum_{\nu=1}^d 
\frac{ Z_{\mu} \left[ {\bold D}^{-1}_{\mu \nu} \left(\vec x - {\sqrt \tau} (1-\beta)  \vec Z\right) \right] Z_{\nu} }
{ 4  } 
 + { \sqrt \tau } \sum_{\mu=1}^d Z_{\mu}   {\cal A}^{[\beta]}_{\mu}\left( \vec x - {\sqrt \tau} (1-\beta) \vec Z\right)
 - \tau V^{[\beta]}\left( \vec x - {\sqrt \tau} (1-\beta) \vec Z\right) }  }
 { \displaystyle \sqrt { (4 \pi )^d {\rm det } \left[ {\bold D} \left( \vec x - {\sqrt \tau} (1-\beta) \vec Z\right) \right] }  } \ \ \ 
\label{dynquantumtaubeta}
\end{eqnarray}
It is useful to introduce the following finite-difference operator 
acting on an arbitrary function $\phi( \vec x)$
\begin{eqnarray}
 e^{\displaystyle  -  {\sqrt \tau} (1-\beta) \sum_{\nu=1}^d Z_{\nu}  \frac{\partial }{ \partial x_{\nu}} } \phi(\vec y) = \phi \left( \vec x - {\sqrt \tau} (1-\beta) \vec Z\right)
 \label{finitediffnux}
\end{eqnarray}
in order to rewrite Eq. \ref{dynquantumtaubeta}
\begin{eqnarray}
&& \rho_{t+\tau}(  \vec x)   =  \int d^d \vec Z    \rho_t\left( \vec x - {\sqrt \tau } \vec Z \right) 
  \nonumber \\
 && \times 
\left( e^{\displaystyle -  {\sqrt \tau} (1-\beta) \sum_{\nu'=1}^d Z_{\nu'}  \frac{\partial }{ \partial x_{\nu'}} } \right)
\left[ 
\frac{  e^{ - \displaystyle  \sum_{\mu=1}^d \sum_{\nu=1}^d 
\frac{ Z_{\mu} \left[ {\bold D}^{-1}_{\mu \nu} \left(\vec x \right) \right] Z_{\nu} }
{ 4  } 
 + { \sqrt \tau } \sum_{\mu=1}^d Z_{\mu}   {\cal A}^{[\beta]}_{\mu}\left( \vec x \right)
 - \tau V^{[\beta]}\left( \vec x \right) }  }
 { \displaystyle \sqrt { (4 \pi )^d {\rm det } \left[ {\bold D} \left( \vec x \right) \right] }  }
 \right]
\label{markovchainccontinuityrescal}
\end{eqnarray}

On can then plug the series expansion expansion up to order $\tau$ of
the finite-difference operator
\begin{eqnarray}
 e^{\displaystyle -  {\sqrt \tau} (1-\beta) \sum_{\nu'=1}^d Z_{\nu'}  \frac{\partial }{ \partial x_{\nu'}} }  = 
 1-  {\sqrt \tau} (1-\beta) \sum_{\nu'=1}^d Z_{\nu'}  \frac{\partial }{ \partial x_{\nu'}} 
 +  \frac{ \tau (1-\beta)^2}{2} \sum_{\nu'=1}^d \sum_{\nu''=1}^d Z_{\nu'} Z_{\nu''} \frac{\partial^2 }{ \partial x_{\nu'}\partial x_{\nu''}} +o(\tau)
  \label{finitediffnuyexp}
\end{eqnarray}
and the Taylor expansions of the functions
\begin{eqnarray}
 \rho_t\left( \vec x - {\sqrt \tau } \vec Z \right) 
 && =\rho_t( \vec x )  -  { \sqrt \tau } \sum_{\sigma=1}^d Z_{\sigma}  \frac{\partial  \rho_t( \vec x ) }{ \partial x_{\sigma}} 
 +  \frac{ \tau }{2} \sum_{\sigma=1}^d \sum_{\sigma'=1}^d Z_{\sigma} Z_{\sigma'} \frac{\partial^2  \rho_t( \vec x ) }{ \partial x_{\sigma}\partial x_{\sigma'}} +o(\tau)
\nonumber \\
e^{ \displaystyle { \sqrt \tau } \sum_{\mu=1}^d Z_{\mu}   {\cal A}^{[\beta]}_{\mu}\left( \vec y \right)
 - \tau V^{[\beta]}\left( \vec y \right)}
&& = 1  +{ \sqrt \tau } \sum_{\mu'=1}^d Z_{\mu'}   {\cal A}^{[\beta]}_{\mu'}\left( \vec y \right)
+\frac{\tau}{2} \sum_{\mu'=1}^d \sum_{\mu''=1}^d  Z_{\mu'} Z_{\mu''}   {\cal A}^{[\beta]}_{\mu'} \left( \vec y \right) {\cal A}^{[\beta]}_{\mu''} \left( \vec y \right)
 - \tau V^{[\beta]}\left( \vec y \right)  +o(\tau)
  \label{expseries}
\end{eqnarray}
into Eq. \ref{markovchainccontinuityrescal}
to obtain the various contributions
\begin{eqnarray}
&&  \rho_t(  \vec x) + \tau \partial_t \rho_t(  \vec x) + o(\tau) 
 =   \frac{ \tau }{2} \sum_{\sigma=1}^d \sum_{\sigma'=1}^d  \frac{\partial^2  \rho_t( \vec x ) }{ \partial x_{\sigma}\partial x_{\sigma'}} 
 \left[
  \int d^d \vec Z Z_{\sigma} Z_{\sigma'}
 \frac{  e^{ - \displaystyle  \sum_{\mu=1}^d \sum_{\nu=1}^d 
\frac{ Z_{\mu} \left[ {\bold D}^{-1}_{\mu \nu} \left( \vec x \right) \right] Z_{\nu} } { 4  }  }  }
 { \displaystyle \sqrt { (4 \pi )^d {\rm det } \left[ {\bold D} \left( \vec x  \right) \right] }  } 
 \right]
 \nonumber \\
 &&
    -  { \sqrt \tau } \sum_{\sigma=1}^d   \frac{\partial  \rho_t( \vec x ) }{ \partial x_{\sigma}} 
 \int d^d \vec Z 
 \left[ \frac{  e^{ - \displaystyle  \sum_{\mu=1}^d \sum_{\nu=1}^d 
\frac{ Z_{\mu} \left[ {\bold D}^{-1}_{\mu \nu} \left( \vec x \right) \right] Z_{\nu} } { 4  }  }  }
 { \displaystyle \sqrt { (4 \pi )^d {\rm det } \left[ {\bold D} \left( \vec x  \right) \right] }  } 
 \left(  Z_{\sigma}  +{ \sqrt \tau } \sum_{\mu'=1}^d Z_{\sigma} Z_{\mu'}   {\cal A}_{\mu'}\left( \vec x \right)
  \right)
 \right]  
 \nonumber \\
 &&
    +   \tau  (1-\beta)\sum_{\sigma=1}^d   \frac{\partial  \rho_t( \vec x ) }{ \partial x_{\sigma}} 
       \sum_{\nu'=1}^d   \frac{\partial }{ \partial x_{\nu'}} 
 \left[  \int d^d \vec Z Z_{\sigma}Z_{\nu'}
  \frac{  e^{ - \displaystyle  \sum_{\mu=1}^d \sum_{\nu=1}^d 
\frac{ Z_{\mu} \left[ {\bold D}^{-1}_{\mu \nu} \left( \vec x \right) \right] Z_{\nu} } { 4  }  }  }
 { \displaystyle \sqrt { (4 \pi )^d {\rm det } \left[ {\bold D} \left( \vec x  \right) \right] }  } 
 \right] 
 \nonumber \\
 &&
 + \rho_t( \vec x )  \int d^d \vec Z 
 \left[ \frac{  e^{ - \displaystyle  \sum_{\mu=1}^d \sum_{\nu=1}^d 
\frac{ Z_{\mu} \left[ {\bold D}^{-1}_{\mu \nu} \left( \vec x \right) \right] Z_{\nu} } { 4  }  }  }
 { \displaystyle \sqrt { (4 \pi )^d {\rm det } \left[ {\bold D} \left( \vec x  \right) \right] }  } 
 \left(  1  +{ \sqrt \tau } \sum_{\mu'=1}^d Z_{\mu'}   {\cal A}^{[\beta]}_{\mu'}\left( \vec x \right)
+\frac{\tau}{2} \sum_{\mu'=1}^d \sum_{\mu''=1}^d  Z_{\mu'} Z_{\mu''}   {\cal A}^{[\beta]}_{\mu'} \left( \vec x \right) {\cal A}^{[\beta]}_{\mu''} \left( \vec x \right)
 - \tau V^{[\beta]}\left( \vec x \right)  \right)
 \right]
  \nonumber \\
 &&
 -  \sqrt{\tau} (1-\beta) \rho_t( \vec x )     \sum_{\nu'=1}^d   \frac{\partial }{ \partial x_{\nu'}}
\left[   \int d^d \vec Z 
  \frac{  e^{ - \displaystyle  \sum_{\mu=1}^d \sum_{\nu=1}^d 
\frac{ Z_{\mu} \left[ {\bold D}^{-1}_{\mu \nu} \left( \vec x \right) \right] Z_{\nu} } { 4  }  }  }
 { \displaystyle \sqrt { (4 \pi )^d {\rm det } \left[ {\bold D} \left( \vec x  \right) \right] }  } 
 \left(  Z_{\nu'}  +{ \sqrt \tau } \sum_{\mu'=1}^d Z_{\nu'} Z_{\mu'}   {\cal A}_{\mu'}\left( \vec x \right)
  \right)
 \right] 
  \nonumber \\
 &&
 + \frac{ \tau (1-\beta)^2}{2}
 \rho_t( \vec x ) \sum_{\nu'=1}^d \sum_{\nu''=1}^d  \frac{\partial^2 }{ \partial x_{\nu'}\partial x_{\nu''}} 
  \left[  \int d^d \vec Z Z_{\nu'} Z_{\nu''}
    \frac{  e^{ - \displaystyle  \sum_{\mu=1}^d \sum_{\nu=1}^d 
\frac{ Z_{\mu} \left[ {\bold D}^{-1}_{\mu \nu} \left( \vec x \right) \right] Z_{\nu} } { 4  }  }  }
 { \displaystyle \sqrt { (4 \pi )^d {\rm det } \left[ {\bold D} \left( \vec x  \right) \right] }  } 
 \right]
 +o(\tau) 
 \label{markovchainccontinuityseriesexp}
\end{eqnarray}

The evaluation of the various Gaussian integrals in $\vec Z$ 
\begin{eqnarray}
\int d^d \vec Z  
   \frac{  e^{ - \displaystyle  \sum_{\mu=1}^d \sum_{\nu=1}^d 
\frac{ Z_{\mu} \left[ {\bold D}^{-1}_{\mu \nu} \left( \vec x \right) \right] Z_{\nu} } { 4  }  }  }
 { \displaystyle \sqrt { (4 \pi )^d {\rm det } \left[ {\bold D} \left( \vec x  \right) \right] }  } 
 &&=1
\nonumber \\
\int d^d \vec Z  Z_{\nu'} Z_{\nu''}
   \frac{  e^{ - \displaystyle  \sum_{\mu=1}^d \sum_{\nu=1}^d 
\frac{ Z_{\mu} \left[ {\bold D}^{-1}_{\mu \nu} \left( \vec x \right) \right] Z_{\nu} } { 4  }  }  }
 { \displaystyle \sqrt { (4 \pi )^d {\rm det } \left[ {\bold D} \left( \vec x  \right) \right] }  } 
  && = 2 {\bold D}_{\nu' \nu''}(\vec x)
\label{gaussianintegrals}
\end{eqnarray}
leads to the following dynamics at order $\tau$
\begin{eqnarray}
&&   \partial_t \rho_t(  \vec x) 
 =    \sum_{\sigma=1}^d \sum_{\sigma'=1}^d  \frac{\partial^2  \rho_t( \vec x ) }{ \partial x_{\sigma}\partial x_{\sigma'}}  {\bold D}_{\sigma \sigma'}(\vec x) 
    -  2  \sum_{\sigma=1}^d   \frac{\partial  \rho_t( \vec x ) }{ \partial x_{\sigma}} 
 \left[ \sum_{\mu'=1}^d {\bold D}_{\sigma \mu'}(\vec x)   {\cal A}^{[\beta]}_{\mu'}\left( \vec x \right)  \right]
    +  2 (1-\beta)  \sum_{\sigma=1}^d   \frac{\partial  \rho_t( \vec x ) }{ \partial x_{\sigma}} 
       \sum_{\nu'=1}^d   \frac{\partial {\bold D}_{\sigma \nu'}(\vec x)}{ \partial x_{\nu'}} 
 \nonumber \\
 &&
 + \rho_t( \vec x )  
 \bigg(   \sum_{\mu'=1}^d \sum_{\mu''=1}^d   {\bold D}_{\mu' \mu''}(\vec x)    {\cal A}^{[\beta]}_{\mu'} \left( \vec x \right) {\cal A}^{[\beta]}_{\mu''} \left( \vec x \right)
 -  V^{[\beta]}\left( \vec x \right)  
\nonumber \\
 &&  -   2 (1-\beta)     \sum_{\nu'=1}^d   \frac{\partial }{ \partial x_{\nu'}}
\left[   \sum_{\mu'=1}^d {\bold D}_{\nu' \mu'}(\vec x)  {\cal A}^{[\beta]}_{\mu'}\left( \vec x \right) \right] 
+  (1-\beta)^2
  \sum_{\nu'=1}^d \sum_{\nu''=1}^d  \frac{\partial^2 {\bold D}_{\nu' \nu''}(\vec x)}{ \partial x_{\nu'}\partial x_{\nu''}} 
\bigg)
 \label{markovchainccontinuityseriesexpeval}
\end{eqnarray}
that should be identified with the Fokker-Planck dynamics of Eq. \ref{FPexpandrho} 
that reads
\begin{eqnarray}
\partial_t \rho_t (\vec x)  = -  H \rho_t (\vec x) 
&&  =  \sum_{\mu=1}^d \sum_{\nu=1}^d {\bold D}_{\mu \nu} (\vec x)
    \frac{\partial^2 \rho_t (\vec x)}{\partial x_{\mu} \partial x_{\nu}} 
+  \sum_{\mu=1}^d \left[ - F_{\mu} ( \vec x) + \sum_{\nu=1}^d  \frac{\partial {\bold D}_{\mu \nu} (\vec x)}{ \partial x_{\nu}} \right]   \frac{\partial \rho_t (\vec x)}{ \partial x_{\mu}}
- \rho_t (\vec x) \sum_{\mu=1}^d \frac{\partial F_{\mu} ( \vec x)}{ \partial x_{\mu}}  
\label{FPexpandrhobis}
\end{eqnarray}
with the following conclusions :

(i) The terms involving the double derivatives $\frac{\partial^2 \rho_t (\vec x)}{\partial x_{\mu} \partial x_{\nu}} $ always coincide.

(ii) The identification of the terms involving the single derivative $\frac{\partial \rho_t (\vec x)}{ \partial x_{\mu}} $ in Eqs \ref{markovchainccontinuityseriesexpeval} and \ref{FPexpandrhobis}
\begin{eqnarray}
  F_{\mu} ( \vec x) - \sum_{\nu=1}^d  \frac{\partial {\bold D}_{\mu \nu} (\vec x)}{ \partial x_{\nu}} 
 =      2    \sum_{\nu=1}^d {\bold D}_{\mu \nu}(\vec x)   {\cal A}^{[\beta]}_{\nu}\left( \vec x \right) 
    -  2 (1-\beta)   
       \sum_{\nu=1}^d   \frac{\partial {\bold D}_{\mu \nu}(\vec x)}{ \partial x_{\nu}} 
\label{FPexpandrho1}
\end{eqnarray}
determines the Lagrangian vector potential  ${\cal A}^{[\beta]}_{\sigma}\left( \vec x \right)  $
in terms of the Fokker-Planck force $F_{\mu} ( \vec x) $ and the diffusion matrix ${\bold D}(\vec r) $
\begin{eqnarray}
 {\cal A}^{[\beta]}_{\sigma}\left( \vec x \right)   
&& = \frac{1}{2} \sum_{\mu=1}^d {\bold D}^{-1}_{\sigma \mu} 
\left[  F_{\mu} ( \vec x) 
+ \left( 1 - 2 \beta \right)  \sum_{\nu=1}^d  \frac{\partial {\bold D}_{\mu \nu} (\vec x)}{ \partial x_{\nu}} \right]
\label{FPexpandrho1af}
\end{eqnarray}
or equivalently in terms of the vector potential ${\cal A}_{\sigma} ( \vec x ) $ of Eq. \ref{vectorpotmatrix}  
that appear in the quantum Hamiltonian $H$ of Eq. \ref{Hamiltonianmatrix}
\begin{eqnarray}
 {\cal A}^{[\beta]}_{\sigma}\left( \vec x \right)   
&& = {\cal A}_{\sigma} ( \vec x ) 
+ \left( \frac{1}{2} - \beta \right)\sum_{\mu=1}^d {\bold D}^{-1}_{\sigma \mu}  \sum_{\nu=1}^d  \frac{\partial {\bold D}_{\mu \nu} (\vec x)}{ \partial x_{\nu}} 
\label{FPexpandrho1aba}
\end{eqnarray}

(iii) the identification of the terms involving $\rho_t(\vec x)$ 
in Eqs \ref{markovchainccontinuityseriesexpeval} and \ref{FPexpandrhobis}
\begin{eqnarray}
-  \sum_{\mu=1}^d \frac{\partial F_{\mu} ( \vec x)}{ \partial x_{\mu}}  
&& =   \sum_{\mu=1}^d \sum_{\nu=1}^d   {\bold D}_{\mu \nu}(\vec x)    {\cal A}^{[\beta]}_{\mu} \left( \vec x \right) {\cal A}^{[\beta]}_{\nu} \left( \vec x \right)
 -  V^{[\beta]}\left( \vec x \right)  
  -   2 (1-\beta)     \sum_{\nu=1}^d   \frac{\partial }{ \partial x_{\nu}}
\left[   \sum_{\mu=1}^d {\bold D}_{\nu \mu}(\vec x)  {\cal A}^{[\beta]}_{\mu}\left( \vec x \right) \right] 
\nonumber \\
&& +  (1-\beta)^2
  \sum_{\mu=1}^d \sum_{\nu=1}^d  \frac{\partial^2 {\bold D}_{\mu \nu}(\vec x)}{ \partial x_{\mu}\partial x_{\nu}} 
\label{FPexpandrho2}
\end{eqnarray}
determines the Lagrangian scalar potential $V^{[\beta]}\left( \vec x \right) $
in terms of the Lagrangian vector potential ${\cal A}^{[\beta]}_.(.) $
if one uses Eq. \ref{FPexpandrho1} to replace the Fokker-Planck force $F_{\mu} ( \vec x) $ 
\begin{eqnarray}
    V^{[\beta]}\left( \vec x \right) 
&& =  \sum_{\mu=1}^d \sum_{\nu=1}^d   {\bold D}_{\mu \nu}(\vec x)    {\cal A}^{[\beta]}_{\mu} \left( \vec x \right) {\cal A}^{[\beta]}_{\nu} \left( \vec x \right)
  +   2 \beta     \sum_{\nu=1}^d   \frac{\partial }{ \partial x_{\nu}}
\left[   \sum_{\mu=1}^d {\bold D}_{\nu \mu}(\vec x)  {\cal A}^{[\beta]}_{\mu}\left( \vec x \right) \right] 
 +  \beta^2
  \sum_{\mu=1}^d \sum_{\nu=1}^d  \frac{\partial^2 {\bold D}_{\mu \nu}(\vec x)}{ \partial x_{\mu}\partial x_{\nu}} 
\label{FPexpandrho2v}
\end{eqnarray}
Note that this value of the Lagrangian scalar potential $V^{[\beta]}\left( \vec x \right) $
ensures that the Fokker-Planck dynamics conserves the normalization at order $\tau$,
i.e. ensures that the Markov kernel $W^{[\beta]}_{\tau}(\vec x \vert \vec y) $ 
satisfies the normalization of Eq. \ref{pathintegraltaunormadef}
 at order $o(\tau)$ for any $\vec y$.
 

 
 \subsection{ Simplifications in special cases }

 \subsubsection { Advantages of the standard mid-point discretization $\beta=\frac{1}{2}$ of quantum path-integrals }

 For the standard mid-point discretization $\beta=\frac{1}{2}$ of quantum mechanic,
 the Lagrangian vector potential $ {\cal A}^{[\beta=\frac{1}{2}]}_{\sigma}\left( \vec x \right)$
 coincides with the vector potential ${\cal A}_{\sigma} ( \vec x ) $ of Eq. \ref{vectorpotmatrix}  
that appear in the quantum Hamiltonian $H$ of Eq. \ref{Hamiltonianmatrix}
 \begin{eqnarray}
 {\cal A}^{[\beta=\frac{1}{2}]}_{\sigma}\left( \vec x \right)   
 = {\cal A}_{\sigma} ( \vec x ) 
\equiv \sum_{\nu=1}^d {\bold D}^{-1}_{\mu \nu} \frac{  F_{\nu} ( \vec x )}{2 }
\label{FPexpandrho1afdemi}
\end{eqnarray}
so that the Lagrangian scalar potential $V^{[\beta=\frac{1}{2}]}\left( \vec x \right) $ of Eq. \ref{FPexpandrho2v}
\begin{eqnarray}
    V^{[\beta=\frac{1}{2}]}\left( \vec x \right) 
&& =  \sum_{\mu=1}^d \sum_{\nu=1}^d   {\bold D}_{\mu \nu}(\vec x)    {\cal A}_{\mu} \left( \vec x \right) {\cal A}_{\nu} \left( \vec x \right)
  +        \sum_{\nu=1}^d   \frac{\partial }{ \partial x_{\nu}}
\left[   \sum_{\mu=1}^d {\bold D}_{\nu \mu}(\vec x)  {\cal A}^{[\beta]}_{\mu}\left( \vec x \right) \right] 
 +  \frac{1}{4}
  \sum_{\mu=1}^d \sum_{\nu=1}^d  \frac{\partial^2 {\bold D}_{\mu \nu}(\vec x)}{ \partial x_{\mu}\partial x_{\nu}} 
  \nonumber \\
  && = \sum_{\mu=1}^d \sum_{\nu=1}^d \frac{ F_{\mu} (\vec x) {\bold D}^{-1}_{\mu \nu}(\vec x) F_{\nu} (\vec x)}{ 4 } + \frac{1 }{2} \sum_{\mu=1}^d  \frac{\partial F_{\mu} ( \vec x)}{ \partial x_{\mu}}
   +  \frac{1}{4}
  \sum_{\mu=1}^d \sum_{\nu=1}^d  \frac{\partial^2 {\bold D}_{\mu \nu}(\vec x)}{ \partial x_{\mu}\partial x_{\nu}} 
   \nonumber \\
  && 
  \equiv V(\vec x) +V^{add}(\vec x)
\label{FPexpandrho2vdemi}
\end{eqnarray}
differs from the scalar potential $V(\vec x) $ of Eq. \ref{scalarpotmatrix}
that appear in the quantum Hamiltonian $H$ of Eq. \ref{Hamiltonianmatrix}
only by the single additional contribution
\begin{eqnarray}
V^{add}(\vec x) = \frac{1}{4}
  \sum_{\mu=1}^d \sum_{\nu=1}^d  \frac{\partial^2 {\bold D}_{\mu \nu}(\vec x)}{ \partial x_{\mu}\partial x_{\nu}} 
\label{vectorpotadd}
\end{eqnarray}

From the point of view of non-hermitian quantum mechanics in electromagnetic potentials,
this mid-point choice $\beta=\frac{1}{2}$ 
is thus clearly the best choice to remain the closest to the standard quantum-classical correspondence recalled in Eq. \ref{usualQM}.

From the point of view of Markov processes, this mid-point choice $\beta=\frac{1}{2}$
is useful to make the link with the lattice model considered in the main text
corresponding to a diagonal diffusion matrix ${\bold D}_{\mu \nu}(\vec x)=D_{\mu}(\vec x) \delta_{\mu \nu} $,
where the force and the diffusion coefficients are defined on the links and not on the sites
(see the transition rates of Eq. \ref{rates}), and
where the regular finite contribution
of the trajectory information computed in Eq. \ref {informationjumpempiajlatticeratesrescalseriesfinalalphafinite}
directly involves the vector potential of Eq. \ref{FPexpandrho1afdemi}
and the scalar potential of \ref{FPexpandrho2vdemi}.
This mid-point choice is also the most appropriate
 to discuss fluctuation theorems involving backward trajectories \cite{jarzynski}.


 \subsubsection { Simplifications for the initial-point discretization $\beta=0$  } 
 
 For the initial-point discretization $\beta=0$, 
 the Lagrangian vector potential of Eqs \ref{FPexpandrho1af} \ref{FPexpandrho1aba}
 \begin{eqnarray}
 {\cal A}^{[\beta=0]}_{\sigma}\left( \vec x \right)   
&& = \frac{1}{2} \sum_{\mu=1}^d {\bold D}^{-1}_{\sigma \mu} 
\left[  F_{\mu} ( \vec x) 
+   \sum_{\sigma=1}^d  \frac{\partial {\bold D}_{\mu \sigma} (\vec x)}{ \partial x_{\sigma}} \right]
 = {\cal A}_{\sigma} ( \vec x ) 
+ \frac{1}{2} \sum_{\mu=1}^d {\bold D}^{-1}_{\sigma \mu}  \sum_{\sigma=1}^d  \frac{\partial {\bold D}_{\mu \sigma} (\vec x)}{ \partial x_{\sigma}} 
\label{FPexpandrho1afito}
\end{eqnarray}
is different from the vector potential ${\cal A}_{\sigma} ( \vec x ) $ of Eq. \ref{vectorpotmatrix}  
that appear in the quantum Hamiltonian $H$ of Eq. \ref{Hamiltonianmatrix}
whenever the diffusion matrix $ {\bold D}(\vec x) $ is space-dependent.
The Lagrangian vector potential of Eq. \ref{FPexpandrho2v} reduces to the quadratic term 
in the vector potential ${\cal A}^{[\beta=0]}_{.} \left( . \right) $
\begin{eqnarray}
    V^{[\beta=0]}\left( \vec x \right) 
&& =  \sum_{\mu=1}^d \sum_{\nu=1}^d   {\bold D}_{\mu \nu}(\vec x)    {\cal A}^{[0]}_{\mu} \left( \vec x \right) {\cal A}^{[0]}_{\nu} \left( \vec x \right) 
\label{FPexpandrho2vito}
\end{eqnarray}
 but differs from the scalar potential $V(\vec x) $ of Eq. \ref{scalarpotmatrix}
that appear in the quantum Hamiltonian $H$ of Eq. \ref{Hamiltonianmatrix} via many terms.

The quadratic form of Eq. \ref{FPexpandrho2vito} corresponds to the fact that
the Markov kernel of Eq. \ref{pathintegraltau} for $\beta=0$
 \begin{eqnarray}
&&  W^{[\beta=0]}_{\tau}(\vec x \vert \vec y) 
 =  \frac{  e^{ - \displaystyle  \sum_{\mu=1}^d \sum_{\nu=1}^d 
\frac{ (x_{\mu}-y_{\mu}) \left[ {\bold D}^{-1}_{\mu \nu} ( \vec y) \right] (x_{\nu}-y_{\nu}) }
{ 4 \tau } 
 + \sum_{\mu=1}^d (x_{\mu} -y_{\mu} )  {\cal A}^{[0]}_{\mu}( \vec y)
 - \tau  \sum_{\mu=1}^d \sum_{\nu=1}^d   {\bold D}_{\mu \nu}(\vec x)    {\cal A}^{[0]}_{\mu} \left( \vec y \right) {\cal A}^{[0]}_{\nu} \left( \vec y \right)  }  }
 { \displaystyle \sqrt { (4 \pi \tau)^d {\rm det } \left[ {\bold D} ( \vec y) \right] }  }
\label{pathintegraltauIto}
\end{eqnarray}
is then normalized for any finite time-step $\tau$
 \begin{eqnarray}
 \int d^d \vec x W^{[\beta=0]}_{\tau}(\vec x \vert \vec y)  =  1 
\label{pathintegraltauItonorma}
\end{eqnarray}
and not just at order $o(\tau)$ as in Eq. \ref{pathintegraltaunormadef}
for the other discretization choices $\beta \ne 0$.


\subsubsection{ Simplifications for the diagonal case ${\bold D}_{\mu \nu}(\vec x)=D_{\mu}(\vec x) \delta_{\mu \nu} $ for any discretization $\beta$  }

For the diagonal case ${\bold D}_{\mu \nu}(\vec x)=D_{\mu}(\vec x) \delta_{\mu \nu} $ considered in the main text,
the Markov kernel of Eq. \ref{pathintegraltaudiag} becomes
\begin{eqnarray}
  W^{[\beta]}_{\tau}(\vec x \vert \vec y) 
=  \frac{  e^{ - \displaystyle  \sum_{\mu=1}^d  \frac{ (x_{\mu}-y_{\mu})^2  }
{ 4 \tau  D_{\mu } \left(\beta  \vec x + (1-\beta)   \vec y\right)}
 + \sum_{\mu=1}^d (x_{\mu} -y_{\mu} )  {\cal A}^{[\beta]}_{\mu}\left( \beta  \vec x + (1-\beta)   \vec y\right)
 - \tau V^{[\beta]}\left( \beta  \vec x + (1-\beta)   \vec y\right) }  }
 { \displaystyle \sqrt { (4 \pi \tau)^d \prod_{\nu=1}^d   D_{\nu } \left( \beta  \vec x + (1-\beta)   \vec y\right)  }  }
\label{pathintegraltaudiag}
\end{eqnarray}
where the vector potential of Eq. \ref{FPexpandrho1af} reduces to
\begin{eqnarray}
 {\cal A}^{[\beta]}_{\mu}\left( \vec x \right)   
 = \frac{F_{\mu} ( \vec x) + (1- 2 \beta )  \frac{\partial  D_{\mu} (\vec x)}{ \partial x_{\mu}}}{2 D_{\mu} ( \vec x)}  
\label{FPexpandrho1afdiag}
\end{eqnarray}
while the scalar potential of Eq. \ref{FPexpandrho2v}
reads
\begin{eqnarray}
    V^{[\beta]}\left( \vec x \right) 
&& =  \sum_{\mu=1}^d    D_{\mu }(\vec x)   [ {\cal A}^{[\beta]}_{\mu} \left( \vec x \right) ]^2
  +   2 \beta     \sum_{\mu=1}^d   \frac{\partial }{ \partial x_{\mu}}
\left[   D_{\mu}(\vec x)  {\cal A}^{[\beta]}_{\mu}\left( \vec x \right) \right] 
 +  \beta^2  \sum_{\mu=1}^d  \frac{\partial^2 D_{\mu }(\vec x)}{ \partial x_{\mu}^2} 
 \nonumber \\
 && =  \sum_{\mu=1}^d   \frac{ \left[ F_{\mu} ( \vec x) + (1- 2 \beta )  \frac{\partial  D_{\mu} (\vec x)}{ \partial x_{\mu}} \right]^2}{4 D^2_{\mu} ( \vec x)}   
  +    \beta     \sum_{\mu=1}^d   \frac{\partial }{ \partial x_{\mu}}
\left[   F_{\mu} ( \vec x) + (1- 2 \beta )  \frac{\partial  D_{\mu} (\vec x)}{ \partial x_{\mu}}   \right] 
 +  \beta^2  \sum_{\mu=1}^d  \frac{\partial^2 D_{\mu }(\vec x)}{ \partial x_{\mu}^2} 
\label{FPexpandrho2vdiag}
\end{eqnarray}


\section{ Large deviations at Level 2.5 for the Markov chain with finite $\tau$ and in the limit $\tau \to 0$  }

\label{app_level2.5chain}

In this Appendix, we first reformulate the Level 2.5 for the Markov chains with finite $\tau$ 
in order to be able to analyze the limit $\tau \to 0$ towards diffusion processes.

\subsection{Large deviations at Level 2.5 for the Markov chain with finite $\tau$ in terms of the empirical kernel ${\tilde W}_{\tau}(\vec x, \vec y) $} 

In order to analyze later the limit $\tau \to 0$,
it is first convenient to rephrase the general analysis concerning general Markov chains 
summarized in subsection \ref{intro_chain} as follows.

\subsubsection{ Empirical density $ \rho(\vec x)  $ and empirical kernel ${\tilde W}_{\tau} (\vec x \vert \vec y) 
=\frac{ \rho^{(2)} ( \vec x, \vec y)}{\rho(\vec y)}$ with their constraints }

It will be convenient to replace the 2-point density $\rho^{(2)}(\vec x, \vec y) $ by the empirical kernel of Eq. \ref{modeleeff}
\begin{eqnarray}
{\tilde W}_{\tau}(\vec x, \vec y)  \equiv \frac{ \rho^{(2)}(\vec x, \vec y)  }{\rho( \vec y)  } 
\label{modeleeffempirical}
\end{eqnarray}
Then the two constraints of Eqs \ref{rho2ptovery} and \ref{rho2ptoverx} are replaced by
\begin{eqnarray}
1 && = \int d^d \vec x {\tilde W}_{\tau}(\vec x, \vec y) \ \ \ \text { for any } \vec y
\nonumber \\
\rho(\vec x) &&  
= \int d^d \vec y {\tilde W}_{\tau}(\vec x, \vec y) \rho( \vec y) \ \ \ \text { for any }  \vec x
\label{rho2forwempi}
\end{eqnarray}
i.e. the empirical kernel ${\tilde W}_{\tau}(\vec x, \vec y) $ should be normalized 
and should have the empirical normalized 1-point density $\rho(.)   $ as steady state.
So the constraints of Eq. \ref{constraints2.5chain} become
\begin{eqnarray}
 C_{2.5}  [  \rho(.) ; {\tilde W}_{\tau}(.,.) ]
  = \delta \left( \int d^d \vec x \rho(\vec x)  - 1 \right) 
   \left[\prod_{\vec y} \delta \left(  \int d^d \vec x {\tilde W}_{\tau}(\vec x, \vec y) -1   \right)  \right]
  \left[ \prod_{\vec x} \delta \left( \int d^d \vec y {\tilde W}_{\tau}(\vec x, \vec y) \rho( \vec y) - \rho(\vec x) \right) \right]
\label{constraints2.5chainwempi}
\end{eqnarray}


\subsubsection{Rate function $I_{2.5}\left[ \rho(.) ;  {\tilde W}(.,.) \right] $ in terms of the empirical density $ \rho(\vec x)  $ and the empirical kernel ${\tilde W} (\vec x \vert \vec y) $ } 

When the empirical 2-point density $\rho^{(2)}(\vec x, \vec y) $ is replaced by the empirical kernel of Eq. \ref{modeleeffempirical} :

(i) the trajectory information ${\cal I}^{traj} [x(0 \leq t \leq T) ] =  {\cal I}[  \rho^{(2)}(.,.)] $ of Eq. \ref{itrajempimc}
that involves the true Markov kernel $ W_{\tau}(\vec x, \vec y)$
becomes 
\begin{eqnarray}
 {\cal I}\left[ \rho(.) ; {\tilde W}_{\tau}(.,.) \right]
= - \frac{1}{ \tau } \int d^d \vec y \rho(\vec y) \int d^d \vec x   {\tilde W}_{\tau} (\vec x, \vec y) \ln \left[W_{\tau}(\vec x, \vec y) \right]
\label{itrajempitildew}
\end{eqnarray}

(ii) the entropy $S^{[2.5]}\left[\rho^{(2)}(.,.)  \right] $ of trajectories with a given empirical 2-point density $\rho^{(2)}(.,.) $
of Eq. \ref{entropyempi} becomes 
in terms of the empirical density $ \rho(\vec x)  $ and the empirical kernel ${\tilde W} (\vec x \vert \vec y) $
\begin{eqnarray}
S^{[2.5]}\left[ \rho(.) ; {\tilde W}_{\tau}(.,.)  \right]  &&  
 =  - \frac{1}{ \tau } \int d^d \vec y \rho( \vec y) \int d^d \vec x {\tilde W}_{\tau}(\vec x, \vec y)  \ln \left[ {\tilde W}_{\tau}(\vec x, \vec y)   \right]
  \label{entropyempitildew}
\end{eqnarray}

(iii) the rate function at level 2.5 of Eq. \ref{rate2.5chaindiff} 
corresponds to the difference between Eqs \ref{itrajempitildew}
and  \ref{entropyempitildew}
\begin{eqnarray}
  I_{2.5} \left[ \rho(.) ;  {\tilde W}(.,.) \right]
  = {\cal I} \left[ \rho(.) ;  {\tilde W}(.,.) \right] - S^{[2.5]}\left[\rho(.) ;  {\tilde W}_{\tau}(.,.)  \right]
  =   \frac{1}{ \tau } \int d^d \vec y \rho(\vec y) \int d^d \vec x   {\tilde W}_{\tau} (\vec x, \vec y) \ln 
 \left[ \frac{{\tilde W}_{\tau} (\vec x, \vec y)}{ W_{\tau}(\vec x, \vec y) } \right]
 \label{rate2.5empitildew}
\end{eqnarray}


\subsection{ Parametrization of the empirical kernel ${\tilde W}_{\tau} (\vec x \vert \vec y) 
$ in the limit of small time-step $\tau \to 0$ }

In the limit $\tau \to 0$, the normalized empirical density $\rho(\vec x)  $ of Eq. \ref{rho1pt} does not need any rescaling,
while the empirical kernel ${\tilde W}_{\tau} ( \vec x, \vec y) $ will have 
a scaling form analog to the true Markov kernel of Eq. \ref{pathintegraltau}
\begin{eqnarray}
&&  {\tilde W}^{[\beta]}_{\tau}(\vec x \vert \vec y) 
 \label{pathintegraltautilde}
 \\
&& \equiv  \frac{  e^{ - \displaystyle  \sum_{\mu=1}^d \sum_{\nu=1}^d 
\frac{ (x_{\mu}-y_{\mu}) \left[ {\tilde{\bold D}}^{-1}_{\mu \nu} \left( \beta  \vec x + (1-\beta)   \vec y\right) \right] (x_{\nu}-y_{\nu}) }
{ 4 \tau } 
 + \sum_{\mu=1}^d (x_{\mu} -y_{\mu} ) {\tilde {\cal A}}^{[\beta]}_{\mu}\left( \beta  \vec x + (1-\beta)   \vec y\right)
 - \tau {\tilde V}^{[\beta]}\left( \beta  \vec x + (1-\beta)   \vec y\right) }  }
 { \displaystyle \sqrt { (4 \pi \tau)^d {\rm det } \left[ {\tilde{\bold D}} \left( \beta  \vec x + (1-\beta)   \vec y\right) \right] }  }
\nonumber
\end{eqnarray}
involving the empirical diffusion matrix ${\tilde{\bold D}} $ and its inverse ${\tilde{\bold D}} $,
the empirical vector potential ${\tilde {\cal  A}}^{[\beta]}_{\mu}(.) $ 
and the empirical scalar potential ${\tilde V}^{[\beta]}(.) $ instead of their true counterparts.

The two constraints of Eq. \ref{rho2forwempi} should be valid at order $\tau$
so that one can use the calculations of Appendix \ref{app_PathIntegral} as follows :

(i) the first constraint of Eq. \ref{rho2forwempi} at order $\tau$
concerning the normalization of the empirical kernel $ {\tilde W}_{\tau}(\vec x \vert \vec y) $ of Eq. \ref{pathintegraltautilde}
  determines the empirical scalar potential ${\tilde V}^{[\beta]} $
  in terms of the empirical vector potential ${\tilde {\cal  A}}^{[\beta]}_{\mu}(.) $
 and the empirical diffusion matrix ${\tilde{\bold D}} $
via the formula analog to Eq. \ref{FPexpandrho2v}
\begin{eqnarray}
    {\tilde V}^{[\beta]}\left( \vec x \right) 
&& =  \sum_{\mu=1}^d \sum_{\nu=1}^d  
{\tilde {\bold D}}_{\mu \nu}(\vec x)    {\tilde{\cal A}}^{[\beta]}_{\mu} \left( \vec x \right) 
{\tilde {\cal A}}^{[\beta]}_{\nu} \left( \vec x \right)
  +   2 \beta     \sum_{\nu=1}^d   \frac{\partial }{ \partial x_{\nu}}
\left[   \sum_{\mu=1}^d {\tilde{\bold D}}_{\nu \mu}(\vec x) {\tilde {\cal A}}^{[\beta]}_{\mu}\left( \vec x \right) \right] 
 +  \beta^2
  \sum_{\mu=1}^d \sum_{\nu=1}^d  \frac{\partial^2 {\tilde{\bold D}}_{\mu \nu}(\vec x)}{ \partial x_{\mu}\partial x_{\nu}} 
\label{FPexpandrho2vtilde}
\end{eqnarray}

(ii) the second constraint of Eq. \ref{rho2forwempi} at order $\tau$
means that the empirical density $\rho(.)$
should be the steady state produced by the empirical kernel $ {\tilde W}_{\tau}(\vec x \vert \vec y) $ of Eq. \ref{pathintegraltautilde}.
The calculations of Eq. \ref{markovchainccontinuityseriesexpeval}
yields that it is useful to introduce the empirical force via the analog of Eq. \ref{FPexpandrho1}
\begin{eqnarray}
   {\tilde F}_{\mu} ( \vec x)  
 =      2    \sum_{\nu=1}^d {\tilde {\bold D}}_{\mu \nu}(\vec x)   {\tilde {\cal A}}^{[\beta]}_{\nu}\left( \vec x \right) 
    +  (2 \beta-1)  
       \sum_{\nu=1}^d   \frac{\partial  {\tilde{\bold D}}_{\mu \nu}(\vec x)}{ \partial x_{\nu}} 
\label{FPexpandrho1tilde}
\end{eqnarray}
in order to write the empirical current $j_{\mu}(\vec x) $ associated to the empirical density $\rho(.) $
\begin{eqnarray}
j_{\mu}(\vec x)  \equiv \rho( \vec x ) {\tilde F}_{\mu}(\vec x)  - \sum_{\nu=1}^d  {\tilde {\bold D}}_{\mu \nu}(\vec x)  \frac{\partial  \rho( \vec x ) }{ \partial x_{\nu}} 
 \label{defjtildeempi}
\end{eqnarray}
that should be divergenceless
\begin{eqnarray}
  \sum_{\mu=1}^d \frac{\partial j_{\mu}(\vec x)}{\partial x_{\mu} } =0
\label{divjempitilde}
\end{eqnarray}

In the limit of small time-step $\tau \to 0$,
the constraints of Eq. \ref{constraints2.5chainwempi} become in terms of the empirical density $\rho(.)$
and in terms of the empirical force ${\tilde F}_.(.) $ and the empirical diffusion matrix ${\tilde {\bold D}}_{..}(.) $
that parametrize the empirical kernel of Eq. \ref{pathintegraltautilde}
\begin{eqnarray}
 c_{2.5}  [  \rho(.) ; {\tilde F}_.(.)  ; {\tilde {\bold D}}_{..}(.) ]
  = \delta \left( \int d^d \vec x \rho(\vec x)  - 1 \right) 
  \left[ \prod_{\vec x} \delta \left( \sum_{\mu=1}^d \frac{\partial }{\partial x_{\mu} } \left[ \rho( \vec x ) {\tilde F}_{\mu}(\vec x)  - \sum_{\nu=1}^d  {\tilde {\bold D}}_{\mu \nu}(\vec x)  \frac{\partial  \rho( \vec x ) }{ \partial x_{\nu}}\right] \right) \right]
\label{constraints2.5chainwempid}
\end{eqnarray}


\subsection{Rate function $  I_{2.5} \left[ . \right]$ at level 2.5
in the limit of small time-step $\tau \to 0$ } 

Plugging the logarithms of the true kernel $W^{[\beta]}_{\tau}(\vec x, \vec y) $ of Eq. \ref{pathintegraltau} 
and of the empirical kernel ${\tilde W}^{[\beta]}_{\tau}(\vec x, \vec y) $ of Eq. \ref{pathintegraltautilde}
into the rate function of Eq. \ref{rate2.5empitildew} yields
using the new integration variables $\vec Z= \frac{\vec x - \vec y}{\sqrt \tau }$ 
and $\vec r= \beta  \vec x + (1-\beta)   \vec y$
\begin{eqnarray}
&&  I_{2.5} \left[ \rho(.) ;  {\tilde {\cal A}}^{[\beta]}_.(.)  ; {\tilde{\bold D}}(.) \right]
  =   \frac{1}{ \tau } \int d^d \vec y \rho(\vec y) \int d^d \vec x   {\tilde W}_{\tau} (\vec x, \vec y) 
  \left[ \ln \left[{\tilde W}_{\tau}(\vec x, \vec y) \right]
  - \ln \left[W_{\tau}(\vec x, \vec y) \right]  \right]
   \label{rate2.5empitildeww}
 \\
&&  =
 \frac{1}{ \tau } \int d^d \vec r \int d^d \vec Z  \rho \left( \vec r - {\sqrt \tau } \beta \vec Z\right) 
  \frac{  e^{ - \displaystyle  \sum_{\mu=1}^d \sum_{\nu=1}^d 
\frac{ Z_{\mu} \left[ {\tilde{\bold D}}^{-1}_{\mu \nu} (\vec r) \right] Z_{\nu} }
{  4 } 
 + {\sqrt \tau } \sum_{\mu=1}^d Z_{\mu}  {\tilde {\cal A}}^{[\beta]}_{\mu}(\vec r)
 - \tau {\tilde V}^{[\beta]}(\vec r) }  }
 { \displaystyle \sqrt { (4 \pi )^d {\rm det } \left[ {\tilde{\bold D}} (\vec r) \right] }  }
\nonumber \\
&&  \bigg[
  \frac{1}{2} \ln \left( \frac{ {\rm det } \left[ {\bold D} (\vec r) \right]}{{\rm det } \left[{\tilde {\bold D}} (\vec r) \right] }\right)
 +   \sum_{\mu'=1}^d \sum_{\nu'=1}^d 
 \left( {\bold D}^{-1}_{\mu' \nu'}(\vec r)-{\tilde {\bold D}}^{-1}_{\mu' \nu'}(\vec r)  \right)\frac{ Z_{\mu'}  Z_{\nu'} }{ 4  }
 +{\sqrt \tau } \sum_{\mu'=1}^d Z_{\mu'}  \left(  {\tilde {\cal A}}^{[\beta]}_{\mu'} (\vec r)
 - {\cal A}^{[\beta]}_{\mu'} (\vec r) \right)
 + \tau \left( V^{[\beta]}(\vec r) - {\tilde V}^{[\beta]}(\vec r)   \right)
\bigg]
\nonumber
\end{eqnarray}

One can now write the series expansion in $\tau$ of the various terms
\begin{eqnarray}
&&  I_{2.5} \left[ \rho(.) ;  {\tilde {\cal A}}^{[\beta]}_.(.)  ;{\tilde{\bold D}}(.) \right]
  =
 \frac{1}{ \tau } \int d^d \vec r \int d^d \vec Z  
  \frac{  e^{ - \displaystyle  \sum_{\mu=1}^d \sum_{\nu=1}^d 
\frac{ Z_{\mu} \left[ {\tilde{\bold D}}^{-1}_{\mu \nu} (\vec r) \right] Z_{\nu} }{  4 }  }  }
 { \displaystyle \sqrt { (4 \pi )^d {\rm det } \left[ {\tilde{\bold D}} (\vec r) \right] }  }
  \nonumber \\
&&
\left(\rho ( \vec r ) -  {\sqrt \tau } \beta
\sum_{\sigma=1}^d Z_{\sigma}  \frac{\partial \rho ( \vec r ) }{ \partial r_{\sigma}} 
 +  \frac{ \tau \beta^2}{2} \sum_{\sigma=1}^d \sum_{\sigma'=1}^d
 Z_{\sigma} Z_{\sigma'} \frac{\partial^2 \rho ( \vec r ) }{ \partial r_{\sigma}\partial r_{\sigma'}} +o(\tau)
 \right)
  \nonumber \\
&&
\left( 1+ {\sqrt \tau }\sum_{\mu''=1}^d Z_{\mu''}   {\tilde {\cal  A}}^{[\beta]}_{\mu''}\left( \vec r\right)
+ \frac{\tau }{2}\sum_{\mu''=1}^d \sum_{\nu''=1}^d 
Z_{\mu''}  Z_{\nu''} {\tilde {\cal  A}}^{[\beta]}_{\mu'}\left( \vec r\right)
{\tilde {\cal  A}}^{[\beta]}_{\nu''}\left( \vec r\right)
 - \tau {\tilde V}^{[\beta]}\left( \vec r\right)
 +o(\tau)
 \right)
\nonumber \\
&&  \bigg[
  \frac{1}{2} \ln \left( \frac{ {\rm det } \left[ {\bold D} (\vec r) \right]}{{\rm det } \left[{\tilde {\bold D}} (\vec r) \right] }\right)
 +   \sum_{\mu'=1}^d \sum_{\nu'=1}^d 
 \left( {\bold D}^{-1}_{\mu' \nu'}(\vec r)-{\tilde {\bold D}}^{-1}_{\mu' \nu'}(\vec r)  \right)\frac{ Z_{\mu'}  Z_{\nu'} }{ 4  }
 +{\sqrt \tau } \sum_{\mu'=1}^d Z_{\mu'}  \left(  {\tilde {\cal A}}^{[\beta]}_{\mu'} (\vec r)
 - {\cal A}^{[\beta]}_{\mu'} (\vec r) \right)
 + \tau \left( V^{[\beta]}(\vec r) - {\tilde V}^{[\beta]}(\vec r)   \right)
\bigg]
 \nonumber \\
&& \opsimeq_{\tau \to 0} \frac{  I_{2.5}^{[-1]}\left[ \rho(.) ;  {\tilde {\bold D}}_{.}(.) \right]}{\tau} 
+ \frac{ 0}{\sqrt \tau}
 +  I_{2.5}^{[0]}\left[ \rho(.) ; {\tilde {\cal A}}^{[\beta]}_.(.)  ; {\tilde {\bold D}}_{.}(.) \right]
 \label{rate2.5empitildewwexp}
\end{eqnarray}
where the last line involve the contributions that survive in the limit $\tau \to 0$ :

(i) the term of order $\frac{1}{\tau}$ has for coefficient
\begin{eqnarray}
 I_{2.5}^{[-1]}\left[ \rho(.)  ; {\tilde{\bold D}}(.) \right]
&&=  \int d^d \vec r \rho ( \vec r )
\int d^d \vec Z  
  \frac{  e^{ - \displaystyle  \sum_{\mu=1}^d \sum_{\nu=1}^d 
\frac{ Z_{\mu} \left[ {\tilde{\bold D}}^{-1}_{\mu \nu} (\vec r) \right] Z_{\nu} }{  4 }  }  }
 { \displaystyle \sqrt { (4 \pi )^d {\rm det } \left[ {\tilde{\bold D}} (\vec r) \right] }  } 
\nonumber \\
&& \times
 \bigg[  \frac{1}{2} \ln \left( \frac{ {\rm det } \left[ {\bold D} (\vec r) \right]}{{\rm det } \left[{\tilde {\bold D}} (\vec r) \right] }\right)
 +   \sum_{\mu'=1}^d \sum_{\nu'=1}^d 
 \left( {\bold D}^{-1}_{\mu' \nu'}(\vec r)-{\tilde {\bold D}}^{-1}_{\mu' \nu'}(\vec r)  \right)
 \frac{ Z_{\mu'}  Z_{\nu'} }{ 4  }
\bigg]
\nonumber \\
&&
=    \int d^d \vec r \rho ( \vec r )
 \bigg[  
 - \frac{1}{2} \ln \left(  {\rm det } \left[ {\bold D}^{-1}(\vec r) {\tilde  {\bold D}} (\vec r) \right] \right)
 +   \sum_{\mu'=1}^d \sum_{\nu'=1}^d 
 \left( {\bold D}^{-1}_{\mu' \nu'}(\vec r)-{\tilde {\bold D}}^{-1}_{\mu' \nu'}(\vec r)  \right)
 \frac{ {\tilde  {\bold D}}_{\nu' \mu'} (\vec r) }{ 2  }
 \bigg]
 \nonumber \\
&&
=    \frac{1}{2}   \int d^d \vec r \rho ( \vec r )
 \bigg[  
-  {\rm Tr} \left(  \ln \left[ {\bold D}^{-1}(\vec r) {\tilde  {\bold D}} (\vec r) \right] \right)
 +  {\rm Tr}   \left( {\bold D}^{-1}(\vec r) {\tilde  {\bold D}} (\vec r) \right) - d
 \bigg]
\label{itrajempitildeexpcalculm1}
\end{eqnarray}

(ii) the term of order $\frac{1}{\sqrt\tau}$ has a vanishing coefficient
because the corresponding contributions in Eq. \ref{rate2.5empitildewwexp}
are odd with respect to the variables $Z_.$ and thus disappear after the Gaussian integration.

(iii)  the finite term $I_{2.5}^{[0]}\left[ \rho(.) ; {\tilde {\cal A}}^{[\beta]}_.(.)  ; {\tilde{\bold D}}(.) \right] $
will de discussed later.


\subsubsection{ Conditions on the empirical matrix ${\tilde{\bold D}}(.) $
in order to have a finite rate function 
$I_{2.5}\left[ \rho(.) ; {\tilde {\cal A}}_.(.)  ;{\tilde{\bold D}}(.) \right] $ 
in the limit $\tau \to 0$ }

In order to have a finite rate function 
$I_{2.5}\left[ \rho(.) ; {\tilde {\cal A}}_.(.)  ;{\tilde{\bold D}}(.) \right] $ in Eq. \ref{rate2.5empitildewwexp}
in the limit $\tau \to 0$,
the ratio $\frac{  I_{2.5}^{[-1]}\left[ \rho(.) ;  {\tilde{\bold D}}(.) \right]}{\tau} $ 
involving the coefficient $ I_{2.5}^{[-1]}\left[ \rho(.)  ; {\tilde{\bold D}}(.) \right] $ of Eq. \ref{itrajempitildeexpcalculm1}
must remain finite as $\tau \to 0$, so
the empirical diffusion matrix $ {\tilde{\bold D}}(.)$
can be different from the true diffusion matrix ${\bold D}(.)$ only at order $\sqrt \tau$.
It is thus convenient to parametrize the empirical 
 diffusion matrix $ {\tilde{\bold D}}(.)$ in terms of the 
 new symmetric matrix ${\tilde {\bold G}}(\vec r)$
 as
\begin{eqnarray}
  {\tilde {\bold D}}_{\mu \nu} (\vec r) = {\bold D}_{\mu \nu} (\vec r) 
  \left[1+{\sqrt \tau} {\tilde {\bold G}}_{\mu \nu}(\vec r) \right]
  \ \ \ {\rm with } \ \ {\tilde {\bold G}}_{\mu \nu}(\vec r)={\tilde {\bold G}}_{\nu \mu}(\vec r) 
\label{empidsqrt}
\end{eqnarray}
Then the ratio $\frac{  I_{2.5}^{[-1]}\left[ \rho(.) ;  {\tilde{\bold D}}(.) \right]}{\tau} $ 
involving the coefficient $ I_{2.5}^{[-1]}\left[ \rho(.)  ; {\tilde{\bold D}}(.) \right] $ of Eq. \ref{itrajempitildeexpcalculm1}
becomes
\begin{eqnarray}
\frac{ I_{2.5}^{[-1]}\left[ \rho(.)  
 ; {\tilde{\bold D}}(.)  = {\bold D}(.)  [ {\bold 1}+{\sqrt \tau} {\tilde{\bold G}}(.)]  \right]}{\tau}
&&
=   \frac{1}{2 \tau}   \int d^d \vec r \rho ( \vec r )
 \bigg[  
-  {\rm Tr} \left(  \ln \left[ {\bold 1}+{\sqrt \tau} {\tilde{\bold G}}(\vec r) \right] \right)
 +  {\rm Tr}  
 \left[ {\bold 1}+{\sqrt \tau} {\tilde{\bold G}}(\vec r) \right] - d
 \bigg]
\nonumber \\
&& =  \frac{1}{2 \tau}   \int d^d \vec r \rho ( \vec r )
 \bigg[  
-  {\rm Tr} \left( {\sqrt \tau} {\tilde{\bold G}}(\vec r) - \frac{\tau}{2} {\tilde{\bold G}}^2(\vec r)+ o(\tau) \right)
 + {\sqrt \tau}  {\rm Tr}   \left[ {\tilde{\bold G}}(\vec r) \right] 
 \bigg]
\nonumber \\
\nonumber \\
&& =  \frac{1}{4}   \int d^d \vec r \rho ( \vec r ) {\rm Tr} \left(   {\tilde{\bold G}}^2(\vec r) \right)
 +\frac{o(\tau)}{\tau}
\nonumber \\
&&
=\frac{1}{4}   \int d^d \vec r \rho ( \vec r ) \sum_{\mu=1}^d \sum_{\nu=1}^d
 {\tilde{\bold G}}^2_{\mu \nu}(\vec r) 
+\frac{o(\tau)}{\tau}
\label{itrajempitildeexpcalculm1zeta}
\end{eqnarray}

Plugging 
the rescaling of the the empirical diffusion matrix of Eq. \ref{empidsqrt} into 
the rate function of Eq. \ref{rate2.5empitildewwexp}, one obtains 
\begin{eqnarray}
&&  I_{2.5} \left[ \rho(.) ;  {\tilde {\cal A}}_.(.) 
 ; {\tilde{\bold D}}(.)  = {\bold D}(.)  [ {\bold 1}+{\sqrt \tau} {\tilde{\bold G}}(.)]\right]
  =  \frac{1}{4}   \int d^d \vec r \rho ( \vec r ) \sum_{\mu=1}^d \sum_{\nu=1}^d
 {\tilde{\bold G}}^2_{\mu \nu}(\vec r) 
  + 
   I_{2.5}^{[0]}\left[ \rho(.) ; {\tilde {\cal A}}^{[\beta]}_.(.)  ; {\bold D}(.) \right] +\frac{o(\tau)}{\tau}
 \label{rate2.5empitildewwexprescal}
\end{eqnarray}
where the finite contribution $I_{2.5}^{[0]}\left[ . \right] $
introduced in Eq. \ref{rate2.5empitildewwexp}
has to be computed only for the leading order of Eq. \ref{empidsqrt}
where the empirical diffusion matrix coincides
with the true diffusion matrix ${\tilde{\bold D}}(.)  = {\bold D}(.) $ as done in the next subsection.


\subsubsection{ Finite contribution $I_{2.5}^{[0]}\left[ . \right] $
when the empirical diffusion matrix coincides
with the true diffusion matrix ${\tilde{\bold D}}(.)  = {\bold D}(.) $}

For ${\tilde{\bold D}}(.)  = {\bold D}(.) $,
the finite contribution $I_{2.5}^{[0]}\left[ . \right] $
of Eq. \ref{rate2.5empitildewwexp} reduces to
\begin{eqnarray}
&&  I_{2.5}^{[0]} \left[ \rho(.) ;  {\tilde {\cal A}}^{[\beta]}_.(.)  ; {\bold D}(.)\right]
  =
 \int d^d \vec r \rho ( \vec r ) \left(   V^{[\beta]}(\vec r) - {\tilde V}^{[\beta]}(\vec r)  \right)
 \int d^d \vec Z  
  \frac{  e^{ - \displaystyle  \sum_{\mu=1}^d \sum_{\nu=1}^d 
\frac{ Z_{\mu} \left[{\bold D}^{-1}_{\mu \nu} (\vec r) \right] Z_{\nu} }{  4 }  }  }
 { \displaystyle \sqrt { (4 \pi )^d {\rm det } \left[ {\bold D} (\vec r) \right] }  }
\nonumber \\
 && +  \int d^d \vec r \left(\rho ( \vec r )\sum_{\sigma=1}^d   {\tilde {\cal  A}}^{[\beta]}_{\sigma}\left( \vec r\right)
-  \beta\sum_{\sigma=1}^d   \frac{\partial \rho ( \vec r ) }{ \partial r_{\sigma}} 
 \right)
  \sum_{\mu'=1}^d   \left(  {\tilde {\cal A}}^{[\beta]}_{\mu'} (\vec r) - {\cal A}^{[\beta]}_{\mu'} (\vec r) \right)
\int d^d \vec Z  
  \frac{  e^{ - \displaystyle  \sum_{\mu=1}^d \sum_{\nu=1}^d 
\frac{ Z_{\mu} \left[{\bold D}^{-1}_{\mu \nu} (\vec r) \right] Z_{\nu} }{  4 }  }  }
 { \displaystyle \sqrt { (4 \pi )^d {\rm det } \left[ {\bold D} (\vec r) \right] }  }Z_{\sigma}Z_{\mu'}
 \nonumber \\
 &&  =
 \int d^d \vec r \rho ( \vec r ) \left(   V^{[\beta]}(\vec r) - {\tilde V}^{[\beta]}(\vec r)  \right)
  +  \int d^d \vec r \left(\rho ( \vec r )\sum_{\nu=1}^d   {\tilde {\cal  A}}^{[\beta]}_{\nu}\left( \vec r\right)
-  \beta\sum_{\nu=1}^d   \frac{\partial \rho ( \vec r ) }{ \partial r_{\nu}} 
 \right)
  \sum_{\mu=1}^d   \left(  {\tilde {\cal A}}^{[\beta]}_{\mu} (\vec r) - {\cal A}^{[\beta]}_{\mu} (\vec r) \right)
  2 {\bold D}_{\nu \mu } (\vec r) 
  \ \ \ \ \ 
 \label{rate2.5empitildewwexpzeroleading}
\end{eqnarray}
When ${\tilde{\bold D}}(.)  = {\bold D}(.) $,
the difference between $V^{[\beta]}(\vec r) $ of Eq. \ref{FPexpandrho2v}
and ${\tilde V}^{[\beta]}(\vec r) $ of Eq. \ref{FPexpandrho2vtilde} reduces to
\begin{eqnarray}
    V^{[\beta]}\left( \vec r \right) -   {\tilde V}^{[\beta]}\left( \vec r \right) 
&& =  \sum_{\mu=1}^d \sum_{\nu=1}^d   {\bold D}_{\mu \nu}(\vec r)  
\left[   {\cal A}^{[\beta]}_{\mu} \left( \vec r \right) {\cal A}^{[\beta]}_{\nu} \left( \vec r \right)
- {\tilde{\cal A}}^{[\beta]}_{\mu} \left( \vec r \right) 
{\tilde {\cal A}}^{[\beta]}_{\nu} \left( \vec r \right)\right]
 \nonumber \\
 && +   2 \beta     \sum_{\nu=1}^d   \frac{\partial }{ \partial r_{\nu}}
\left[   \sum_{\mu=1}^d {\bold D}_{\nu \mu}(\vec r) 
\left(  {\cal A}^{[\beta]}_{\mu}\left( \vec r \right) -{\tilde {\cal A}}^{[\beta]}_{\mu}\left( \vec r \right) \right) \right]  
\label{FPexpandrho2vdifference}
\end{eqnarray}
so that Eq. \ref{rate2.5empitildewwexpzeroleading}
becomes
\begin{eqnarray}
&&  I_{2.5}^{[0]} \left[ \rho(.) ;  {\tilde {\cal A}}^{[\beta]}_.(.)  ; {\bold D}(.)\right]
  =
 \int d^d \vec r \rho ( \vec r )
  \sum_{\mu=1}^d \sum_{\nu=1}^d   {\bold D}_{\mu \nu}(\vec r)  
\left[   {\cal A}^{[\beta]}_{\mu} \left( \vec r \right) {\cal A}^{[\beta]}_{\nu} \left( \vec r \right)
- {\tilde{\cal A}}^{[\beta]}_{\mu} \left( \vec r \right) 
{\tilde {\cal A}}^{[\beta]}_{\nu} \left( \vec r \right)\right]
 \nonumber \\ &&
+  2 \beta \int d^d \vec r \rho ( \vec r )     \sum_{\nu=1}^d   \frac{\partial }{ \partial r_{\nu}}
\left[   \sum_{\mu=1}^d {\bold D}_{\nu \mu}(\vec r) 
\left(  {\cal A}^{[\beta]}_{\mu}\left( \vec r \right) -{\tilde {\cal A}}^{[\beta]}_{\mu}\left( \vec r \right) \right) \right] 
  \nonumber \\ &&
  + 2 \int d^d \vec r \rho ( \vec r )\sum_{\nu=1}^d   {\tilde {\cal  A}}^{[\beta]}_{\nu}\left( \vec r\right)
  \sum_{\mu=1}^d   {\bold D}_{\nu \mu } (\vec r) 
   \left(  {\tilde {\cal A}}^{[\beta]}_{\mu} (\vec r) - {\cal A}^{[\beta]}_{\mu} (\vec r) \right)  
   \nonumber \\ &&
  +  2 \beta \int d^d \vec r   \sum_{\nu=1}^d   \frac{\partial \rho ( \vec r ) }{ \partial r_{\nu}} 
  \sum_{\mu=1}^d  {\bold D}_{\nu \mu } (\vec r)   \left(   {\cal A}^{[\beta]}_{\mu} (\vec r) - {\tilde {\cal A}}^{[\beta]}_{\mu} (\vec r) \right)
  \label{rate2.5empitildewwexpzeroleadingcalcul}
\end{eqnarray}
The two terms with coefficients $(2 \beta)$ can be written as a total derivative inside the integral
and thus disappear, while the two other contributions can be rewritten together 
using the symmetry of the diffusion matrix to obtain
\begin{eqnarray}
&&  I_{2.5}^{[0]} \left[ \rho(.) ;  {\tilde {\cal A}}^{[\beta]}_.(.)  ; {\bold D}(.)\right]
  =
 \int d^d \vec r \rho ( \vec r )
  \sum_{\mu=1}^d \sum_{\nu=1}^d   {\bold D}_{\mu \nu}(\vec r)  
\left[   {\cal A}^{[\beta]}_{\mu} \left( \vec r \right) {\cal A}^{[\beta]}_{\nu} \left( \vec r \right)
+ {\tilde{\cal A}}^{[\beta]}_{\mu} \left( \vec r \right) {\tilde {\cal A}}^{[\beta]}_{\nu} \left( \vec r \right)
-  2{\cal A}^{[\beta]}_{\mu} (\vec r) {\tilde {\cal  A}}^{[\beta]}_{\nu}\left( \vec r\right) 
\right]
\nonumber \\
&& =
 \int d^d \vec r \rho ( \vec r )
  \sum_{\mu=1}^d \sum_{\nu=1}^d   {\bold D}_{\mu \nu}(\vec r)  
\left[   - {\cal A}^{[\beta]}_{\mu} \left( \vec r \right) 
\left(  {\tilde {\cal  A}}^{[\beta]}_{\nu}\left( \vec r \right)- {\cal A}^{[\beta]}_{\nu} \left( \vec r \right) \right)
+ \left({\tilde{\cal A}}^{[\beta]}_{\mu} \left( \vec r \right)
-  {\cal A}^{[\beta]}_{\mu} (\vec r) \right) {\tilde {\cal  A}}^{[\beta]}_{\nu}\left( \vec r\right) 
\right]
\nonumber \\
&& =
 \int d^d \vec r \rho ( \vec r )
  \sum_{\mu=1}^d \sum_{\nu=1}^d   {\bold D}_{\mu \nu}(\vec r)  
 \left({\tilde{\cal A}}^{[\beta]}_{\mu} \left( \vec r \right)-  {\cal A}^{[\beta]}_{\mu} (\vec r) \right) 
\left( {\tilde {\cal  A}}^{[\beta]}_{\nu}\left( \vec r\right) - {\cal  A}^{[\beta]}_{\nu}\left( \vec r\right)\right)
  \label{rate2.5empitildewwexpzeroleadingcal}
\end{eqnarray}
When ${\tilde{\bold D}}(.)  = {\bold D}(.) $,
the difference between the empirical vector potential ${\tilde {\cal A}}^{[\beta]}_{.} (.) $
written in terms of the empirical force ${\tilde F}_.(.) $ of Eq. \ref{FPexpandrho1tilde}
the true vector potential ${\cal A}^{[\beta]}_{.} (.) $ of Eq. \ref{FPexpandrho1af}
written in terms of the true force $F_{.} ( .) $ reduce to
\begin{eqnarray}
{\tilde {\cal A}}^{[\beta]}_{\mu}\left( \vec r \right)   - {\cal A}^{[\beta]}_{\mu}\left( \vec r \right)   
&& = \frac{1}{2} \sum_{\sigma=1}^d {\bold D}^{-1}_{\mu \sigma } 
\left[ {\tilde F}_{\sigma} ( \vec r)-  F_{\sigma} ( \vec r) \right]
\label{FPexpandrho1afdiff}
\end{eqnarray}
so that Eq. \ref{rate2.5empitildewwexpzeroleadingcal}
reads in terms of the empirical force $ {\tilde F}_.(.)$
\begin{eqnarray}
&&  I_{2.5}^{[0]} \left[ \rho(.) ;  {\tilde F}_.(.)  ; {\bold D}(.)\right]
   =
\frac{1}{4} \int d^d \vec r \rho ( \vec r )
  \sum_{\sigma=1}^d \sum_{\sigma'=1}^d   {\bold D}^{-1}_{\sigma \sigma'}(\vec r)  
 \left[ {\tilde F}_{\sigma} ( \vec r)-  F_{\sigma} ( \vec r) \right]
 \left[ {\tilde F}_{\sigma'} ( \vec r)-  F_{\sigma'} ( \vec r) \right]
  \label{rate2.5empitildewwexpzeroleadingcalf}
\end{eqnarray}
One can now replace the empirical force ${\tilde F}_{\mu} ( \vec r) $
in terms of the empirical current $j_{\mu}(\vec r)  $ of Eq. \ref{defjtildeempi}
using ${\tilde{\bold D}}(.)  = {\bold D}(.) $
\begin{eqnarray}
 {\tilde F}_{\mu}(\vec r) 
= \frac{ j_{\mu}(\vec r)  + \sum_{\nu=1}^d  {\bold D}_{\mu \nu}(\vec r)  \frac{\partial  \rho( \vec r ) }{ \partial r_{\nu}} }{ \rho( \vec r ) }
 \label{Ftildejempi}
\end{eqnarray}
to rewrite Eq. \ref{rate2.5empitildewwexpzeroleadingcalf}
as
\begin{eqnarray}
  I_{2.5}^{[0]} \left[ \rho(.) ;  j_.(.) \right]
&&   =
 \int d^d \vec r 
  \sum_{\mu=1}^d \sum_{\nu=1}^d   \frac{ {\bold D}^{-1}_{\mu \nu}(\vec r)   } { 4 \rho ( \vec r ) }
 \left[ j_{\mu}(\vec r) -  F_{\mu} ( \vec r) \rho ( \vec r ) + \sum_{\sigma=1}^d  {\bold D}_{\mu \sigma}(\vec r)  \frac{\partial  \rho( \vec r ) }{ \partial r_{\sigma} }\right]
 \left[  j_{\nu}(\vec r) -  F_{\nu} ( \vec r) \rho ( \vec r ) + \sum_{\sigma'=1}^d  {\bold D}_{\nu \sigma'}(\vec r)  \frac{\partial  \rho( \vec r ) }{ \partial r_{\sigma'} } \right]
\nonumber \\
&& \equiv i_{2.25}[\rho(.) ;  j_.(.) ]
  \label{rate2.5empitildewwexpzeroleadingcalj}
\end{eqnarray}
where one recognizes the rate function $i_{2.25}[\rho(.) ;  j_.(.) ] $
for diffusion processes involving a general diffusion matrix,
 while the special case where the diffusion matrix is diagonal
 was given in Eq. \ref{rate2.25diff}
 of the main text.
 

\subsubsection{ Conclusion }

In summary, the rate function of Eq. \ref{rate2.5empitildewwexprescal} reads using Eq. \ref{rate2.5empitildewwexpzeroleadingcalj}
\begin{eqnarray}
&&  I_{2.5} \left[ \rho(.) ;  j_.(). ; 
 ; {\tilde{\bold D}}(.)  = {\bold D}(.)  [ {\bold 1}+{\sqrt \tau} {\tilde{\bold G}}(.)]\right]
  =  \frac{1}{4}   \int d^d \vec r \rho ( \vec r ) \sum_{\mu=1}^d \sum_{\nu=1}^d
 {\tilde{\bold G}}^2_{\mu \nu}(\vec r) 
  +    i_{2.25}[\rho(.) ;  j_.(.) ] +\frac{o(\tau)}{\tau}
 \label{rate2.5empitildewwexprescalfin}
\end{eqnarray}
This result is thus the analog of Eq. \ref{i2.5latticerescalseriesalpha} in the main text concerning the lattice model :
besides the rate function $i_{2.25}[\rho(.) ;  j_.(.) ] $ at level 2.25 for the diffusion process,
there are gaussian fluctuations for the matrix ${\tilde{\bold G}} $ that parametrizes
  the subleading fluctuations of the empirical diffusion matrix 
  ${\tilde{\bold D}}(.) = [ {\bold 1}+{\sqrt \tau} {\tilde{\bold G}}(.)]$
  around the true diffusion matrix ${\bold D}(.) $.



\section{ Trajectory observables for the Markov chain in the limit $\tau \to 0$ }

\label{app_trajobschain}

In this Appendix, we first consider arbitrary trajectory observables of the Markov chain
in order to describe how the limit $\tau \to 0$ can be analyzed
before discussing examples.

\subsection{ Arbitrary trajectory observables of the Markov chain }

The general trajectory observable ${\cal O}^{Traj}\left[ x(0 \leq t \leq T) \right] ={\cal O} \left[ \rho^{(2)}(.,.) \right] $ of Eqs \ref{observablerho2} \ref{observablerho2empi} becomes
using the new integration variables $\vec Z= \frac{\vec x - \vec y}{\sqrt \tau }$
 and $\vec r= \beta  \vec x + (1-\beta)   \vec y$
\begin{eqnarray}
&& {\cal O} \left[ \rho(.) ; {\tilde W}_{\tau} (.,.)\right]
 =   \int d^d \vec x  \int d^d \vec y \Omega^{(2)}(\vec x , \vec y) 
 {\tilde W}_{\tau}(\vec x , \vec y) \rho(\vec y)
\nonumber \\
&& =  \int d^d \vec r \int d^d \vec Z \Omega^{(2)}\left(
 \vec r + {\sqrt \tau } (1-\beta) \vec Z,  \vec r - {\sqrt \tau } \beta \vec Z\right) 
 \frac{  e^{ - \displaystyle  \sum_{\mu=1}^d \sum_{\nu=1}^d 
\frac{ Z_{\mu} \left[ {\tilde{\bold D}}^{-1}_{\mu \nu} (\vec r) \right] Z_{\nu} }
{  4 } 
 + {\sqrt \tau } \sum_{\mu=1}^d Z_{\mu}  {\tilde {\cal A}}^{[\beta]}_{\mu}(\vec r)
 - \tau {\tilde V}^{[\beta]}(\vec r) }  }
 { \displaystyle \sqrt { (4 \pi )^d {\rm det } \left[ {\tilde{\bold D}} (\vec r) \right] }  }
 \rho \left( \vec r - {\sqrt \tau } \beta \vec Z\right)
 \ \ \ \ 
\label{observableQempiw}
\end{eqnarray}

It is thus convenient to introduce the series expansion of $ \Omega^{(2)}\left( \vec x= \vec r + {\sqrt \tau } (1-\beta) \vec Z,  \vec y=\vec r - {\sqrt \tau } \beta \vec Z\right) $ with respect to the variables $\vec Z$
\begin{eqnarray}
\Omega^{(2)}\left( \vec x= \vec r + {\sqrt \tau } (1-\beta) \vec Z,  \vec y=\vec r - {\sqrt \tau } \beta \vec Z\right) 
\equiv \Omega_0(\vec r) + \sum_{\mu_1=1}^d Z_{\mu_1} \Omega_{\mu_1}(\vec r) 
+  \sum_{\mu_1=1}^d \sum_{\mu_2=1}^d Z_{\mu_1} Z_{\mu_2} \Omega_{\mu_1 \mu_2}(\vec r) +...
\label{observableDVz}
\end{eqnarray}

Plugging Eq. \ref{observableDVz} into Eq. \ref{observableQempiw}
yields
\begin{eqnarray}
&& {\cal O}\left[\rho(.) ; {\tilde {\cal A}}^{[\beta]}_.(.)  ;{\tilde{\bold D}}(.) \right]
=   \int d^d \vec r  
\int d^d \vec Z  \rho \left( \vec r - {\sqrt \tau } \beta \vec Z\right) 
  \frac{  e^{ - \displaystyle  \sum_{\mu=1}^d \sum_{\nu=1}^d 
\frac{ Z_{\mu} \left[ {\tilde{\bold D}}^{-1}_{\mu \nu} (\vec r) \right] Z_{\nu} }
{  4 } 
 + {\sqrt \tau } \sum_{\mu=1}^d Z_{\mu}  {\tilde {\cal A}}^{[\beta]}_{\mu}(\vec r)
 - \tau {\tilde V}^{[\beta]}(\vec r) }  }
 { \displaystyle \sqrt { (4 \pi )^d {\rm det } \left[ {\tilde{\bold D}} (\vec r) \right] }  }
 \nonumber \\
&&
\times \left[ \Omega_0(\vec r) + \sum_{\mu_1=1}^d Z_{\mu_1} \Omega_{\mu_1}(\vec r) 
+  \sum_{\mu_1=1}^d \sum_{\mu_2=1}^d Z_{\mu_1} Z_{\mu_2} \Omega_{\mu_1 \mu_2}(\vec r) +...\right]
 \nonumber \\
&&
=   \int d^d \vec r \int d^d \vec Z  
  \frac{  e^{ - \displaystyle  \sum_{\mu=1}^d \sum_{\nu=1}^d 
\frac{ Z_{\mu} \left[ {\tilde{\bold D}}^{-1}_{\mu \nu} (\vec r) \right] Z_{\nu} }{  4 }  }  }
 { \displaystyle \sqrt { (4 \pi )^d {\rm det } \left[ {\tilde{\bold D}} (\vec r) \right] }  }
  \nonumber \\
&&
\left(\rho ( \vec r ) -  {\sqrt \tau } \beta
\sum_{\sigma=1}^d Z_{\sigma}  \frac{\partial \rho ( \vec r ) }{ \partial r_{\sigma}} 
 +  \frac{ \tau \beta^2}{2} \sum_{\sigma=1}^d \sum_{\sigma'=1}^d
 Z_{\sigma} Z_{\sigma'} \frac{\partial^2 \rho ( \vec r ) }{ \partial r_{\sigma}\partial r_{\sigma'}} +o(\tau)
 \right)
  \nonumber \\
&&
\left( 1+ {\sqrt \tau }\sum_{\mu''=1}^d Z_{\mu''}   {\tilde {\cal  A}}^{[\beta]}_{\mu''}\left( \vec r\right)
+ \frac{\tau }{2}\sum_{\mu''=1}^d \sum_{\nu''=1}^d 
Z_{\mu''}  Z_{\nu''} {\tilde {\cal  A}}^{[\beta]}_{\mu'}\left( \vec r\right)
{\tilde {\cal  A}}^{[\beta]}_{\nu''}\left( \vec r\right)
 - \tau {\tilde V}^{[\beta]}\left( \vec r\right)
 +o(\tau)
 \right)
  \nonumber \\
&&
\times \left[ \Omega_0(\vec r) + \sum_{\mu_1=1}^d Z_{\mu_1} \Omega_{\mu_1}(\vec r) 
+  \sum_{\mu_1=1}^d \sum_{\mu_2=1}^d Z_{\mu_1} Z_{\mu_2} \Omega_{\mu_1 \mu_2}(\vec r) +...\right]
\label{itrajempitildeww}
\end{eqnarray}

In order to identify the terms that will survive in the limit $\tau \to 0$ after the Gaussian integration over the variables $\vec Z$, 
one needs to 
discuss how the functions $\Omega_0(\vec r) $, $\Omega_{\mu_1}(\vec r) $, $ \Omega_{\mu_1 \mu_2}(\vec r) $ ...
depend on $\tau$ for the specific observable of Eq. \ref{observableDVz}
one is interested in.
In the previous Appendix, we have described in detail the example of the rate function at level 2.5
with the identification of the singular and regular contributions as $\tau \to 0$
in Eq. \ref{rate2.5empitildewwexp}.
In the following, we discuss other examples in order to make the link with
the trajectory observables of Eqs \ref{Obsdiff} \ref{Obsdiffempi} that one considers for diffusion processes.


\subsection{ Examples of trajectory observables associated to elementary contributions}


\subsubsection{ Observables ${\cal O}_0 $ associated to $\Omega^{(2)}\left( \vec x= \vec r + {\sqrt \tau } (1-\beta) \vec Z,  \vec y=\vec r - {\sqrt \tau } \beta \vec Z\right)
=\Omega_0(\vec r) $ that remain finite for $\tau \to 0$} 

For observables ${\cal O}_0 $ where there is only the function  $\Omega_0(\vec r) $ in Eq. \ref{observableDVz}
that remains finite  for $\tau \to 0$,
then Eq. \ref{observableQempiw} reduces to 
\begin{eqnarray}
 {\cal O}_0\left[\rho(.) \right]
 =  \int d^d \vec r O_0(\vec r) \rho ( \vec r ) +o(\sqrt \tau)
\label{itrajempitildeww0}
\end{eqnarray}
So in the continuous-time limit $\tau \to 0$,
these observables correspond to the form of Eqs \ref{Obsdiff} \ref{Obsdiffempi}
that depend only on the empirical density $\rho(.) $ with the correspondance $O_0(\vec r)=\omega(\vec r) $ as expected
\begin{eqnarray}
 {\cal O}_0\left[\rho(.) \right]= \int d^d \vec x   O_0( \vec x)\rho (\vec x) 
 = \frac{1}{T} \int_0^T dt O_0(\vec x(t)) = O_0^{Traj}[\vec x (0 \leq t \leq T) ]  
\label{Obsdiffempiw0}
\end{eqnarray}


\subsubsection{ Observables ${\cal O}_1 $ associated to $\Omega^{(2)}\left( \vec x= \vec r + {\sqrt \tau } (1-\beta) \vec Z,  \vec y=\vec r - {\sqrt \tau } \beta \vec Z\right)= \sum_{\mu_1=1}^d Z_{\mu_1} \frac{\lambda_{\mu_1}(\vec r)}{\sqrt \tau}  $ 
where $\lambda_{\mu_1}(\vec r) $ remains finite for $\tau \to 0$} 

Let us now consider observables ${\cal O}_1 $ where there is only the linear contribution in Eq. \ref{observableDVz}, where $\Omega_{\mu_1}(\vec r)=\frac{\lambda_{\mu_1}(\vec r)}{\sqrt \tau} $ with $\lambda_{\mu_1}(\vec r) $ remaining finite  for $\tau \to 0$
\begin{eqnarray}
\Omega^{(2)}\left( \vec x= \vec r + {\sqrt \tau } (1-\beta) \vec Z,  \vec y=\vec r - {\sqrt \tau } \beta \vec Z\right) 
=  \sum_{\mu_1=1}^d Z_{\mu_1} \frac{\lambda_{\mu_1}(\vec r)}{\sqrt \tau}
=  \sum_{\mu_1=1}^d \left( \frac{x_{\mu_1} - y_{\mu_1} }{\tau} \right) \lambda_{\mu_1}(\vec r)
\label{observableDVz1}
\end{eqnarray}
that involves the discrete-time analog $\left( \frac{ \vec x_ - \vec y }{\tau} \right) $ of the velocity $\frac{d  \vec x (t)}{dt} $.

Then Eq. \ref{observableQempiw} reduces to 
\begin{eqnarray}
&& {\cal O}_1\left[\rho(.) ; {\tilde {\cal A}}^{[\beta]}_.(.)  ;{\tilde{\bold D}}(.) \right]
=   \int d^d \vec r \sum_{\mu_1=1}^d \lambda_{\mu_1}(\vec r)
\int d^d \vec Z  
  \frac{  e^{ - \displaystyle  \sum_{\mu=1}^d \sum_{\nu=1}^d 
\frac{ Z_{\mu} \left[ {\tilde{\bold D}}^{-1}_{\mu \nu} (\vec r) \right] Z_{\nu} }{  4 }  }  }
 { \displaystyle \sqrt { (4 \pi )^d {\rm det } \left[ {\tilde{\bold D}} (\vec r) \right] }  } 
  \nonumber \\&& 
\left( \frac{Z_{\mu_1}}{\sqrt \tau}\rho ( \vec r )  
+  
\sum_{\sigma=1}^d Z_{\sigma} Z_{\mu_1}
\left( \rho ( \vec r ) {\tilde {\cal  A}}^{[\beta]}_{\sigma}\left( \vec r\right)
-  \beta  \frac{\partial \rho ( \vec r ) }{ \partial r_{\sigma}} \right)
 +\frac{o(\sqrt \tau)}{\sqrt \tau}
 \right)
  \nonumber \\
&&
=  \int d^d \vec r \sum_{\mu_1=1}^d \lambda_{\mu_1}(\vec r)
\left[ \rho ( \vec r ) \sum_{\sigma=1}^d 2 {\tilde{\bold D}}_{\sigma \mu_1}{\tilde {\cal  A}}^{[\beta]}_{\sigma}\left( \vec r\right)
-  2 \beta \sum_{\sigma=1}^d  {\tilde{\bold D}}_{\sigma \mu_1} \frac{\partial \rho ( \vec r ) }{ \partial r_{\sigma}} \right]
 +\frac{o(\sqrt \tau)}{\sqrt \tau}
\label{itrajempitildeww1cal}
\end{eqnarray}
The replacement of the empirical vector potential ${\tilde {\cal A}}^{[\beta]}_.(.) $
 in terms of the empirical force ${\tilde F}_{.} ( .) $ of Eq. \ref{FPexpandrho1tilde} yields
 \begin{eqnarray}
&& {\cal O}_1\left[\rho(.) ; {\tilde F}_.(.)  ;{\tilde{\bold D}}(.) \right]
=  \int d^d \vec r \sum_{\mu=1}^d \lambda_{\mu}(\vec r)
\left[ \rho ( \vec r ) \left( {\tilde F}_{\mu} ( \vec r) +(1-2 \beta)  
       \sum_{\sigma=1}^d   \frac{\partial  {\tilde{\bold D}}_{\mu \sigma}(\vec r)}{ \partial r_{\sigma}}\right)
-  2 \beta \sum_{\sigma=1}^d  {\tilde{\bold D}}_{\sigma \mu} \frac{\partial \rho ( \vec r ) }{ \partial r_{\sigma}} \right]
\nonumber \\
&& =  \int d^d \vec r \sum_{\mu=1}^d \lambda_{\mu}(\vec r)
\left[ \rho ( \vec r )  {\tilde F}_{\mu} ( \vec r) 
-   \sum_{\sigma=1}^d  {\tilde{\bold D}}_{\sigma \mu} \frac{\partial \rho ( \vec r ) }{ \partial r_{\sigma}}
+(1-2 \beta)  \sum_{\sigma=1}^d   \frac{\partial  }{ \partial r_{\sigma}}
\left(\rho ( \vec r ) {\tilde{\bold D}}_{\mu \sigma}(\vec r) \right)
\right]
\label{itrajempitildeww1}
\end{eqnarray}
For the mid-point discretization $\beta=\frac{1}{2}$, one recognizes the empirical current $j_{\mu}(\vec x) $ of Eq. \ref{defjtildeempi}
  \begin{eqnarray}
\text {For $\beta=\frac{1}{2}$ } :  {\cal O}_1\left[\rho(.) ; {\tilde F}_.(.)  ;{\tilde{\bold D}}(.) \right]
 =  \int d^d \vec r \sum_{\mu=1}^d \lambda_{\mu}(\vec r)
\left[ \rho ( \vec r )  {\tilde F}_{\mu} ( \vec r) 
-   \sum_{\sigma=1}^d  {\tilde{\bold D}}_{\sigma \mu} \frac{\partial \rho ( \vec r ) }{ \partial r_{\sigma}}
\right] =  \int d^d \vec r \sum_{\mu=1}^d \lambda_{\mu}(\vec r) j_{\mu} ( \vec r) 
\label{itrajempitildeww1demi}
\end{eqnarray}
In addition, the empirical 
 diffusion matrix $ {\tilde{\bold D}}(.)$ coincides with the true diffusion matrix ${\tilde{\bold D}}(.) $ at leading order
 of Eq. \ref{empidsqrt}.
 
 So in the continuous-time limit $\tau \to 0$,
these observables correspond to the form of Eqs \ref{Obsdiff} \ref{Obsdiffempi}
that depend only on the empirical current 
  \begin{eqnarray}
\text {For $\beta=\frac{1}{2}$ } :  {\cal O}_1\left[j_.(.)   \right]
  =  \int d^d \vec x \sum_{\mu=1}^d \lambda_{\mu}(\vec x) j_{\mu} ( \vec x) 
  = \frac{1}{T} \int_0^T dt  {\vec  \lambda}(\vec x) . \frac{d  \vec x (t)}{dt} = O^{Traj}_1[\vec x (0 \leq t \leq T) ]
\label{itrajempitildeww1demij}
\end{eqnarray}


\subsubsection{ Observables ${\cal O}_2 $ associated to $\Omega^{(2)}\left( \vec x= \vec r + {\sqrt \tau } (1-\beta) \vec Z,  \vec y=\vec r - {\sqrt \tau } \beta \vec Z\right)
=  \sum_{\mu_1=1}^d \sum_{\mu_2=1}^d Z_{\mu_1} Z_{\mu_2} \Omega_{\mu_1 \mu_2}(\vec r)$ with $\Omega_{\mu_1 \mu_2}(\vec r) $ remaining finite for $\tau \to 0$} 

Let us now consider observables ${\cal O}_2 $ where there is only the quadratic contribution in Eq. \ref{observableDVz}, where $ \Omega_{\mu_1 \mu_2}(\vec r) $ remains finite  for $\tau \to 0$
\begin{eqnarray}
\Omega^{(2)}_2\left( \vec x= \vec r + {\sqrt \tau } (1-\beta) \vec Z,  \vec y=\vec r - {\sqrt \tau } \beta \vec Z\right)  
&&=  \sum_{\mu_1=1}^d \sum_{\mu_2=1}^d Z_{\mu_1} Z_{\mu_2} \Omega_{\mu_1 \mu_2}(\vec r)
 \nonumber \\ 
 && 
=  \sum_{\mu_1=1}^d \sum_{\mu_2=1}^d \tau \left( \frac{x_{\mu_1} - y_{\mu_1} }{\tau} \right) 
\left( \frac{x_{\mu_2} - y_{\mu_2} }{\tau} \right) \Omega_{\mu_1 \mu_2}(\vec r) 
\label{observableDVz2}
\end{eqnarray}
that involves the discrete-time analog $\left( \frac{ \vec x_ - \vec y }{\tau} \right) $ of the velocity $\frac{d  \vec x (t)}{dt} $.

Then Eq. \ref{observableQempiw} reduces to 
 \begin{eqnarray}
&& {\cal O}_2\left[\rho(.)   ;{\tilde{\bold D}}(.) \right]
=   \int d^d \vec r \rho ( \vec r ) \sum_{\mu_1=1}^d \sum_{\mu_2=1}^d  \Omega_{\mu_1 \mu_2}(\vec r) 
\int d^d \vec Z  
  \frac{  e^{ - \displaystyle  \sum_{\mu=1}^d \sum_{\nu=1}^d 
\frac{ Z_{\mu} \left[ {\tilde{\bold D}}^{-1}_{\mu \nu} (\vec r) \right] Z_{\nu} }{  4 }  }  }
 { \displaystyle \sqrt { (4 \pi )^d {\rm det } \left[ {\tilde{\bold D}} (\vec r) \right] }  } Z_{\mu_1} Z_{\mu_2}
+o(\sqrt \tau)
  \nonumber \\
&&= \int d^d \vec r \rho ( \vec r ) \sum_{\mu_1=1}^d \sum_{\mu_2=1}^d  \Omega_{\mu_1 \mu_2}(\vec r) 
 2 {\tilde{\bold D}}_{\mu_1 \mu_2} (\vec r)
+o(\sqrt \tau)
\label{itrajempitildeww2cal}
\end{eqnarray}
 Since the empirical 
 diffusion matrix $ {\tilde{\bold D}}(.)$ coincides with the true diffusion matrix ${\tilde{\bold D}}(.) $ at leading order
 of Eq. \ref{empidsqrt}, one obtains that Eq. \ref{itrajempitildeww2}
 only depends on the empirical density $\rho(.)$ in the continuous-time limit $\tau to 0$
 \begin{eqnarray}
&&{\cal O}_2\left[\rho(.)  \right]
= \int d^d \vec r \rho ( \vec r ) \sum_{\mu_1=1}^d \sum_{\mu_2=1}^d  \Omega_{\mu_1 \mu_2}(\vec r) 
 2 {\bold D}_{\mu_1 \mu_2} (\vec r)
+o(\sqrt \tau)
\label{itrajempitildeww2}
\end{eqnarray}
and thus becomes analog to the observables of Eq. \ref{Obsdiffempiw0}.

 As example, let us consider the case $\Omega_{\mu_1 \mu_2}(\vec r) =\delta_{\mu_1,\mu_2} $ 
 where Eq. \ref{observableDVz2}
 reduces to
 \begin{eqnarray}
\Omega^{(2)}_2\left( \vec x= \vec r + {\sqrt \tau } (1-\beta) \vec Z,  \vec y=\vec r - {\sqrt \tau } \beta \vec Z\right) 
&& = \sum_{\mu=1}^d Z_{\mu}^2=  \sum_{\mu=1}^d   \frac{(x_{\mu} - y_{\mu})^2 }{\tau}  
\label{observableDVz2delta}
\end{eqnarray}
 and where Eq. \ref{itrajempitildeww2} 
  \begin{eqnarray}
&&{\cal O}_2\left[\rho(.)  \right]
= \int d^d \vec r \rho ( \vec r ) \sum_{\mu=1}^d  2 {\bold D}_{\mu \mu} (\vec r)
+o(\sqrt \tau)
\label{itrajempitildeww2ex}
\end{eqnarray}
 is the analog of Eq. \ref{observableAJempilatticerescaldis} obtained in the main text
 for the lattice model.
 
 This is an example of the so-called "substitution rules" (see the recent review \cite{Vivien2023} and references therein) :
 more generally, many observables of the Markov chain involving even powers of the variables $\vec Z$ 
will only depend on the empirical density $\rho(.)$ of the diffusion process in the limit $\tau \to 0$,
while many observables of the Markov chain involving odd powers of the variables $\vec Z$ 
will only depend on the empirical current $\vec j(\vec x)$ of the diffusion process in the limit $\tau \to 0$,
as described for a single power in Eq. \ref{itrajempitildeww1demi}.
These simplifications come from the Gaussian statistics of the variables $\vec Z$ 
governed by the diffusion matrix ${\bold D}(.)$ since the empirical diffusion matrix ${\tilde{ \bold D}}(.)$
cannot fluctuate anymore in the limit $\tau \to 0$.
Note however that for a general observable of Eq. \ref{observableDVz},
one needs to 
discuss how the functions $\Omega_0(\vec r) $, $\Omega_{\mu_1}(\vec r) $, $ \Omega_{\mu_1 \mu_2}(\vec r) $ ...
depend on $\tau$ in order to identify
the various terms that will survive in the limit $\tau \to 0$ after the Gaussian integration over the variables $\vec Z$
in Eq. \ref{itrajempitildeww}, and to obtain the singular and regular contributions in the limit $\tau \to 0$,
as described in detail for the example of the rate function at level 2.5 of Eq. \ref{rate2.5empitildewwexp}
in the previous Appendix.



\begin{thebibliography}{99}



\bibitem{oono}
Y. Oono,
Progress of Theoretical Physics Supplement 99, 165 (1989).

\bibitem{ellis}
R.S. Ellis, Physica D 133, 106 (1999).

\bibitem{review_touchette}
H. Touchette, Phys. Rep. 478, 1 (2009); \\
H. Touchette, Modern Computational Science 11: Lecture Notes from the 3rd International Oldenburg Summer School, BIS-Verlag der Carl von Ossietzky Universitat Oldenburg, (2011).


\bibitem{derrida-lecture}
B. Derrida, J. Stat. Mech. P07023 (2007).

\bibitem{harris_Schu}
R J Harris and G M Sch\"utz,
J. Stat. Mech.  P07020 (2007).

\bibitem{searles}
E.M. Sevick, R. Prabhakar, S. R. Williams and  D. J. Searles,
Ann. Rev. of Phys. Chem.  Vol 59, 603 (2008). 

\bibitem{harris}
H. Touchette and R.J. Harris, chapter "Large deviation approach to nonequilibrium systems"
of the book "Nonequilibrium Statistical Physics of Small Systems: Fluctuation Relations and Beyond", Wiley (2013).

\bibitem{mft}
L. Bertini, A. De Sole, D. Gabrielli, G. Jona-Lasinio and C. Landim
Rev. Mod. Phys. 87, 593 (2015).

\bibitem{sollich_review}
R. L. Jack and P. Sollich, The European Physical Journal Special Topics  224, 2351 (2015).

\bibitem{lazarescu_companion}
A. Lazarescu, J. Phys. A: Math. Theor. 48 503001 (2015).

\bibitem{lazarescu_generic}
A. Lazarescu, J. Phys. A: Math. Theor. 50 254004 (2017).

\bibitem{jack_review}
R. L. Jack, Eur. Phy. J. B  93, 74 (2020).


\bibitem{fortelle_thesis}
A. de La Fortelle, PhD Thesis (2000)
"Contributions to the theory of large deviations and applications" INRIA Rocquencourt.


\bibitem{vivien_thesis}
V. Lecomte, PhD Thesis (2007)
"Thermodynamique des histoires et fluctuations hors d'\'equilibre"
Universit\'e Paris 7.

\bibitem{chetrite_thesis}
R. Ch\'etrite, PhD Thesis (2008)
"Grandes d\'eviations et relations de fluctuation dans certains mod\`eles de syst\`emes
hors d'\'equilibre"  ENS Lyon.

\bibitem{wynants_thesis}
B. Wynants, arXiv:1011.4210, PhD Thesis (2010), "Structures of Nonequilibrium Fluctuations", Catholic University of Leuven.

\bibitem{chabane_thesis}
L. Chabane, PhD Thesis (2021) "From rarity to typicality : the improbable journey of a large deviation",
Universit\'e Paris-Saclay.

\bibitem{duBuisson_thesis}
J. du Buisson, PhD Thesis (2022) "Dynamical large deviations of diffusions" Stellenbosch University, South Africa
arXiv:2210.09040.

\bibitem{chetrite_HDR}
R. Ch\'etrite, HDR Thesis (2018)
"P\'er\'egrinations sur les ph\'enom\`enes al\'eatoires dans la nature",
 Laboratoire J.A. Dieudonn\'e, Universit\' e de Nice.





\bibitem{peliti}
C. Giardina, J. Kurchan and L. Peliti, Phys. Rev. Lett. 96, 120603 (2006).


\bibitem{lecomte_chaotic}
V. Lecomte, C. Appert-Rolland and F. van Wijland,
Phys. Rev. Lett. 95 010601 (2005).

\bibitem{lecomte_thermo}
V. Lecomte, C. Appert-Rolland and F. van Wijland,
J. Stat. Phys. 127 51-106 (2007).

\bibitem{lecomte_formalism}
V. Lecomte, C. Appert-Rolland and F. van Wijland,
Comptes Rendus Physique 8, 609 (2007).

\bibitem{lecomte_glass}
J.P. Garrahan, R.L. Jack, V. Lecomte, E. Pitard, K. van Duijvendijk and F. van Wijland,
Phys. Rev. Lett. 98, 195702 (2007).

\bibitem{kristina1}
J.P. Garrahan, R.L. Jack, V. Lecomte, E. Pitard, K. van Duijvendijk and F. van Wijland, 
J. Phys. A 42, 075007 (2009).

\bibitem{kristina2}
K. van Duijvendijk, R.L. Jack and F. van Wijland, 
Phys. Rev. E 81, 011110 (2010).

\bibitem{jack_ensemble}
R. L. Jack and P. Sollich, Prog. Theor. Phys. Supp. 184, 304 (2010).

\bibitem{simon1}
D. Simon, J. Stat. Mech. (2009) P07017.



\bibitem{simon2}
V. Popkov, G. M. Schuetz and D. Simon, J. Stat. Mech. P10007 (2010).

\bibitem{tailleur}
 C. Giardina, J. Kurchan, V. Lecomte and J. Tailleur, J. Stat. Phys. 145, 787 (2011).

\bibitem{simon3}
D. Simon, J. Stat. Phys. 142,  931 (2011).

\bibitem{Gunter1}
V. Popkov and G. M. Schuetz, J. Stat. Phys 142,  627 (2011).


\bibitem{Gunter2}
V. Belitsky and G. M. Schuetz, J. Stat. Phys. 152, 93 (2013).


\bibitem{Gunter3}
O. Hirschberg, D. Mukamel and G. M. Schuetz, J. Stat. Mech. P11023 (2015).

\bibitem{Gunter4}
G. M. Schuetz, From Particle Systems to Partial Differential Equations II, Springer Proceedings in Mathematics and Statistics Volume 129, pp 371-393, P. Gonçalves and A.J. Soares (Eds.), (Springer, Cham, 2015).


\bibitem{chetrite_canonical}
R. Ch\'etrite and H. Touchette,
Phys. Rev. Lett. 111, 120601 (2013).

\bibitem{chetrite_conditioned}
R. Ch\'etrite and H. Touchette
 Ann. Henri Poincare 16, 2005 (2015).

\bibitem{chetrite_optimal}
R. Ch\'etrite and H. Touchette, J. Stat. Mech. P12001 (2015).



\bibitem{touchette_circle}
P. T. Nyawo and H. Touchette, Phys. Rev. E 94, 032101 (2016).

\bibitem{touchette_langevin}
H. Touchette, Physica A 504, 5 (2018).

\bibitem{touchette_occ}
F. Angeletti and H. Touchette, Journal of Mathematical Physics 57, 023303 (2016).

\bibitem{touchette_occupation}
P. T. Nyawo and H. Touchette, Europhys. Lett. 116, 50009 (2016); \\
P. T. Nyawo and H. Touchette, Phys. Rev. E 98, 052103 (2018).

\bibitem{garrahan_lecture}
J.P. Garrahan, Physica A 504, 130 (2018).

\bibitem{c_ring}
C. Monthus, J. Stat. Mech. (2019) 023206.

\bibitem{c_detailed}
C. Monthus, J. Phys. A: Math. Theor. 52, 485001 (2019).

\bibitem{Vivo}
E. Roldan and P. Vivo, Phys. Rev. E 100, 042108 (2019).


\bibitem{chemical}
 A. Lazarescu, T. Cossetto, G. Falasco and  M. Esposito,
J. Chem. Phys. 151, 064117 (2019).



\bibitem{derrida-conditioned}
B. Derrida and T. Sadhu, Journal of Statistical Physics 176, 773 (2019); \\
B. Derrida and T. Sadhu, 
Journal of Statistical Physics 177, 151 (2019).

\bibitem{derrida-ring}
K. Proesmans and B. Derrida, J. Stat. Mech. (2019) 023201.

\bibitem{bertin-conditioned}
N. Tizon-Escamilla, V. Lecomte and E. Bertin, J. Stat. Mech. (2019) 013201.

\bibitem{touchette-reflected}
J. du Buisson and H. Touchette, Phys. Rev. E 102, 012148 (2020).

\bibitem{touchette-reflectedbis} 
E. Mallmin, J. du Buisson and H. Touchette,  J. Phys. A: Math. Theor. 54 295001 (2021).

\bibitem{c_lyapunov}
C. Monthus, J. Stat. Mech. (2021) 033303.

\bibitem{previousquantum2.5doob}
F. Carollo, J. P. Garrahan, I. Lesanovsky and C. Perez-Espigares, Phys. Rev. A 98, 010103 (2018).

 \bibitem{quantum2.5doob}
F. Carollo, R.L. Jack and J.P. Garrahan, Phys. Rev. Lett. 122, 130605 (2019).

 \bibitem{quantum2.5dooblong} 
F. Carollo, J.P. Garrahan and R.L. Jack, J. Stat. Phys. 184, 13 (2021).



\bibitem{c_ruelle}
C. Monthus, J. Stat. Mech. (2021) 063301.


\bibitem{lapolla}
A. Lapolla, D. Hartich and A. Godec, Phys. Rev. Research 2, 043084 (2020).

\bibitem{c_east}
C. Monthus, Eur. Phys. J. B 95, 32 (2022).

\bibitem{chabane}
L. Chabane, A. Lazarescu and G. Verley, Journal of Statistical Physics 187,  6 (2022).

\bibitem{us_gyrator}
A. Mazzolo and C. Monthus, Phys. Rev. E 107, 014101 (2023).

\bibitem{duBuisson_gyrator}
 J. du Buisson and H. Touchette, Phys. Rev. E 107, 054111 (2023).
 
 \bibitem{c_largedevpearson}
C. Monthus, J. Stat. Mech. (2023) 083204

 
 

\bibitem{Hollander}
 F. den Hollander, "Large deviations", American Mathematical Society, 2000.

\bibitem{fortelle_chain}
G. Fayolle and A. de La Fortelle,
Problems of Information Transmission 38, 354 (2002).

\bibitem{Polettini}
 M. Polettini, J. Phys. A: Math. Theor. 48 (2015) 365005.


\bibitem{c_largedevdisorder}
C. Monthus, Eur. Phys. J. B 92, 149 (2019) in the
topical issue " Recent Advances in the Theory of Disordered Systems"
edited by F. Igloi and H. Rieger.


\bibitem{c_reset}
C. Monthus, J. Stat. Mech. (2021) 033201.

 \bibitem{c_inference}
C. Monthus, J. Stat. Mech. (2021) 063211.

\bibitem{carugno}
G. Carugno, P. Vivo and F. Coghi, J. Phys. A: Math. Theor. 55 (2022) 295001.


 \bibitem{c_microcanoEnsembles}
 C. Monthus, Eur. Phys. J. B 95, 139 (2022).




\bibitem{fortelle_jump}
A. de La Fortelle, 
Problems of Information Transmission 37 , 120 (2001).



\bibitem{maes_canonical}
C. Maes and K. Netocny, Europhys. Lett. 82, 30003 (2008).

\bibitem{maes_onandbeyond}
C. Maes, K. Netocny and B. Wynants, Markov Proc. Rel. Fields. 14, 445 (2008).


\bibitem{chetrite_formal}
A. C. Barato and R. Ch\'etrite, J. Stat. Phys. 160, 1154 (2015).

\bibitem{BFG1}
L. Bertini, A. Faggionato and D. Gabrielli, 
Ann. Inst. Henri Poincare Prob. and Stat. 51, 867 (2015).

\bibitem{BFG2}
L. Bertini, A. Faggionato and D. Gabrielli, 
Stoch. Process. Appli. 125, 2786 (2015).


\bibitem{c_interactions}
C. Monthus, J. Phys. A: Math. Theor. 52, 135003 (2019).


\bibitem{c_open}
C. Monthus, J. Phys. A: Math. Theor. 52, 025001 (2019).

\bibitem{barato_periodic}
A. C. Barato, R. Ch\'etrite, J. Stat. Mech. (2018) 053207.

\bibitem{chetrite_periodic}
L. Chabane, R. Ch\'etrite, G. Verley, J. Stat. Mech. (2020) 033208.

\bibitem{c_LargeDevAbsorbing} 
C. Monthus, J. Stat. Mech. (2022) 013206.


\bibitem{c_susyboundarydriven}
C. Monthus, J. Stat. Mech. (2023) 063206.


\bibitem{maes_diffusion}
C. Maes, K. Netocny and B.  Wynants,
Physica A 387, 2675 (2008).


\bibitem{engel}
J. Hoppenau, D. Nickelsen and A. Engel,
 New J. Phys. 18 083010 (2016).
 



\bibitem{c_runandtumble}
C. Monthus, J. Stat. Mech. (2021) 083212.


\bibitem{c_jumpdrift}
C. Monthus, J. Stat. Mech. (2021) 083205


\bibitem{c_SkewDB} 
C. Monthus, J. Stat. Mech. (2021) 103202.


\bibitem{c_Inverse}
C. Monthus, arXiv:2308.12638.

\bibitem{c_SmallNoise}
C. Monthus, arXiv:2309.15542.



\bibitem{feynman}
R. P. Feynman and A. R. Hibbs,
``Quantum Mechanics and Path Integrals" McGraw-Hill, New-York (1965).

\bibitem{Morette}
C. Morette, Phys. Rev. 81, 848 (1951)

\bibitem{DeWitt}
B.S. DeWitt, Rev. Mod. Phys. 29 , 377 (1957).

\bibitem{GrahamPath}
R. Graham, Z. Phys. B Cond. Mat. 26, 281 (1977).

\bibitem{GrahamNoneq}
R. Graham,  Z. Phys. B Cond. Mat. 26, 397  (1977).

\bibitem{weiss}
U. Weiss, Z. Physik B 30, 429 (1978) 

\bibitem{langouche}
F. Langouche, D. Roekaerts and E. Tirapegui,
J. Phys. A: Math. Gen. 13,  449 (1980) 

\bibitem{dekker}
H. Dekker, Phys. Lett 76A, 8 (1980)

\bibitem{jarzynski}
V. Y. Chernyak, M. Chertkov and C. Jarzynski, J. Stat. Mech. (2006) P08001.

\bibitem{Vivien2017} 
L.F. Cugliandolo and V. Lecomte, J. Phys. A: Math.
Theor. 50 , 345001 (2017).

\bibitem{Vivien2019} 
L.F. Cugliandolo, V. Lecomte, and F. Van Wijland,  J. Phys. A: Math. Theor. 52 , 50LT01 (2019).

\bibitem{curved}
M. Ding and X. Xing, Quantum 6, 694 (2022)

\bibitem{Vivien2023} 
T. Arnoulx de Pirey, L. F. Cugliandolo, V. Lecomte, F. van Wijland, Advances in Physics (2023).



\end{thebibliography}
\end{document}